\documentclass[
preprint,
 amsmath,amssymb,
 aps,
]{revtex4-2}
\usepackage[utf8]{inputenc}
\usepackage[T1]{fontenc}
\usepackage[english]{babel}
\usepackage{graphicx}
\usepackage{dcolumn}
\usepackage{bm}
\usepackage{xcolor}
\usepackage{appendix}
\usepackage{amsmath}

\newcommand{\idel}{\tilde{\alpha}} 
\newcommand{\bsigxz}{\bar{\sigma}_{xz}}

\newcommand{\bq}{{\bf q}}
\newcommand{\bs}{{\bf s}}

\newcommand{\bfind}{\boldsymbol{\Lambda}}
\newcommand{\bu}{{\bf u}}
\newcommand{\be}{{\bf e}}
\newcommand{\br}{{\bf r}}
\newcommand{\req}{r^{(\text{eq})}}

\newcommand{\ii}{\mathrm{i}}
\newcommand{\ksp}{{\bf K}} 
\newcommand{\kspw}{\tilde{\boldsymbol{\Gamma}}^w} 
\newcommand{\kcom}{\tilde{\boldsymbol{\Gamma}}} 
\newcommand{\hmat}{\boldsymbol{\mathcal{H}}}
\newcommand{\kmat}{\boldsymbol{\mathcal{K}}} 

\newcommand{\mob}{\boldsymbol{\mathcal{M}}}  
\newcommand{\res}{\boldsymbol{\mathcal{R}}}  

\newcommand{\vwall}{v_{\text{wall}}}
\newcommand{\bvwall}{{\bf v}_{\text{wall}}}
\newcommand{\bx}{\hat{\bf x}}
\newcommand{\by}{\hat{\bf y}}
\newcommand{\bz}{\hat{\bf z}}

\newcommand{\bv}{\boldsymbol{v}}

\newcommand{\bF}{\boldsymbol{F}}

\newcommand{\vol}{\mathbb{V}}
\newcommand{\brr}{{\bf R}}
\newcommand{\bqq}{{\bf Q}}
\newcommand{\buu}{{\bf U}}
\newcommand{\bvp}{{\bf v}}

\begin{document}

\title{Fast spectral solver for viscoelastic structures under oscillatory flow in free space or wall-bounded domains: applications to quartz crystal microbalance and force spectroscopy}

\author{Pablo Palacios Alonso}
\affiliation{Departamento de Fisica de la Materia Condensada, Universidad Autonoma de Madrid, Campus de Cantoblanco, Madrid 28049
}
\author{Ra\'ul P\'erez Pel\'aez}
\affiliation{Departamento de Fisica de la Materia Condensada, Universidad Autonoma de Madrid, Campus de Cantoblanco, Madrid 28049}

\author{Rafael Delgado-Buscalioni$^*$}
\affiliation{Departamento de Fisica de la Materia Condensada, Universidad Autonoma de Madrid, Campus de Cantoblanco, Madrid 28049 
Condensed Matter Institute, IFIMAC\\
rafael.delgado@uam.es}


\begin{abstract}
We present a fast spectral solver for the linear response of viscoelastic structures under oscillatory flow either in free space or close to a flat moving wall. The scheme works in the frequency domain (using phasors) and couples the oscillatory Stokes equation with rigid or flexible structures, modeled by viscoelastic networks of immersed boundary kernels. The fluid-structure coupling can be solved by two routes. One route calculates the hydrodynamic mobility matrix required to solve the equation for the structure deformation rate in matrix form. The second route iteratively solves the coupled fluid-structure equations: fluid-induced forces on the structures create a tension field which is then transferred to the fluid, until convergence. The resulting fixed-point problem is solved iteratively using the Anderson acceleration method. The mobility route is optimal when dealing with one or few structures, while the iterative scheme is preferred for denser dispersions. In any case, the flow resulting from the body forces is solved by a recently developed scheme [J. Fluid. Mech. {\bf 1010} A57, 2025] which is spectral in space and time and deals with doubly periodic open domains (either free-space or wall-bounded) where meshing is restricted to the region of interest around the structures. We test the present scheme in two applied contexts: quartz-crystal-microbalance (QCM) of spheres, suspended, adsorbed or tethered to viscoelastic linkers; and force spectroscopy (via atomic force microscopy) reproducing the power spectra of vibrating microparticles near a solid boundary. In all cases, comparisons with analytical, numerical and experimental results show excellent agreement. We conclude by discussing new routes the scheme opens in force spectroscopy and QCM analyses of soft objects.
\end{abstract}

\maketitle


\section{\label{sec:intro} Introduction}

Forcing the oscillation of an unknown system is probably the oldest, most natural and accurate way humans discovered to unveil elastic and viscous responses.  This is the basis of many experimental techniques using all types of forcing fields to excite oscillations of soft matter in liquid, covering all scales from Angstroms (Raman spectra) up to microns. Above the atomic scales,  we can mention quartz crystal microbalance (QCM)  \cite{Johannsmann2021}, force spectroscopy by atomic force microscopy (AFM) (either in tapping mode\cite{AFMhydro_APL2008,CalzadoMartin2016} or in thermal environment \cite{Heenan2021}), alternating magnetic fields (AMF)  \cite{2021_covid_harmonics} including AC susceptibility \cite{2022_susceptibility_empirical_biosensing}), alternating electric fields (AEF) \cite{Aranda2008,Vlahovska2015}, electrochemical impedance \cite{2023Lanzanas} and more. Some techniques use non-linear couplings between the excited oscillatory fields (or variables) to create steady currents and net driving forces, like in ultrasound manipulation \cite{balboaUltrasound13,Bruus2012}, AC electrokinetics \cite{Ramos_2021}, dielectrophoresis \cite{Bruus2007},  optic-matter interaction\cite{huang2022primeval}; or to extract information from the hysteresis area \cite{2025_palacios,2024_Elena} or to create heat \cite{2017-liposomes,2011_Carrey_empiricalEQ_hyperthermia}.  Here we focus on experimental techniques based on linear response,  where the main goal is to evaluate the \emph{intrinsic} complex susceptibility $\boldsymbol{\chi}$ of the analyte. The susceptibility relates two conjugated variables: force (${\bf F}$) and response ${\bf u}=\boldsymbol{\chi}^{-1}\, {\bf F}$. Note that ${\bf G}=\boldsymbol{\chi}^{-1}$ is the Green function encoding the in-phase (elastic, conservative) and delayed (viscous,  frictional, dissipative) response of the material. To mention some examples, the tapping-mode AFM is used in a broad range of scales: from the identification and characterization of cancerous cells \cite{CalzadoMartin2016,Puerto-Belda2024} (apparently less rigid than healthy cells at low frequencies), to the cohesion energy of viruses \cite{MartinGonzalez2021} down to the activity of tethered enzymes \cite{Heenan2021,Patel2021} analyzed by AFM force spectroscopy. Also, QCM is broadly used to characterize all forms of soft matter: cells \cite{2020QCM_cells}, viruses \cite{Bingen2008,Feroz2017,Adamczyk2021} colloids \cite{Sadowska2020}, proteins \cite{Bingen2008}, membranes, polymers and more \cite{Johannsmann2015,BuscalioniSM21}. QCM has an outstanding sensitivity, being able, for instance, to distinguish signals from extremely alike "sister" proteins such as avidin, streptavidin and neutravidin on different substrates \cite{Wolny2010}.  Particularly in QCM and AFM, the experiment often involves strong sample-substrate interaction.  The presence of a wall breaks the spatial symmetry by introducing a reference direction.  In this sense, QCM and AFM perturb the sample in different directions and also in different frequency ranges. AFM vibrates in the KHz regime and usually in normal-to-wall direction, while QCM resonators oscillate parallel to their plane and in the MHz range. While these techniques work close to the wall (AFM) or adjacent to it (QCM), others (such as AMF or AEF \cite{Aranda2008}) permits to carry out oscillatory perturbations in the bulk free space \cite{2024_Elena}. In whatever case biological analytes and the majority of  soft matter experiments  take place in liquid environment (usually water). And this fact introduces a serious complication which is currently hampering a rigorous quantitative interpretation of many experimental signals in terms of the real or "intrinsic" molecular properties of the samples.  Paradigmatic cases are AFM vibrational spectra analyses and QCM, but the problem percolates to many other techniques such as AC electrokinetics \cite{Ramos_2021} and even optic-matter interaction \cite{louis2023unravelling}. Liquids present a viscous (i.e. delayed) response which spreads over long-distances, completely modifying the phase-lag between the force ${\bf F}$ and the sample's response ${\bu}$. In such case, the overall susceptibility mixes the intrinsic structure viscoelasticity \emph{and} the viscous fluid response. To understand the sample's intrinsic response, one first needs to solve the hydrodynamic contribution and "subtract"  it from the overall experimental signal.  Quantitative matching with the experiments will then allow extracting the analyte molecular information, via an integrated theoretical-experimental approach. Determining intrinsic molecular properties of bio-structures \emph{in liquid} is attracting a great deal of attention both, at molecular and coarse-grained simulations levels \cite{2025Monago}. This problem covers a large spatio-temporal range: force spectroscopy is used to analyze proteins  \cite{Patel2021} up to micron-size objects \cite{Puerto-Belda2024}, and similarly QCM covers ranges from few nanometers to microns \cite{Johannsmann2015} and the hydrodynamic contribution is usually dominant when dealing with quasi-neutrally buoyant objects \cite{Gillissen2018,BuscalioniSM21,Buscalioni_langmuir23,Leshansky2023,Leshansky2024}. 

\begin{figure}
   \centering
    \includegraphics[width=0.8\linewidth]{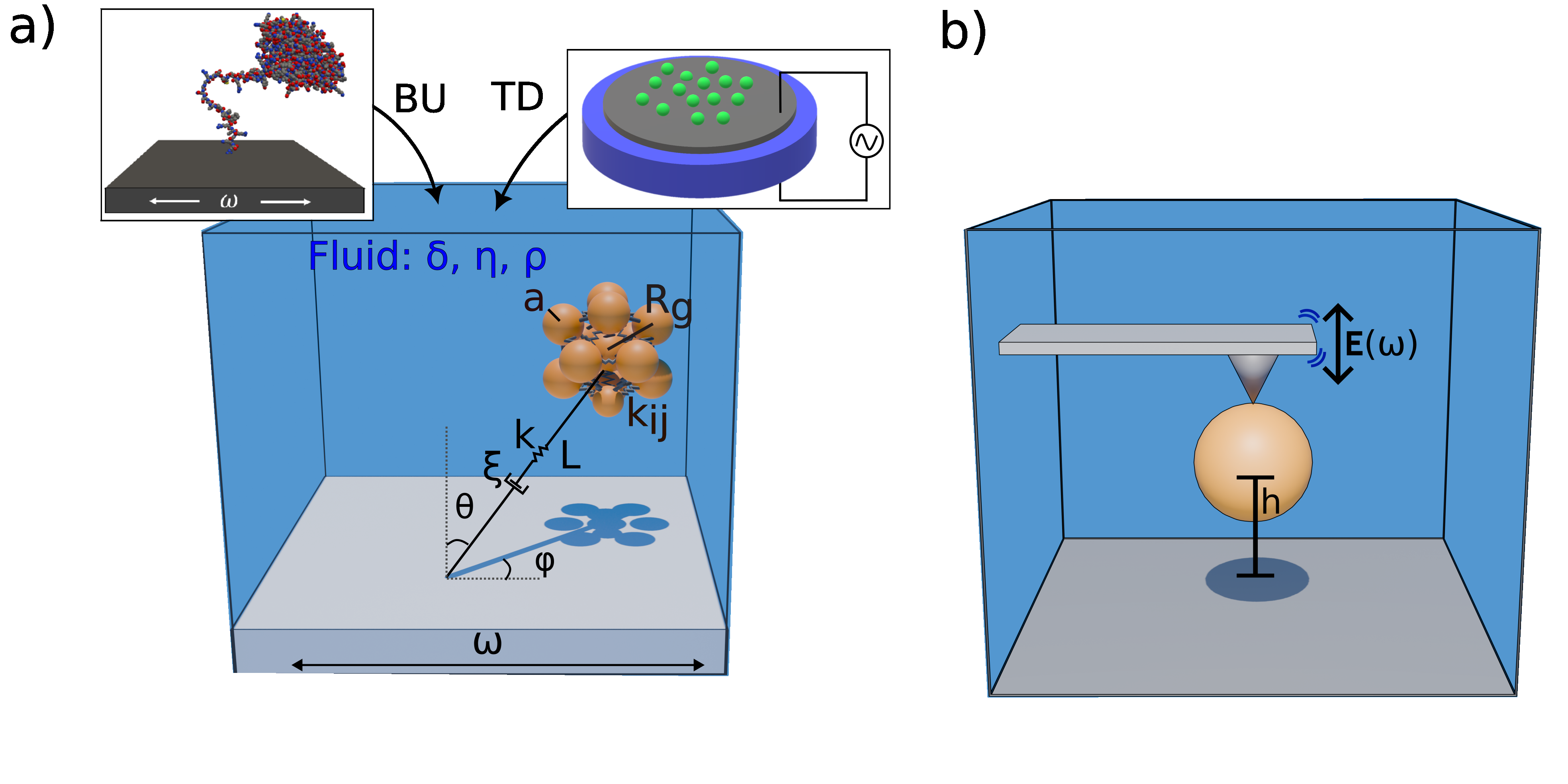}
\caption{Two types of experiments reproduced by the present solver. (Left) Quarzt crystal microbalance experiments and (right) vibrational spectra obtained from atomic force microscopy (AFM). The left panel schematically represent three different levels of description for the same system: a FtsZ protein having a flexible peptide chain along its intrinsically disordered region (IDR). (Top-left) The microscale, all-atom model (water molecules not shown and colors indicate partial charges of the residues) can provide information to the mesoscale model (center) by a Bottom-Up (BU)  formalism. The mesoscale level is based on a coarse-grained (CG) model with a globular part  (radius $R_g$) formed by a multiblob structure (12 blobs in the vertices and one in the center of an isoshedron). The later is connected to the highly flexible IDR, modelled by a viscoelastic spring . The CG model is coupled to the oscillatory solvent flow via immersed-boundary continuum fluid dynamics \cite{2025_spectral_solver}. The CG parameters are the intrisinc friction of the chain connected to the substrate $\xi$, the chain stiffness $k$ and average length $L$, the chain orientation angles, as well as the elastic network of springs between the beads forming the rigid globule. At the  macroscale (top-right corresponding to the QCM experiment)  a Top-Down (TD) approach provides connection with the meso-level, via quantitative reproduction of experimental signals (frequency and dissipation shifts, directly related to the complex-phasor of the averaged excess surface tangential stress $\bar{\sigma}_{xz}$).  In the QCM device, oscillations of the quartz crystal substrate are excited by an oscillatory voltage, via the inverse-piezo-electric effect, leading to the substrate horizontal oscillation at angular frequency $\omega_n \in 10 \pi \, n \, 10^{6}\, \mathrm{rad/s}$ and nanometer amplitude ($n=\left\{1,3,5,..\right\}$ is called the \emph{overtone})}.
   \label{fig:levels}
\end{figure}

Quite often, a lack of first-principles connection between theory and experiments has led to the development of ad hoc phenomenological models with little connection with real physicochemical quantities. Models built from first-principles have quantitative predictive power, and ideally they open new routes to indirect measurements of molecular properties otherwise "hidden" to the eyes of the raw experimental signals.  This is the idea of concepts such as "integrated theory-experiment", "deep-science" or "digital twins".  In this quest, the interpretation of vibrational spectra in force spectroscopy of micro and nanostructures in liquids is still far from being solved \cite{Puerto-Belda2024}.  Advancements in AMF allow for predictive experiment-theory coupling \cite{2023_Yoshida_simultaneous,2025_palacios}. In the case of QCM, recent efforts have started to reveal the response of rigid samples \cite{Buscalioni_langmuir23,Leshansky2023,Leshansky2024}, yet soft viscoelastic analytes present a particularly difficult challenge. The \emph{intrinsic} viscoelasticity of molecular structures in liquid is still a relatively unexplored field \cite{2025Monago}. And at this point it is important to recall that the concept of dissipation (and thus susceptibility) is intimately related to the coarse-grained (CG) nature of the chosen observable, i.e., to the level of description \cite{Zwanzig1961,Hijon_2010}. Importantly, any given experiment measures the response (or susceptibility) of some CG variable, such as the net wall stress in QCM against the substrate velocity, the normal-to-wall (z) height of the whole analyte in AFM against the applied force; the total magnetic moment in AMF or the electric dipole in AEF against the applied fields, etc.  Compared with simulations, these experiments sample a high-level CG variable, containing the least detailed information with the highest entropy but, importantly, from the \emph{natural} system.  Having a clear scientific question allows us to determine the "relevant" scale one needs to focus on.  The integration of theory and simulations permits us to pose questions at finer scales than the experiment sampling. For example, we would like to measure the stress response of the analyte-substrate contact, or its Young modulus, or its affinity to some substrate, or the precise size of the magnetic particle corona \cite{2025_palacios}, etc.  And to ensure cause-consequence, first-principles simulations need to deploy CG building blocks, which should be \emph{finer} than the scale the question demands. For example, contact details must be resolved to capture the adhesion energy of some virus \cite{GarciaArribas2024},  describe a vesicle using a viscoelastic network to evaluate its Young modulus or its intrinsic viscosity, or model the viscoelasticity of a peptide chain to interpret the spatial configuration of a tethered protein (Fig. \ref{fig:levels}).  First-principle mesoscopic models can be connected with finer-scale descriptions such as all-atom molecular dynamics (MD) \cite{Hijon_2010}. This hierarchy in description levels, covering what we call "micro" (e.g. MD), "meso" (CG), and "macro-level" (experiment) is illustrated in Fig. \ref{fig:levels}. In this work, we focus on the meso-level, which is essential to bridge information from the experiment to the scientific question,  and, ideally, with the micro-scale, by connecting with all-atom simulations via dynamic coarse-graining techniques  \cite{2025Monago,Hijon_2010,2004Voth}.  While the micro-meso link is beyond the present contribution, we will sketch the route.

In the spirit of the mesoscale,  we use a hybrid Eulerian-Lagrangian scheme which directly deals with the frequency-based description. One of the benefits of not using time-stepping is to avoid solving long-transient times to reach the limit cycle where measurements are made. A key operation in the Eulerian-Lagrangian communication consists of spreading the forces induced by the immersed structure to the fluid. To this end, we use a novel fully-spectral solver for the oscillatory Stokes hydrodynamics in doubly-periodic (open) boundaries \cite{2025_spectral_solver}.  Similarly to a previous steady Stokes version \citep{2023aleksdp}, it uses Fast Fourier transforms (FFTs) in the $xy$ plane and Chebyshev transforms in the aperiodic $z$ direction. Aside from being spectrally accurate in \emph{time and space}, the whole scheme has been implemented for graphical processor units (GPU) using a single 3D-FFT for each Fourier–Chebyshev transformation \citep{Thesis-Pelaez,2023aleksdp}.  Importantly, the computational mesh is restricted to a finite relevant domain $z\in [-H,H]$ where forces are applied, being connected to the analytical solution of the exterior flow using plane wave expansion. This allows us to treat dynamics in the free space, or bounded by a wall at $z=-H$. Overall, we have estimated (see Ref. \cite{2025_spectral_solver} and hereafter) that the open-boundary, spectral, frequency-based scheme typically brings about 100-fold speed-up with respect to a standard immersed boundary scheme (time-based and in closed domain  \cite{Prapp2020}). In doing so, we pay the price of lacking non-linear terms in the present description. This is a fair price, as many techniques, such as QCM, AFM, AMF, quite often work in the linear response. Moreover, a solver strictly in the linear regime avoids the need to control the forcing amplitude when evaluating hydrodynamic mobilities and structural viscoelastic kernels. In particular, here we directly access their Fourier components, but their time-dependency can be easily obtained by inverse transforms. Furthermore,  the linear solver might be used to perform numerical asymptotic expansions for non-linear interactions, which stands as a promising generalization route.

Section  \ref{sec:theo}  starts by presenting the theoretical framework of the present scheme for fluid-structure coupling. The computational methodology, in Section \ref{sec:method}, proposes two possible routes: mobility evaluation and iterative coupling. Section \ref{sec:qcm} applies the iterative coupling framework to QCM, first presenting the theoretical background and then several validation tests involving rigid nanoparticles in the hundred-nanometer range and analytical theory for viscoelastic linkers, with application to protein experiments. Then, in Section \ref{sec:afm}  we use the mobility route to analyze AFM force spectroscopy, recalling the theoretical basis for thermal vibrational spectra and then using the mobility solver to reproduce recent experiments \cite{AFM_spectra} with $27$ micron-sized particles vibrating near a boundary.  The Appendices provide a detailed description of the optimal placement of the Lagrangian markers, the adaptation of the Anderson iterative method to the present phasor equations, and numerical convergence.  Concluding remarks \ref{sec:con} enumerate some of the new research opportunities this scheme opens and lines of improvement.

\section{Theoretical framework \label{sec:theo}}

Here we focus on describing the viscoelastic stress of structures immersed in a liquid under an oscillatory perturbation with angular frequency $\omega$. We will use a Lagrangian formulation for the structures, described by viscoelastic networks of discrete beads, and an Eulerian treatment of the fluid, which is considered Newtonian, incompressible, and isothermal. The coarse-grained (CG) description of the discrete phase (solutes) consists on the position $\br_i$ and velocities $\bu_i$ of  $i\in\left\{1,...,N\right\}$ beads, while the fluid, of fixed density $\rho$, is fully described by its velocity  $\bv(\br)$ and pressure $p(\br)$ fields.  Bottom-up derivations for the coupled equation of motion for immersed structures starting from the atomic level (Liouville equation) can be found in several works \cite{2015PepAleks,2019Camargo}. The micro-meso connection of friction kernels is well known in the case of melts \cite{Hijon_2010,Li2014}( i.e. no hydrodynamic solvent) using bottom-up derivations based on irreversible thermodynamics \cite{Zwanzig1961} and Mori-Zwanzig projection operators  \cite{Grabert1982,Zwanzig2001,Hijon_2010} and linear response theory \cite{Zwanzig2001}.  Using this background, new theoretical approaches are considering CG models of liquid-immersed structures including internal and hydrodynamic dissipation \cite{2025Monago}.  Our approach will directly start by introducing the relevant CG equations at the mesoscopic level, whose elements (transport coefficients) are, in any case, compatible and computable from the known micro-meso links, both for conservative (Hessian matrix) and dissipative (Green-Kubo) parts of the susceptibility.  

We consider Newton's equation for a bead $i$ of mass $m_i$ and velocity $\bu_i$,
\begin{equation}
\label{eq:mi}
    m_i \dot \bu_i = \bF^{s}_i + \bF^{f}_i + \bF^{e}_i
\end{equation}
where, without loss of generality, the force is decomposed in the internal (structural) forces  $\bF^{s}$, the drag force due to the fluid $\bF^f$, and external forces $\bF^e$ due to external fields and structure-substrate interaction in the form of contact forces,  molecular linkers or tethers. 

The fluid phase might fill the free space (no walls) or be semi-bounded by a rigid wall at $z=-H$. Here we consider the no-slip boundary condition (BC) at the wall, so $\bv(z=-H)=\bvwall$ with wall velocity $\bvwall=\vwall\,\bx$ parallel to the plane. More generally, the fluid BC's could be generalized by imposing an arbitrary linear operator $\mathcal{B}[\bv(z=-H)]=0$ and also by adding another wall at some height $z=H$.  To solve the fluid dynamics, we follow the induced-force formalism \cite{Mazur1974} (which is the basis of the immersed boundary (IB) method \cite{2002_Peskin_IBM}) and treat the discrete phase boundaries as explicit source terms in the equation for the fluid momentum density,  
\begin{equation}
\rho \partial_t \bv = -\nabla \cdot \boldsymbol{\sigma} + \sum_i  S(\br -\br_i) \bfind_i,
\label{eq:fl}
\end{equation}
where the stress tensor $\boldsymbol{\sigma}$ is defined coordinate-wise ($\partial_\alpha \equiv \partial/\partial x_\alpha$) in terms of the pressure $p$, the dynamic viscosity $\eta$, and the fluid velocity field $\bv$ as
\begin{equation}
\sigma_{\alpha \beta} = p \,\delta_{\alpha \beta} - \eta (\partial_\alpha v_\beta + \partial_\beta  v_\alpha).
\label{eq:sigma}
\end{equation}
and $S(\br)$ is a regularized delta which corresponds to a distribution (units of inverse volume) of compact support (Peskin or Gaussian \cite{2025_spectral_solver,uammd}). This kernel $S$ allows to \emph{spread} the force $\bfind_i$ induced by the bead $i$ to the fluid. As in the induced-force formalism \cite{Mazur1974}, this force ensures that the kinetic condition is satisfied, i.e. $J_i[\bv] \equiv \int S(r-\br_i) \bv(\br) d \br^3 = \bu_i$, where the $J$ operator is the  \emph{interpolator}.  Applying $J$ to the momentum equation \ref{eq:fl} this condition implies that
\begin{equation}
\label{eq:mif}
m_i^f \dot\bu_i = \bF^{f}_i + \bfind_i 
\end{equation}
where we have used some key relations in the IB method, namely that for non-overlapping kernels $J_i[S_j]=\delta_{ij}/\vol_i$, where $\vol_i$ is the nominal volume of the bead, such that the local drag force equals $\bF^f_i = -\vol_i J_i[\nabla \cdot \boldsymbol{\sigma}]$ and the fluid mass occupied by the bead is $m_i^f = \rho \vol_i$. Kernel overlappings would introduce two complications: a modification of the mobility matrix and of the blob densities (the particle density becoming a tensor field \cite{vazquez2014multiblob}). The condition of non-overlapping kernels imposes a restriction for the building protocol of the structure, whose beads should be disposed at a certain optimal distance \cite{broms2024accuracy,anna-broms,balboa-blaise} (details provided below). At a more technical level, we note that the interpolator is calculated from a quadrature summation $J[v]= \sum_\xi v(\br_\xi) S(\br_\xi- \br_i)$ where $\br_\xi$ denotes the nodes of the Eulerian mesh. Using a separable form $S(\br) = \prod_\alpha S^P(r_\alpha)$, the Peskin kernels $S^P$ are built in such a way that $\vol_i$ is independent of the location of the bead in the Eulerian mesh. By contrast, Gaussian kernels present small variations on the volume $\vol_i$ over the lattice, yet they allow for higher accuracies at some computational cost \cite{2025_spectral_solver}. Subtracting Eq. \ref{eq:mi} and Eq. \ref{eq:mif} leads to the sought particle-fluid induced force,
\begin{equation}
\label{eq:l}
 \bfind_i= \bF^s_i +\bF^e -m_i^e \dot \bu_i
\end{equation}
where $m^e_i=m_i-m_i^f$ is the excess mass of the bead ($m_i^e=0$ for neutrally buoyant particles). At this point, we connect the structure and the fluid dynamics and ensure that the total momentum of the system is conserved (see Eq. \ref{eq:fl}). As we shall soon show, the fluid-structure coupling requires solving the fluid velocity field $\bv(\br)$ arising from the application of the induced force density field ${\bf f}(\br)\equiv \sum_i S(\br-\br_i) \bfind_i$. This essential part of the present scheme is performed by our frequency-based spectral solver for the oscillatory Stokes equation \cite{2025_spectral_solver}, which allows focusing on the relevant slice of fluid surrounding the structure by  (working with open boundaries) in free space or a semi-infinite domain bounded by a solid plane.
The kinetic condition $J_i[\bv]=\bu_i$ directly provides the first "hydrodynamic route" for the structure dynamics.  At a formal level, the unsteady Stokes equation \ref{eq:fl} can be written as $\mathcal{L}[\bv]={\bf f}(\br)$ and the flow can be decomposed as $\bv(\br,t)=\bv_0(\br,t)+\bvp(\br,t)$, where the basal flow $\bv_0$ satisfies the homogeneous equation $\mathcal{L}[\bv_0]=0$ with inhomogeneous BC's,  $\bv_0(z=0)=\bvwall$ while the perturbative flow $\bvp$ solves $\mathcal{L}[\bvp]={\bf f}(\br)$  with homogeneous BC's, $\bvp(z=0)=0$.  The formal solution for the total flow is $\bv(\br) = \mathcal{L}^{-1} S_j \bfind_j$ (using Einstein notation for repeated indices), and the beads' velocity is thus,
\begin{equation}
    \label{eq:u1}
    \bu_i = \bu_i^{(0)}+J_i \mathcal{L}^{-1} S_j \bfind_j = \bu_i^{(0)}+ \mob_{ij} \bfind_j
\end{equation}
where $\bu_i^{(0)}=J_i[\bv_0]$ is the basal flow velocity interpolated at the $i$ bead and $\mob=J\mathcal{L}^{-1}S$ is the bead mobility matrix which spreads the induced forces creating the perturbative flow.

To close the mechanical problem, we need to use Eq. \ref{eq:l} to connect the induced forces $\bfind$ with the forces supported by the viscoelastic structure.  As shown below, we consider a monochromatic oscillation which implicitly assumes that non-linear terms are negligible. Thus we can use linear viscoelasticity to relate, in matrix form, the structure forces with the small displacements of the network of beads about a relevant configuration, such as their mechanical equilibrium positions. In particular, we write the beads positions as $\br_i(t)= \br_i^{(\text{ref})} + \bq_i(t)$ where $\bq_i(t)$ is a small displacement about some referential $\br_i^{(\text{ref})}$ configuration. A sound choice is  $ \br_i^{(\text{ref})}=\br_i^{(\text{eq})}$, i.e. the mechanical equilibrium configuration at which the net force vanishes on each node vanish $\bF_i^s(\left\{\br^{(\text{eq})}\right\})=0$. To alleviate notation, we use capital letters for the supervectors containing all the beads' positions (e.g $\brr=\left\{\br_1,\dots,\br_N\right\}$) and relate these concepts with the microscopic level. Quite generally, the CG variables are functions of the microscopic degrees of freedom $\bf z$ (atom positions) i.e. $\brr\rightarrow \hat{\brr}({\bf z)}$. The CG free energy $\mathcal{F}(\brr)$  yields their equilibrium distribution $\rho_{mes}(\brr)\propto \exp[-\mathcal{F}(\brr)/k_BT]$ and the conservative part of the CG structural forces $F_{i\alpha}^{sc}=-\partial_{i\alpha} \mathcal{F}$.
Following the micro-meso statistical physics link, this free energy is the constrained average over the microensemble distribution $\rho_{mic}({\bf z})$ \cite{Hijon_2010,2025Monago}, i.e. $\mathcal{F}(\brr)= -k_B T  \ln \langle \delta(\hat\brr({\bf z}) -\brr)\rangle_{mic}$ with $\langle a \rangle_{mic}= \int a({\bf z}) \rho_{mic}({\bf z}) d {\bf z}$. The bead positions at mechanical equilibrium are those providing zero net force, or $\partial_{\brr} \mathcal{F}(\brr^{(\text{eq})})=0$. Hence for small displacements $\bqq$ about equilibrium we approximate  $\mathcal{F}(\brr)$  to its quadratic form $\mathcal{F}(\bqq) = (1/2) \bqq^{T} \kmat \bqq$ (up to additive constant) with the Hessian matrix $\kmat = \nabla_{\brr} \nabla_{\brr} \mathcal{F}|_{\brr=\brr^{e}}$. In explicit form (using Einstein summation notation) $\mathcal{F}(\bqq) = (1/2)\, \mathcal{K}_{i\alpha,j\beta} \, q_{i\alpha} q_{j\beta}$ and
$\mathcal{K}_{ij}^{\alpha \beta} = \partial_{i \alpha} \partial_{j \beta} \mathcal{F}$. Starting from the microscopic level (bottom-up route), the CG Hessian matrix can be obtained from the covariance matrix of the CG bead displacements $\mathcal{C}_{i\alpha,j\beta} \equiv \langle q_{i\alpha}(z) q_{j\beta}(z)\rangle_{mic}$, using the equipartition theorem  $\kmat =k_BT \,\mathcal{C}^{-1} $. The complementary (top-down) route consists of introducing a mesoscopic model with CG properties (such as elastic, bending, dihedral energies) consistent with or amenable to continuum elasticity  (e.g. in the case of vesicles or lipidic viruses \cite{GarciaArribas2024} with the Helfrich model  \cite{Helfrich1973,OuYang1987}).  Here we consider a linear elastic structure described by a network of harmonic springs with pairwise forces $\bF_{ij}^{sc}= -k_{ij} \left(r_{ij}-d_{ij}\right) \,\be_{ij}$ where $\br_{ij}=\br_j-\br_i$ and $\be_{ij}=\br_{ij}/r_{ij}$ is the unit vector joining two nodes of the network. The equilibrium distance for the $ij$ contact is $d_{ij}$ and its stiffness is $k_{ij}$ (real number). Expanding in Taylor series around the equilibrium configuration (${\bf F}^s({\bf r}_{eq})=0$) we derive a linear relation between the conservative part of the structural forces ${\bf F}^{sc}$ and the relative displacement vectors ${\bf q}_{ij}={\bf q}_j-{\bf q}_i$,
\begin{equation}
    \label{eq:fij}
    {\bf F}^{sc}_{ij} =-\ksp_{ij} \,\bq_{ij} +O(q^2)
   \end{equation}
with
\begin{equation}
    \label{eq:k2}
    \ksp_{ij}= k_{ij}\left[\left(1-\frac{d_{ij}}{\req_{ij}}\right) {\bf I}_3 + \frac{d_{ij}}{\req_{ij}}\,\hat{\be}_{ij} \hat{\be}^{T}_{ij} \right]
\end{equation}
While Eq. \ref{eq:k2} depends on relative displacements ${\bf q}_{ij}$ , our computational framework requires a  expression in terms of the absolute displacement vectors ${\bf q}_i$. To make such change of variables, we note that the total conservative structural force on bead $i$ is given by ${\bf F}^{sc}_{i}= \sum_{j\neq i} {\bf F}^{sc}_{ij}$, so, using Eq. \ref{eq:k2}, this force can be expressed as 
$${\bF_i^{sc}} = -\sum_j \kmat_{ij} \bq_j$$
with the elastic matrix (note that $\ksp_{ii}={\bf 0}$),

\begin{equation}
    \kmat_{ij} =  \ksp_{ij}-\sum_{i^{\prime} \ne i} \ksp_{ii^{\prime}} \delta_{ij} 
    \label{eq:kmatrix}
\end{equation}
Let us return to Eq. \ref{eq:k2} to briefly discuss possible scenarios. Note that if $d_{ij} \ne r_{ij}^{(\text{eq})}$ a structure will be pre-stressed even when in mechanical equilibrium: this is the case of vesicles formed by a membrane whose natural curvature radius differs from the vesicle radius \cite{Helfrich1973,OuYang1987,GarciaArribas2024}.  
Also, by modifying the center of the Taylor expansion, one can consider vibrations around an initial configuration under tension  $\brr^{(\text{ref})}=\brr^{(0)}$ due to a fixed external field deforming the structure. To that end simply substitute $\br_{ij}^{(0)} \rightarrow \br_{ij}^{(\text{eq})}$ (and versors $\hat{\be}$) in Eq. \ref{eq:k2}.

We now consider the internal (structural) dissipative forces, which can be generally expressed by a memory kernel. We use a pairwise approximation \cite{Hijon_2010} such that the internal frictional force in bead $i$ due to another node $j$ is
\begin{equation}
    \label{eq:fsd}
    {\bF_{ij}^{sd}(t)= -\int_0^t \boldsymbol{\xi}_{ij}(t-s) \, \bu_{ij}(s) \,ds},
\end{equation}
where the friction kernel $\boldsymbol{\xi}_{ij}$ is a $3\times 3$ matrix \cite{Hijon_2010}. Following the bottom-up route, this kernel can be extracted from the micro-model (MD)  by computing the time correlation between the bead-bead force fluctuations $\xi^{\alpha \beta}_{ij}(t) = \langle \tilde{F}_{i\alpha}(t) \tilde{F}_{j\beta}(0)\rangle_{mic}/k_BT$. Note that  $\boldsymbol{\xi}_{ij}\rightarrow 0$  for $t\rightarrow \infty$ and if the perturbation frequency $\omega$ is slow enough one can use the Markovian approximation $k_BT \xi_{ij}^{\alpha \beta}= \langle \tilde{F}_{i\alpha} \tilde{F}_{j\beta}\rangle \, \delta(t)$ leading to the Green-Kubo expressions $\bar{\boldsymbol{\xi}}_{ij}= \int_0^{\tau} \boldsymbol{\xi}_{ij}(t) dt$ (computed with caution due to the plateau problem \cite{Hijon_2010}).  

Yet, memory does not introduce any computational nuisance when working in the frequency domain as we do here. We consider an oscillatory flow of frequency $\omega$ and take the time Fourier transform in Eqs. \ref{eq:mi}and Eq. \ref{eq:fl}.  Any quantity $\phi(t)$ can be then mapped to a complex number (phasor) $\hat{\phi}$, i.e. $\phi(t) = \mathrm{Re}[\hat{\phi} \, \exp[-\ii \omega t]]$. Unless explicitly stated (if time dependence is indicated), for the sake of simplicity, we will drop the "hat" in the phasor quantities.  For instance, in phasor terms, bead displacements and velocities are related by $\bu_i = -\ii \omega \,\bq_i $ or $\bq_i= \ii \, \bu_i/\omega$. Using the convolution theorem, the dissipative force arising from the memory kernel in Eq. \ref{eq:fsd} becomes greatly simplified,
\begin{equation}
    \label{eq:fsd2}
    {\bF^{sd}_{ij}(\omega)=-\boldsymbol{\xi}_{ij}(\omega) \bu_{ij}(\omega)}
\end{equation}
with $\boldsymbol{\xi}_{ij}(\omega)=(1/k_BT)\,\int_0^{\infty} \langle \tilde{\bF}_i(t) \tilde{\bF}_j(0)\rangle_{mic} \exp[-\ii \omega t] dt$.  This unifies the structural viscoelastic force $\bF^{s}_i=\bF^{sc}_i+\bF^{sd}_i$ in the linear response regime, ${\bF}^{s}= -\kcom \buu$ where $\buu$ is the supervector of beads velocities $\buu=\left\{\bu_i\right\}$ and the complex viscoelastic matrix is defined as $\kcom= \ii \kmat/\omega + \boldsymbol{\Gamma}$ with $\boldsymbol{\Gamma}_{ij} = \boldsymbol{\xi}_{ij}-\sum_{i^{\prime} \ne i} \boldsymbol{\xi}_{ii^{\prime}}\, \delta_{ij} $ (note that $\xi_{ii}={\bf 0}$). 
The external forces could be due to some oscillatory field acting on each bead (written as ${\bf F}_i^e = {\bf E}_i$) or to contact or linker forces.  A link between one bead and the substrate is also treated in a viscoelastic fashion, i.e. ${\bF}_i^e = -\kspw_{i} \left(\bu_i -\bvwall\right)$, where $\kspw_{i}$ is a $3\times 3$ complex matrix which differs from zero if the bead $i$ shares some contact with the substrate, either touching or via some molecular linker (e.g. an oligomer chain).

Taking the time Fourier transforms in Eq. \ref{eq:l}, one readily derives the second relevant equation from the "structure" dynamics,
\begin{eqnarray}
\label{eq:u2}
\bfind_i  &=& \hmat_{ij} \bu_j+\kspw_{i}\, \bvwall+{\bf E}_i\\
 \hmat_{ij} &\equiv& \ii \, m_i^e \omega \,{\bf I}_3 \delta_{ij} - \kspw_{j} \,\delta_{ij} -\kcom_{ij}  
\end{eqnarray}

\section{\label{sec:method} Methods}

Equations \ref{eq:u1} and \ref{eq:u2} constitute a system with two unknowns $\buu$ and $\bfind$.  As we shall soon describe, this set of equations can be solved in two different ways: iteratively, using a fixed point approach, or by direct evaluation of the mobility matrix $\mob$ followed by linear-solve matrix operations. In whatever case, a key step is the evaluation of the perturbative flow (which involves the mobility matrix $\mob$), which we carry out by a novel hydrodynamic solver \cite{2025_spectral_solver}.
We have recently developed a fast, spectral-accurate, frequency-based solver \cite{2025_spectral_solver} for the oscillatory Stokes equation in doubly periodic setups (periodic $x$ and $y$ directions) and aperiodic normal coordinate $z$. The solver is spectral in both space and time. It is frequency-based, so it solves the Stokes equations for the complex phasor fields (flow velocity and pressure). In the reciprocal space, it uses Fourier transforms in $x$ and $y$ directions and the discrete Chebyshev transform in the normal (aperiodic) direction $z$. In both cases, computations are based on Fast Transforms (Fourier and Chebyshev \cite{2025_spectral_solver}): through careful bookkeeping and manipulation, the solver only requires the application of a single 3D FFT call to carry out the Fourier transform in the $x$ and $y$ directions plus the Chebyshev transform in the $z$ direction (see Refs. \citep{Maxian2021} and in \cite{Thesis-Pelaez} for details on FFT).  The computational domain (mesh) is restricted to a slice $z\in [-H,H]$ where all the source terms (forces) are located. One can solve the free space problem by coupling the exterior flow via a plane wave expansion, or place a rigid no-slip wall at $z=-H$ with arbitrary velocity (semi-bounded domain). To fulfill the boundary conditions, we superpose a correction flow over the free (unbounded) solution, and derive the corresponding Dirichlet-to-Neumann maps at the boundaries. The zero-wavelength contribution is treated separately using a Green function formulation. We refer to Ref. \cite{2025_spectral_solver} for all details and the numerous tests involving single blobs described by either Gaussian or Peskin kernels. Being frequency-based, the scheme avoids time-stepping and, importantly is adapted to open systems, as it allows to focus on the relevant domain where the analytes reside. This is particularly important for QCM, where the base flow is present up to heights which are several times the penetration length $\delta\sim 100 \text{nm}$, while a protein domain close to the substrate might be localized in a much smaller domain, of about 5 to 10 nm.  
The tests performed in Ref. \cite{2025_spectral_solver} proved that the hydrodynamic solver typically allows for less than $1\%$ error with respect to analytical derivation for the mobility.

In this contribution, we deploy our oscillatory-Stokes solver to study the interaction between viscoelastic structures and oscillatory flow. We use a multiblob viscoelastic network to describe the immersed structures (Appendix \ref{app:Lagrange} discusses optimal strategies for the location and size of the Lagrangian markers (IB kernels) forming the structures). 
This problem can be solved in at least two ways, which we now discuss. Before that, let us now recapitulate the relevant equations \ref{eq:u1} and \ref{eq:u2} in supervector notation,
\begin{eqnarray}
\label{eq:1}
\buu &=& \buu^{(0)} +\mob \bfind\\
\label{eq:2}
\bfind&=& \hmat \buu + \tilde{\bf E} 
\end{eqnarray}
where the external forces (field plus wall-linkers) are collected in the vector 
\begin{equation}
\label{eq:ext}
\tilde{\bf E}\equiv \kspw \bvwall + {\bf E}    
\end{equation}

\subsection{Solving by direct evaluation of the mobility}

A way to solve the system of equations \ref{eq:1} and \ref{eq:2} is to use the fluid solver to compute each element of the mobility matrix, $\mob$, i.e.

$$
\mathcal{M}_{i\alpha,j\beta} =  J_i \left[\mathcal{L}^{-1} S_j\, \bx_{\beta} \right]\cdot \bx_{\alpha} \;\;\text{for}\;i,j=[1,\dots,N] \;\text{and}\;\alpha,\beta=[x,y,z]
$$
In other words, consecutively, at each particle $i$, we apply a unity force ${\bf F}_i^{\alpha} = \bx_{\alpha}$ (with $\bx_\alpha$ is the unit vector in $\alpha$ direction) and measure the velocity response at particle $j$, in $\beta$ direction. Once the mobility matrix is computed, the resistance matrix $\res = \mob^{-1}$ can be used to obtain $\buu$ and $\bfind$ using the expressions:

\begin{eqnarray}
\label{eq:buu}
\buu &=& \left(\res-\hmat\right)^{-1} \left(\tilde{\bf E} + \res \buu^{(0)}\right)\\
\label{eq:find}
    \bfind &=& \left({\bf I} -\hmat \mob\right)^{-1} \left(\tilde{{\bf E}} +\hmat \buu^{(0)}\right)
\end{eqnarray}

While this method has the advantage of always yielding a solution to the system of equations (and the mobility matrix itself can provide additional insight into the system) it comes with a significant drawback: computing the mobility matrix requires running the fluid solver $3N$ times (once for each bead and spatial direction) and performing on the order of $9N^2$ interpolation operations (indeed, this number can be reduced by roughly half by exploiting symmetry relations.) As a result, the computational cost can become prohibitive for large particle numbers.

In addition, this method requires storing the matrices $\hmat$ and $\mob$ in memory, each containing $9N^2$ complex numbers—something that becomes infeasible on many GPUs for $N > 10^4$. In the case of the matrix $\hmat$, the memory footprint can be greatly reduced by noting that, typically, each bead interacts with only a small fraction of all other particles in the system. This means that most elements of $\hmat$ are zero, and we can therefore take advantage of sparse matrix techniques to store only the non-zero entries\cite{sparse_matrix}. As a result, the memory usage of $\hmat$ scales linearly with $N$ rather than quadratically. However, this strategy cannot be applied to the mobility matrix $\mob$, since hydrodynamic interactions are long-ranged, and thus the majority of its entries are non-zero.

\subsection{Solving by an iterative method}

To overcome the drawbacks of the mobility-based method, we have developed an iterative approach which, although not as straightforward to implement, scales much better both with the number of particles and in terms of memory usage. The core idea of the method is:
\begin{enumerate}
\item Start from an initial condition (for example, the base flow solution $\buu = \buu^{(0)}$)
\item Use Eq. \ref{eq:u2} to compute an initial estimate of the induced forces, $\bfind^{(0)}$
\item Input this estimation into the hydrodynamic solver \cite{2025_spectral_solver} to obtain $\bv^{(1)} = \mathcal{L}^{-1}S \bfind^{(0)}$. 
\item Interpolate to update the velocity as $\bu^{(1)} = \bu^{(0)} + J \bv^{(1)}$ (see Eq. \ref{eq:u1}). 
\item Repeat iteratively until convergence is reached.
\end{enumerate}

This procedure can be compactly written as a fixed-point iteration equation,
\begin{equation}\label{eq:fixed_point}
    \bfind^{n+1} = \hmat\left(\buu^0 + J\mathcal{L}^{-1}S\bfind^{n}\right) + \tilde{\bf E}
\end{equation}
However, a naive fixed-point iteration often diverges when the induced forces are sufficiently stiff. To ensure convergence, fixed-point problems typically require more stable algorithms such as Anderson acceleration \cite{2015_AndersonAcc}, which we have extended to support complex variables (as detailed in the Appendix \ref{app:Anderson}).
The key idea behind Anderson acceleration is to avoid relying solely on the most recent iterate, as done in Eq.~\ref{eq:fixed_point}. Instead, the method builds a new estimate of $\bfind$ taking a linear combination of the most recent $M$ past iterations, where $M$ is an adjustable parameter. The coefficients of this combination are chosen to minimize the residual error of the fixed-point equation, which significantly enhances the convergence rate.

Importantly, the iterative algorithm has the advantage that it no longer requires storing the mobility matrix, so memory usage now scales with $N$ instead of $N^2$. Moreover, regardless of the number of particles, convergence is typically achieved in a few hundred iterations at most (see Appendix \ref{app:Conv}). As a result, for large particle counts, this method requires significantly less computational time than the mobility-based approach. Recall that the mobility-based algorithm requires solving the fluid problem $3N$ times, whereas this iterative method typically requires between 100 and 500 fluid solves. While we find a fast convergence in the vast majority of cases, we have observed particularly slow convergence or even no-convergence if the kernels present significant spatial overlap (e.g. in cases with high density of particles) and/or in some cases, when the elastic constants of the structure are made extremely stiff. This second issue (stiffness) is not so problematic because, as we will show below in examples taken from QCM and AFM realms, the response of an elastic system saturates to the rigid limit once its internal elastic springs surpass a large enough value, which is still computationally tractable. By contrast, it is important to keep the amount of kernel overlapping low.

\section{Application to Quartz Crystal Microbalance (QCM) \label{sec:qcm}}

\subsection{Theory}
The quartz crystal microbalance device applies an AC voltage to a piece of quartz, using the inverse piezoelectric effect to induce tangential Love waves in the crystal. At the center part of the millimeter-size resonator, its surface oscillates in the $x$ direction at tens of MHz frequency. The wall velocity $\vwall =x_{wall} \omega$ is proportional to the amplitude of the oscillation, which is typically small $x_{wall}\sim 2\text{nm}$. The oscillation of the boundary at $z=0$ leads to  a basal flow given by $$\bv^{(0)}(\br)=\vwall\,\exp[-\idel z] \bx$$
where $\idel = (1-\ii)/\delta$ and the fluid penetration length is $\delta=\left(2\nu/\omega\right)^{1/2}$ with $\nu=\eta/\rho$ the fluid kinematic viscosity. The present computational tool permits calculating the QCM impedance $Z$, which is a phasor defined as the ratio between the total tangential stress at the QCM resonator (located at $z=0$) and the wall oscillatory velocity, $Z=\bsigxz/\vwall$.  The fundamental resonator frequency is typically $f_0=5\,\text{MHz}$ and the QCM sensor samples a collection of odd harmonics $f_n=n\,f_0$ with $n=\left\{1,3,5,\dots,13\right\}$. In the linear regime (small load approximation \cite{Johannsmann2015}) the experimental QCM signals, frequency $\Delta f$ and dissipation $\Delta D$ shifts, are related to the complex impedance as \cite{Johannsmann2015} 
$\Delta f = -f_1\text{Im}[Z] f_1/\pi Z_Q)$ and
$\Delta D = -2\text{Re}[Z]/(n \pi Z_Q )$
where $Z_Q=8.8\times 10^{6}\, \text{kg}\,{m^2/s}$ is the crystal-cut impedance. These relations are valid for $\Delta f << f$, which is the case in most QCM experiments. In summary, rather than the adsorbed mass, the QCM device directly senses the net tangential stress acting on the resonator.  The wall stress is the sum of contact forces in $x$ direction acting on the surface plus the net hydrodynamic shear stress. 
$$
\bsigxz=\bsigxz^{c}+\bsigxz^{f}
$$
Where the superscript $c$ denotes contact forces with the structure and $f$ denotes the hydrodynamic contribution. Setting ${\bf E}=0$ in Eq. \ref{eq:ext} (no external fields)  the contact force on the $i$ analyte's bead is $\bF^e_i=-\kspw_i\left(\bu_i-\bvwall\right)$ and the net contact force acting on the wall  in x direction is ${\bf F}^{w}\cdot \bx = - \sum_i {\bf F}^{e}_i \cdot \bx$. The contact stress is then $\bsigxz^c={\bF}^w\cdot \bx/A$, where $A$ is the resonator area, and the net tangential stress on the wall is  

\begin{equation}
  \bar{\sigma}_{xz} = \frac{1}{A}\left( {\bf F}^{w} \cdot \bx + \int_{z=0} \bx \cdot \boldsymbol{\sigma} \cdot \bz\,  d\bs^2 \right)
\end{equation}
where $\bs = x\bx + y \by$  runs over the resonator plane. Here we evaluate the integral in the above expression directly from the numerical solution of the velocity gradient of the perturbative flow, i.e.
$$
\sigma_{xz}^f(z=0)= \bx \cdot \boldsymbol{\sigma}(z=0) \cdot \bz = \eta \left(\frac{\partial v_x}{\partial z}\right)_{z=0}
$$
The evaluation of the integral is particularly fast in the spectral code \cite{2025_spectral_solver}, as it simply corresponds to the zero in-plane wavenumber ${\bf k}=0$ of the Fourier component of the tangential stress $\hat{\sigma}_{xz}$, i.e,
$$
\int_{z=0} \bx \cdot \boldsymbol{\sigma} \cdot \bz \, d\bs^2 = A \hat{\sigma}_{xz}({\bf k}=0)
$$
We note that using a periodic setup in the plane is particularly useful to estimate long-distance hydrodynamic contributions to the QCM impedance at a given finite coverage. A single particle in a periodic plane of sides $L_x$ and $L_y$ corresponds to a periodic lattice of analytes at a coverage proportional to $1/(L_xLy)$. In the infinite plane, a simple expression for the net hydrodynamic stress can be obtained from the hydrodynamic Green function $\boldsymbol{\mathcal{G}}_{\omega}(\br,\br^{\prime})$ whose expression in the reciprocal space for the $xy$ plane was derived by Felderhof \cite{Felderhof2005,2025_spectral_solver}. The perturbative velocity field can be written as,
$$
\bv(\br)= \int \boldsymbol{\mathcal{G}}_{\omega}(\br,\br^{\prime}) \,{\bf f}(\br^{\prime}) d \br^3
$$
so at $z=0$ the total tangential hydrodynamic stress is,
$$\bar{\sigma}^{f}_{xz}=  \frac{\eta}{A} \int_{z=0} \frac{\partial v_x}{\partial z}\, d^2\bs = \frac{\eta}{A} \int_{z=0} \int \partial_z \boldsymbol{\mathcal{G}}_{\omega}(\bs -\bs^{\prime}, 0, z^{\prime})\, {\bf f}(\bs^{\prime},z^{\prime})\, d^3 \br^{\prime} d^2 \bs$$

with ${\bf f}(\br)=\sum_i S(\br-\br_i) \,\bfind_i$.  Expressing the Green function in the in-plane reciprocal space \cite{Felderhof2005,BuscalioniSM21} and taking the zero wavenumber limit, leads to a seemly simple expression for the total hydrodynamic impedance associated with the \emph{perturbative} flow
$$Z^{h}=\frac{1}{A \vwall} \sum_i e^{-\alpha z_i^{\prime}} \, \bfind_i \cdot \bx$$
where $z_i^{\prime}$ indicates the norm. On the other hand, summing over all the beads in $-\ii\,m^e_i \omega \,\bu_i = {\bf F}^{\text{s}}_i + {\bf F}_i^e -\bfind_i$, using $\bF_i^e=-\bF_i^w$ and taking into account that the sum of internal structure forces vanishes $\sum_i \bF^s_i={\bf 0}$, leads to an expression for the total impedance $Z=Z^c+Z^h$,

\begin{equation}
\label{eq:imp1}
    Z=\frac{1}{A \vwall} \sum_i \left[ \ii  m^e_i \omega \, \bu_i\cdot \bx  + \left( e^{-\idel z_i} -1\right)\, \bfind_i \cdot \bx \right].
\end{equation}

The first term of this expression is due to the stress created by the inertia of the adsorbed excess mass. Only in the limit of rigidly adsorbed analytes ($\bu_i\cdot \bx = \vwall$) in vacuum ($\bfind_i=0$ and $m^e=m$) the total impedance coincides with the Sauerbrey expression $Z^{\text{Sau}}=\ii \sum_i m^e_i \omega/A$. For most soft matter analytes the excess mass is, however, small (e.g. $m^e \sim 0.2 \,m^f$ for proteins), and the dominant contribution to the impedance arises from a complicated interplay between hydrodynamic and internal viscoelastic forces, which is reflected in the formal solution of velocities and induced forces indicated in Eqs. \ref{eq:1} and \ref{eq:2}. 
Finally, it should be noted that the "total" impedance introduced by the dispersion (solvent and solute) is $Z_0+Z$ where the  contribution from the basal flow is readily obtained from the basal velocity, being given by $
Z_0 = \eta \idel= (\eta/\delta)\,\left(\ii-1\right)$. Indeed, this contribution is subtracted from the total QCM signals to obtain the relevant $Z$ providing information on the solute phase.

\subsection{Results and validations for QCM \label{sec:res}}

In this section, we will first validate our computational scheme by comparison with finite element calculations for the QCM impedance of suspended and strongly adsorbed rigid structures, recently obtained by Leshansky et al. \cite{Leshansky2023, Leshansky2024} who also derived analytical relations valid for limiting regimes. Then we validate results for viscoelastic tethers by comparison with a theory derived below.

\subsubsection{QCM impedance of suspended rigid particles}

\begin{figure}
    \centering
    \includegraphics[width=0.95\linewidth]{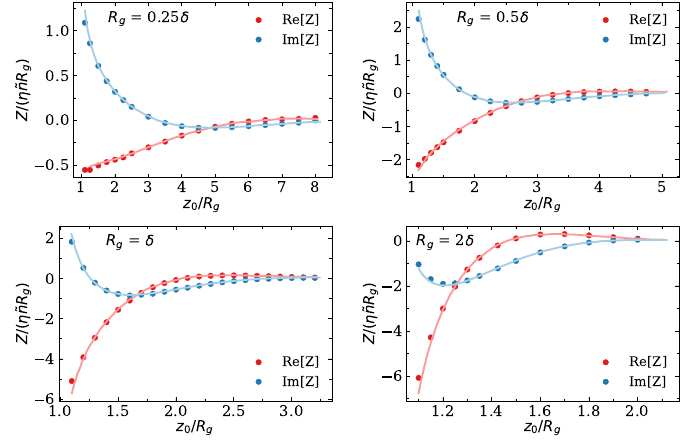}
 \caption{Real (red) and imaginary (blue) components of the impedance at the suspended particle wall as a function of height, calculated using the method described in this work (dots) and compared to the results of Leshansky (2023)\cite{Leshansky2023}.  For the numerical calculations, the sphere, of geometric radius $R_g$ (see appendix \ref{app:Lagrange}) is represented using 92 3-point Peskin kernels. The box side was approximately $L_\parallel\approx8\delta$, and the meshed domain $z\in [-H,H]$ in the unbounded $z$ direction was adapted to  the maximum height in each case as $H=\max(z_0)+2R_g$. The impedance is scaled with $\eta\, \tilde{n} R_g$ where $\tilde{n}$ is the particle surface  
 number density.}
  \label{fgr:Suspended}
\end{figure}

We start by comparing our results for neutrally buoyant rigid spheres suspended near the wall of a QCM at various heights and with different radii, against the theoretical predictions by Leshansky et al. (see Eq. 24 in Ref. \cite{Leshansky2023}). For this purpose, we employ the spherical shell discretization developed in the Appendix \ref{app:Lagrange}, using 92 beads per sphere. Each bead is connected to its nearest neighbors via harmonic springs with an elastic constant of $\kappa = k/m\omega_0^2 = 10^7$, which is sufficiently large to approximate a rigid sphere. For the comparison, we use same non-dimensionalization as in Ref.~\cite{Leshansky2023}, in which the wall impedance $Z$ is divided by the viscosity, the sphere radius, and the surface number density $\tilde n = N/(L_x L_y)$. The results are shown in Fig. \ref{fgr:Suspended}.

The figure shows excellent agreement between our method and the theoretical results from the literature.

\subsubsection{QCM impedance of Adsorbed rigid particles}

\begin{figure}[ht]
\centering
\includegraphics[width=0.5\textwidth]{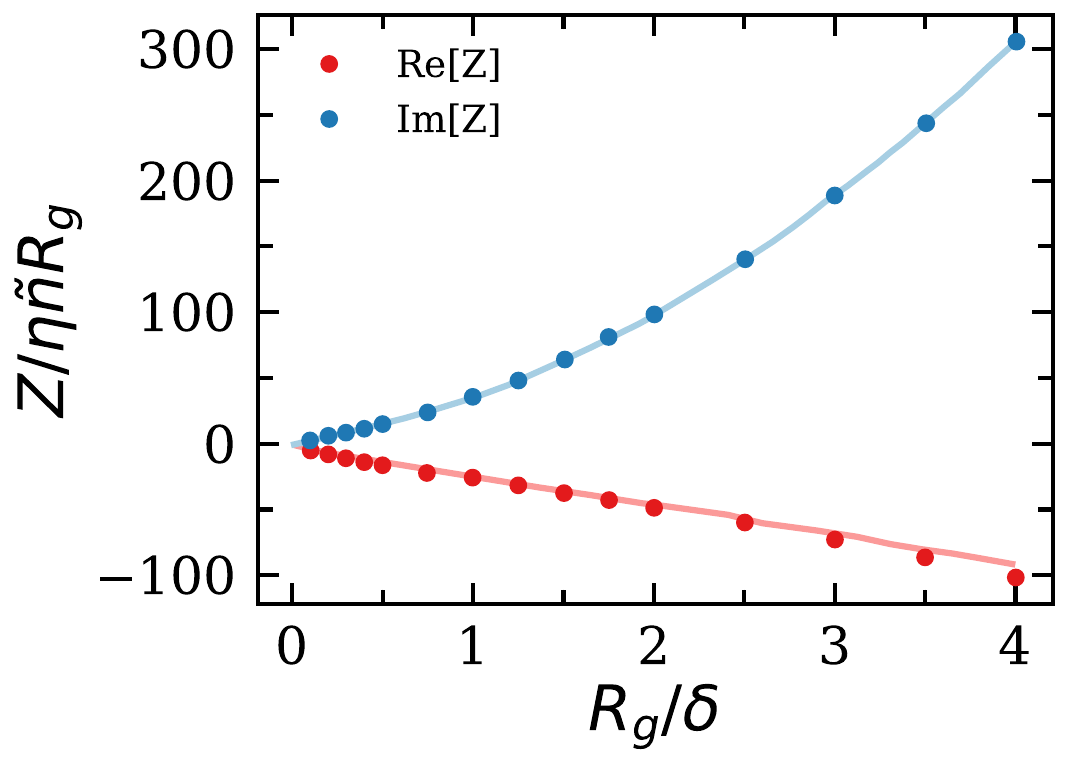}
\caption{Real (red) and imaginary (blue) components of the impedance at the wall for adsorbed rigid particles of geometrical radius $R_g$, calculated using the method described in this work (dots), compared to the theoretical predictions from Ref.~\cite{Leshansky2024}. The number of blobs per sphere is 1415, and the location of the IB blobs follows the protocol explained in the appendix \ref{app:Lagrange} for filled spheres with $T=6$. }
\label{fig:AdsorbedImpedance}
\end{figure}

Next, we validate the implementation of particle-wall interactions by computing the total impedance at the wall in the presence of a rigid sphere fully adsorbed onto it—that is, a sphere oscillating entirely in phase with the wall. The total impedance includes both the fluid contribution and the forces arising from direct contact between the particle and the wall. Results obtained with our numerical method are compared in Fig.~\ref{fig:AdsorbedImpedance} against the theoretical predictions reported in Ref.~\cite{Leshansky2024} (Eqs. 8 and 10). 

During this test, we observed that for particles larger than the penetration depth, spurious internal flows distorted the impedance measurement. To eliminate this artifact, we used the filled-sphere model (see appendix \ref{app:Lagrange}), in which the entire interior of the sphere is also constrained to move rigidly with the wall.  This idea is a discrete-level particularization of more advanced techniques, such as the Immersed Boundary smooth extension \cite{2016Stein}, which can be useful to avoid interior flow complications. 

\subsubsection{Viscoelastic tethers}
A single bead connected to the substrate by a flexible chain will serve to present the main non-dimensional parameters of the viscoelastic case. Besides, this simple setup illustrates the rich and far-from-trivial dynamical response encountered in the QCM signals (frequency and dissipation shifts). For instance, larger dissipation shifts $\Delta D$ (determined by the real part of the impedance) should not necessarily be interpreted as coming from flexible or "dissipative" structures, as usually done in the QCM community.  A globular protein domain linked by a small molecular chain to the substrate is a very common situation in nature \cite{das2013conformations,buske2023evolved} and used in many biomolecular sensors \cite{Prapp2020,Milioni2017}. In nature, proteins are often linked to a membrane via a peptide linker, as is the case with the FtsZ protein  \cite{buske2023evolved,mateos2016monitoring} illustrated in Fig. \ref{fig:levels}.  Here, the linker stiffness is based on the linear spring model of Eq. \ref{eq:k2}. We take the bead-wall equilibrium distance as $r_{1w}^{(eq)}\rightarrow d_{1w}$, where $d_{1w}=\ell$ is the linker length. The link direction is $\be_{1w}= (\br_1-\br_w)/r_{1w}$ where the linked bead position is $\br_1$.
Without loss of generality, we consider that the link at the substrate is located at $\br_w = {\bf 0}$.  For the sake of simplicity, we use a similar treatment for the frictional force which also acts in the linker bonding direction,  $\be_{1w}$. With these assumptions, the linker stiffness matrix (see Eq. \ref{eq:k2}) becomes,
\begin{equation}
\label{eq:kk1}
  \kspw_1 =\left[\frac{\ii k}{\omega}+  \xi_{\parallel}\right]\, {\bf e}_{1w} {\bf e}^T_{1w}=\tilde{\xi}\, {\bf e}_{1w} {\bf e}^T_{1w}
\end{equation}
where ${\bf e}_{1w} {\bf e}^T_{1w}$ is the dyadic product and we have defined the complex parameter $\tilde{\xi}\equiv \ii k/\omega+  \xi$ with units of friction coefficient. In general, the frictional coupling might also contain transversal friction \cite{Hijon_2010} with a friction matrix given by $\boldsymbol{\Gamma}= \xi_{\perp} ({\bf I}_3 -\be \be^T) + \xi_{\parallel} \be \be^T$, or have more complicated forms. 
The bead velocity is directly obtained from Eq. \ref{eq:buu} with $\bvwall =\vwall \bx$.
The hydrodynamic resistance matrix $\res$ deserves some attention as it connects with the induced force. The kinetic condition (or the hydrodynamic route to $\bu_1$) implies that 
\begin{equation}
\label{eq:l1}
    \bu_1 =\bu_1^{(0)}  + \mob \bfind_1
\end{equation}

In general, the mobility can be decomposed as a self contribution $\mob_{ii}$ and a contribution from neighbors $\mob_{ij}^{\dagger}$ with $\mob_{ii}^{\dagger}={\bf 0}$. The self mobility in a semi-bounded domain has contributions from the free-space and from the wall reaction field. The self mobility of a sphere of radius $a$ oscillating in free space was derived by Mazur and Bedeaux \cite{Mazur1974} being $\mob_{\infty}=\xi_{self}^{-1}(\omega)\,{\bf I}_3$ where 
\begin{equation}
    \xi_{MB}=\xi_S\,\left[1+\idel a +(1/3)\,(\idel a)^2\right]
    \label{eq:mb}
\end{equation}
is the Mazur-Bedeaux friction kernel (proportional to the induced force)  \footnote{The inertia contribution $(1/3)\,(\idel a)^2$ corresponds to the fluid-particle induced force (particle excess mass inertia),  one needs to add the acceleration of the displaced fluid to get inertia contribution appearing in the total-drag friction coefficient, $(1/9)\,(\idel a)^2$ \cite{Mazur1974}.}  and $\xi_S=6\pi \eta a$ the Stokes friction for the steady limit.
As initially shown by Felderhof \cite{Felderhof2005} and revisited by others \cite{morrison2018,2025_spectral_solver}, besides the free-space friction, the wall induces an extra drag due to the reaction field, which "reflects back" part of the perturbative flow generated by the particle. 
The mobility associated with the wall's reaction field depends on the height of the bead $z_i$ above the boundary $\mob_{wi}=\mob_{wi}(\omega,z_i)$. This tensor is diagonal with $ \mathcal{M}^w_{\parallel}=\bx \cdot \mob_{wi}  \cdot \bx =\by \cdot \mob_{wi}  \cdot \by $ and $\mathcal{M}^w_{\perp} = \bz \cdot \mob_{wi}  \cdot \bz$. In Ref. \cite{2025_spectral_solver} we collected the analytical expressions for $\mathcal{M}^w_{\perp}(z)$ and $\mathcal{M}^w_{\parallel}(z)$ for a rigid sphere of radius $a$ from the original works \cite{Felderhof2005,Felderhof2009} and also re-derived these mobilities for a Gaussian blob: notably, the rigid and Gaussian blob present quite similar reaction-field mobilities \cite{2025_spectral_solver}.

Instead of using Felderhof's notation for the \emph{reaction} field $R$, we use a notation consistent to the fact that it is a \emph{mobility} contribution from the wall ($w$) reaction flow onto each particle ($i$) and use $\mathcal{M}_{wi}$. Thus, $\mob_{ii} = \mob_{\infty} + \mob_{wi}
$ and we can generally say that for an arbitrary number of beads,
$$
\bu_1 -\bu_1^{(0)}= \left(\mob_{\infty} + \mob_{wi}\right) \, \bfind_1 + \sum_{j>1} \mob^{\dagger}_{1j}\, \bfind_j
$$
The above expression also contains the contributions for the induced forces from the neighbor beads $j>1$. In the present derivation, we just consider a structure with a single bead, so we are neglecting the internal cohesive structural forces introduced by $\mob^{\dagger}_{1j}\bfind_j$ contribution. This term contributes with higher moments (stresslet and rotlet) which we will later numerically analyze.  Neglecting  the $\mob^{\dagger} \bfind$ term and introducing the resistance leads to,
\begin{eqnarray}
\label{eq:l2}
   \bfind_1 &=& \res_{11} \,\left(\bu_1 -\bu_1^{(0)}\right) \;\;\text{with}\\
   \label{eq:l3}
   \res_{11}&=&(\mob_{\infty} + \mob_{w1})^{-1}.
\end{eqnarray}
Equation \ref{eq:l3} closes Eq. \ref{eq:buu} for $\bu_1$, which is used in Eq. \ref{eq:l2} to provide the induced force and subsequently the QCM impedance, via Eq. \ref{eq:imp1}.

We summarize the set of equations in non-dimensional form to introduce the relevant system's parameters. Velocities are nondimensionalized using $\vwall$ and all the impedances (due to inertia $m_i^e \omega $ and
complex friction coefficients $\tilde{\xi}= \ii k/\omega + \xi$) are scaled with the self-friction of a bead $\xi_S = 6\pi \eta \,b$, where $b$ is the bead hydrodynamic radius, determined by the IB kernel $S(\br)$ \cite{Thesis-Pelaez,2025_spectral_solver}.

We define $\boldsymbol{u}\equiv \bu/\vwall$ and $\mu_e\equiv m_1^e \omega/\xi_S$ and $\tilde{\gamma}\equiv \tilde{\xi}/\xi_S=\ii\, \kappa + \gamma$ with
$\kappa \equiv k/(\omega \xi_S)$ and $\gamma\equiv \xi/\xi_S$. We also define the nondimensional resistance
$$
\boldsymbol{\mathbb{R}} \equiv \xi_S^{-1} \res_{11} = \xi_{S}^{-1}\,(\boldsymbol{\mathcal{M}}_{\infty} +  \mob_{w1})^{-1}.
$$
Note that $\boldsymbol{\mathbb{R}}$ is diagonal and $ \boldsymbol{\mathbb{R}}_{\alpha \alpha} =(\xi_{MB}/\xi_S) /(1+\xi_{MB} \mathcal{M}^w_{\alpha \alpha})$, with $\mathcal{M}^w_{xx}=\mathcal{M}^w_{yy}=\mathcal{M}^w_{\parallel}(z)$ and $\mathcal{M}^w_{z}=\mathcal{M}^w_{\perp}(z)$ are the Felderhof's reaction-field mobilities for point particles, generalized for Gaussian blobs in Ref. \cite{2025_spectral_solver}. The nondimensional induced force are then $\boldsymbol{\lambda} = \bfind/(\vwall \xi_S)$ and the nondimensional force balance yields,
\begin{eqnarray}
\label{eq:l0}
   \left(-\ii \mu_1^e \delta_{\alpha \beta} + \tilde{\gamma} e_{\alpha}e_{\beta} + \mathbb{R}_{\alpha \beta} \right)\,u_{\beta} &=& \,\tilde{\gamma}\,e_{\alpha} e_{\beta} \,\delta_{\beta x} \\
   \lambda_{\alpha} &=& \mathbb{R}_{\alpha \beta} \,\left(u_{\beta} - e^{-\idel z} \delta_{\beta x} \right)\\
Z&=&\left[ \ii  \mu_i^e \, u_x  + \left( e^{-\idel z_i} -1\right)\, \lambda_x \right] \tilde{n} \xi_S
\end{eqnarray}
with $\lambda_{x} =\mathbb{R}_{x x} \,\left(u_{x} - e^{-\idel z} \right)$
and $\tilde{n}=1/A$ the number of particles per unit area and $\tilde{n}\xi_S$ is  the Stokes reference impedance. Note that $\tilde{n}\xi_S= (6\delta/a)\,\Theta\,Z_{fluid}$ with $\Theta =\pi a^2\,\tilde{n}$ the surface covered by analytes of radius $a$ and $Z_{fluid}=\eta/\delta$.

\begin{figure}[ht]
\centering
\includegraphics[width=0.85\textwidth]{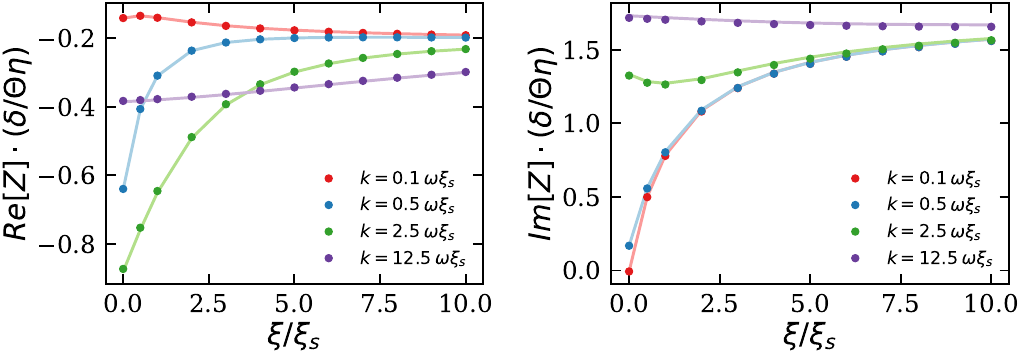}
\caption{Real (\textit{left panel}) and imaginary (\textit{right panel}) components of the wall impedance for a particle connected to the wall via a viscoelastic linker, shown as a function of the damping coefficient for different values of the elastic constant. Results from the numerical code (\textit{dots}) are compared with the predictions of an analytical theory (\textit{lines}). Calculations were performed for neutrally buoyant blobs with a hydrodynamic radius $a = 0.03\delta$, a linker length $l_{\text{link}} = 0.15\delta$, with an longitude and azimuth angle of $45^\circ$. The box size is $L = 50a$, ensuring that the effects of periodic boundary conditions are negligible.}
\label{fgr:ViscoDumbell}
\end{figure}

To analytically solve this set of equations, we use spherical coordinates. The linker vector is determined by the azimuth angle $\varphi$ and polar angle $\theta$, i.e.
$\be_{1w}= (\cos\varphi \sin\theta,\sin\varphi \sin\theta, \cos\theta)$. The linker length is $\ell$ and the basal flow at the bead position is $\bu^{(0)}(\theta)= \vwall\, \exp[-\idel \ell \cos\theta]$. The solution is straightforward, though we used {\tt Mathematica} to deal with the involved matrix operations. Figure \ref{fgr:ViscoDumbell} compares the analytical relation with numerical calculations for $\theta=45^o$ and $\phi=45^o$ (symbols), showing a perfect agreement. For brevity, we provide the impedance averaged over the azimuth angle $\langle Z\rangle_{\varphi} = \int_0^{2\pi} Z(\ell,\varphi,\theta)\, d\varphi$ 
\begin{eqnarray}
    \langle Z\rangle_{\varphi} &=& \frac{e^{-2 \alpha  L \cos\theta }}{\pi r (\mu_e + \ii \mathbb{R}_{xx}) }\, \left(\frac{A+B}{D}\right)\\
    &\text{with}&\\
\nonumber
A&\equiv& 2 \pi  \kappa  \sin ^2(\theta ) (\mu_e+\ii \mathbb{R}_{xx}) (\mu_e+\ii \mathbb{R}_{zz}) \left(-2 \mathbb{R}_{xx} e^{\alpha  L \cos(\theta )}+(\mathbb{R}_{xx}-\ii \mu_e) e^{2 \alpha  L \cos (\theta )}+\mathbb{R}_{xx}\right)\\
\nonumber
B&\equiv & 2 \pi \mu_e \mathbb{R}_{xx} \left(\kappa  \sin ^2(\theta ) (\mu_e+\ii \mathbb{R}_{zz})-  (\mu_e+\ii \mathbb{R}_{xx}) (-\kappa +2 \mu_e+2 \ii \mathbb{R}_{zz})\right)+\kappa  \mu_e \mathbb{R}_{xx} \cos (2 \theta ) (\mu_e+\ii \mathbb{R}_{xx})\\
\nonumber
D &\equiv& 
2\mu_e^2 + \mu_e\left(-2\kappa + 2\ii(\mathbb{R}_{xx} + \mathbb{R}_{zz})\right)
- \ii\left(-2\mathbb{R}_{xx} \mathbb{R}_{zz} + \kappa(\mathbb{R}_{xx} + \mathbb{R}_{zz})\right) - 
\ii \kappa (\mathbb{R}_{xx} - \mathbb{R}_{zz}) \cos(2\theta)
\end{eqnarray}
where $\mu_e = m_e/m_f$ and $r=a/\delta$.

\begin{figure}
    \centering
    \includegraphics[width=0.8\linewidth]{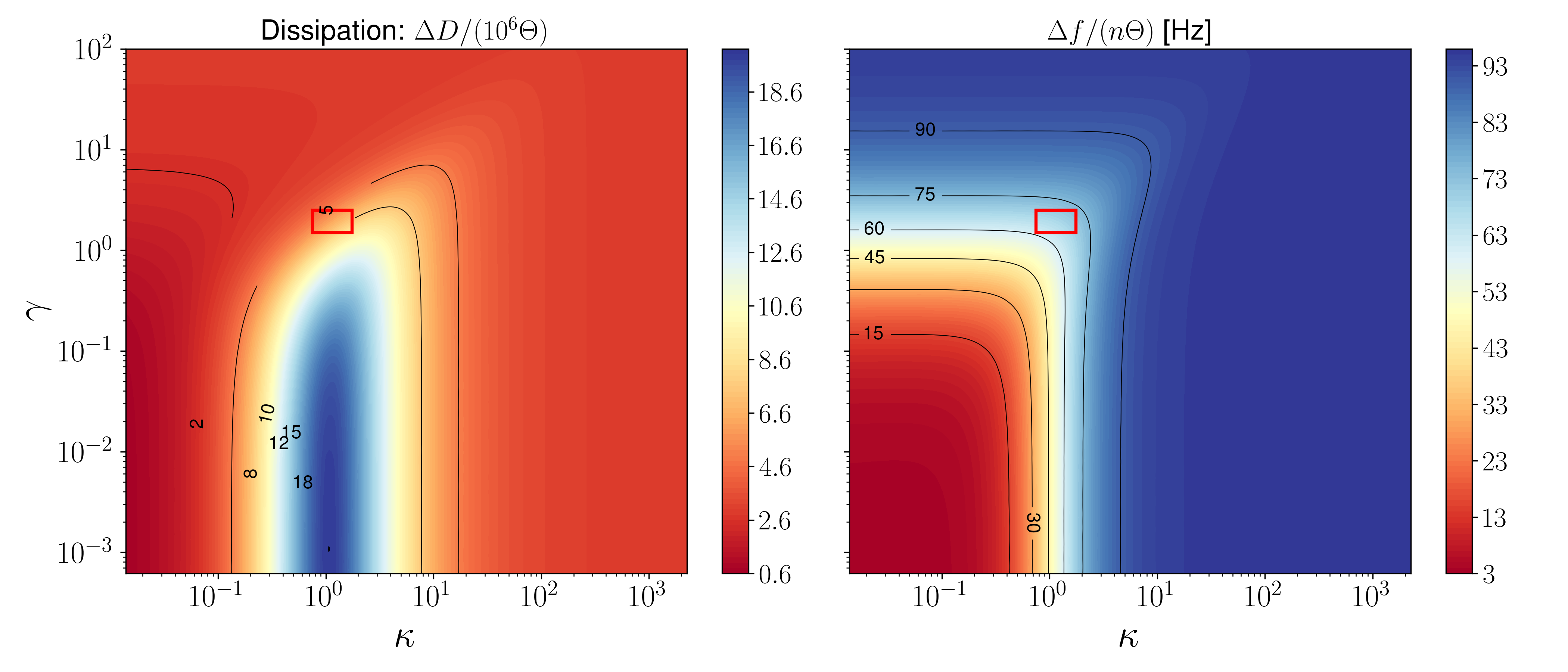}
    \caption{Contour plot for the frequency shift $\Delta F/n$ (Hz) and the disipation $\Delta D\times10^6$ scaled with the coverage $\Theta$, for a viscoelastic linker connected to a bead of $b=2.5\text{nm}$ radius. The linker model corresponds to a Gaussian linker with average lenght $\langle \ell \rangle =10\text{nm}$ and standard deviation $\sigma_\ell= 3\,\text{nm}$. The x-axis corresponds to the elasticity parameter of the linker $\kappa=k/(\omega \xi_S)$ and the y-axis to the internal friction $\gamma=\xi/\xi_S$, the bead mass is $m^e=0.2m_f$ (compatible with a protein). The impedance is averaged according to eq. \ref{eq:zave} for the $n=7$ overtone of a QCM resonator with fundamental frequency $f_0=5\text{MHz}$ in water ($\nu=10^{-6}m^2/s$). The red rectangles indicate the region of parameters for a flexible peptide chain with 60 aminoacids similar to the disordered domain of the FtsZ protein illustrated in Fig. \ref{fig:levels}(a): contour length $L_c=20 \text{nm}$, persistence length $0.35 \text{nm}$ and stiffness $k\approx 0.4 k_B/\text{nm}^2$, which corresponds to $\kappa \approx 1$ for the $n=7$ overtone.}
    \label{fig:contour}
\end{figure}

\begin{figure}
    \centering
    \includegraphics[width=0.8\linewidth]{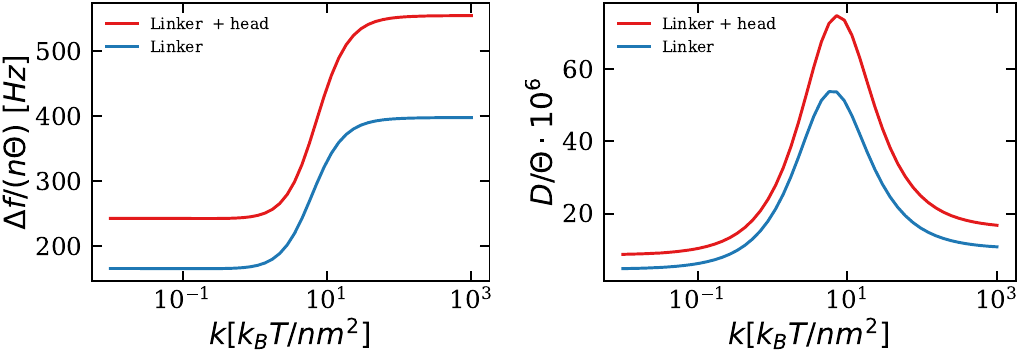}
    \caption{Frequency (left panel) and dissipation (right panel) of the QCM for a sphere tethered to the wall by a linker, as a function of the linker's spring constant. The sphere is modeled either as a single blob (blue line), which excludes stresslet and rotlet contributions, or as an icosahedron of rigidly connected blobs (red line). Calculations were performed for a sphere of radius 2.5 nm in a box of width 45 nm. The linker has a length of 7 nm and is oriented at an azimuthal angle of $\varphi =0^\circ$ and a polar angle of $\theta=45^\circ$, its friction parameter is $\xi=\xi_s$. The density of the sphere is 1.2 times that of water. Results correspond to overtone 7.
}
    \label{fig:Effect_head}
\end{figure}

To provide a connection with QCM experiments, we consider the particular case of a link formed by a small flexible chain. In the dilute regime, where hydrodynamic couplings between structures is negligible (this only happens at very low coverage $\Theta < 0.05$ \cite{2025_spectral_solver,Buscalioni_langmuir23}), the QCM impedance can be set as a linear combination of contributions of individual molecules sampling the equilibrium configurations determined by the probability $P(\ell,\varphi,\theta)$, i.e.
\begin{equation}
    \langle Z\rangle = \int_0^{\infty}\int_0^{2\pi}\int_0^{\pi} P(\ell,\varphi,\theta) Z(\ell,\varphi,\theta) \sin\theta d\varphi d\theta  d\ell
    \label{eq:zave}
\end{equation}
For illustration purposes, we use an azimuth invariant distribution, $P(\ell,\varphi,\theta)= P_{\ell}(\ell) \,\Theta(\theta_c-\theta)$ with a Heaviside modulation on the longitude ($\Theta(s)=1$ if $s>0$ and zero otherwise to take into account the wall, with $\theta_c=\pi/2-\arcsin(2b/\ell)$). The radial distribution $P_{\ell}(\ell)$ is set to a Gaussian distribution with average $\langle \ell \rangle $ and standard deviation $\sigma_\ell$. 
The resulting frequency and dissipation shifts (scaled with the coverage) from this CG model of a short linker (with $\langle \ell \rangle =10\text{nm}$ and  $\sigma_\ell=3\text{nm}$) are shown in the contour plots of Fig.\ref{fig:contour} as a function of the elastic $\kappa$ and viscoelastic $\gamma$ parameters. A resonance effect is clearly revealed around $\kappa \sim 1$, where the dissipation peaks.
This CG description of a linker could well represent a molecular chain presenting a broad distribution of lengths. As expected, a chain homogeneously distributed in the azimuth and polar angles will present an impedance which will depend on the length distribution $P_\ell(\ell)$. In the case of Gaussian chain with persistence length $p$ and contour length $L_c$, one gets $\langle \ell \rangle \approx \sqrt{4 p\,L_c}$ and a stiffness $k\approx 3 k_BT/(pL_c)$. Our model could represent a FtsZ protein tethered to the substrate (membrane) through the N-terminal of its intrinsically disordered region (IDR). The IDR is a flexible and slightly charged
peptide chain of about 60 amino acids, with $L_c=20 \text{nm}$ contour length and persistence length $p\approx 0.35\text{nm}$ characteristic of the peptide bond \cite{buske2022connecting}.
The entropic stiffness would then be $k\approx 0.43\,k_BT/\text{nm}^2$ with 
$P_\ell(\ell) \sim \exp[-k(\ell -\langle \ell \rangle)^2/(2k_BT)]$ and $\sigma_\ell \approx \sqrt{p L_c/3} \sim 1.6\text{nm}$. The Gaussian chain model predicts an
average extension of the linker (IDR portion) of $\approx 5.3\text{nm}$ which agrees with FRET measurements \cite{ohashi2007experimental}. However, the IDR of proteins is charged (as shown in Fig. \ref{fig:levels}(a)) and dependending on environmental conditions they can elongate as extended chains \cite{das2013conformations} presenting a broad (slightly non-Gaussian) distribution of end-to-end distances (as revealed by MD \cite{cohan2019information}) reaching up to $\sigma_l\sim 3\text{nm}$ in the FtsZ case. A recent study indicates specific chain-globule interactions which further broaden the chain's end-to-end distribution \cite{buske2022connecting}. The average end-to-globule-center distance is about 2 nm larger, due to the extra globule radius and from these observations, a reasonable range for the length of our mesoscopic model (Fig. \ref{fig:levels}(b)) would be $\langle \ell \rangle \sim [7-10] \text{nm}$ with $\sigma_l \sim [2-3]\mathrm{nm}$. The intrinsic relaxation rate of the peptide chain is gauged by the self-friction of the globular part, $\xi_S$, and experiments suggest $\xi \sim \xi_S$ (i.e. $\gamma\sim 1$). The protein's typical domain (around $\gamma \sim 1$ and $\kappa = k/(\omega \xi_S)\approx 3$) are indicated by the red rectangles in the contour plots of Fig. \ref{fig:contour}. This figure was built from the analytical model of Eq. \ref{eq:l0}, using $\langle \ell\rangle =10\text{nm}$. In the protein relevant region (rectangles in Fig. \ref{fig:contour}) it predicts $\Delta f/(n\Theta) \sim 80 \text{Hz}$ and $\Delta D\times 10^6/\Theta \sim 10$. These values are somewhat smaller than the experimental values (measured at low protein coverage), which will be reported in a subsequent work. In fact,  although the analytical model of  Eq. \ref{eq:l0} provides a qualitatively correct picture, it does not take into account the stresslet contribution from the protein globule, i.e., the flow induced by the stress supported by the spherical domain against flow deformation. In our IB scheme, the stresslet is created by internal springs forming the "globule blobs" (see the mesoscopic model of Fig. \ref{fig:levels}(b)), which are considered very stiff $\kappa \sim 100$ and with negligible dissipation (see \cite{2025Monago} for a recent micro-meso evaluation of these parameters). The contribution of the globule stresslet to the QCM signals is illustrated in Fig. \ref{fig:Effect_head}. Comparison is made between the analytic model (having a  "hydrodynamic monopole" as head with a single $a=2.5\text{nm}$ blob) and the
mesoscopic model of Fig. \ref{fig:levels}(b) where the $a=2.5\,\text{nm}$ (protein radius) globule is represented by a icosahedral structure of 13 blobs with stiff strings. In both cases, the viscoelastic linker has a fixed $\ell =7\,\text{nm}$ in a flow with penetration length $\delta=95 \text{nm}$ (or $n=7$ in water). The intrinsic friction is set to $\xi=\xi_S$, and we varied $\kappa$. The larger values of $\Delta f$ and $\Delta D$ obtained with the "globule+linker" model (which introduces stresslet contributions) are consistently closer to experimental values. Note that the linker dissipation goes to zero as the linker tension (or stiffness $k$) vanishes (just remaining the Sauerbrey inertia), but the linker significantly contributes to the impedance at high stiffness. As in Fig. \ref{fig:contour}, the dissipation peaks in the resonance region around $k\sim  3 \,k_BT/\text{nm}^2$ (or $\kappa =k/(\xi_S\omega)\approx 8$) for the $n=7$ overtone.  We note that intrinsically disordered proteins tethered to flexible oligochains are located in the range $\kappa \sim [1-10]$ and $\Gamma \sim [1-2]$: An inspection of Fig. \ref{fig:contour} explains the high sensitivity of QCM measurements against the protein linker length \cite{mateos2016monitoring} and also in other molecular linkers such as DNA \cite{Milioni2017,tsortos2016hydrodynamic,Prapp2020}. 

\begin{figure}
    \centering
    \includegraphics[width=0.8\linewidth]{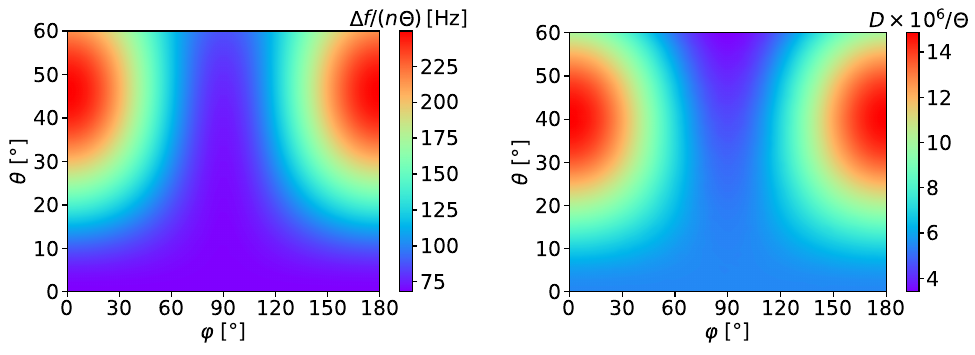}
    \caption{Contour plot of the frequency shift (left) and dissipation shift (right) of the QCM for a sphere discretized as an icosahedron with a central blob, tethered to the wall by a single linker, as a function of the linker’s polar and azimuthal angles. Calculations were performed using $k = 0.32\,k_BT$/nm, $\xi = \xi_s$, a sphere radius of 2.5 nm, a box width of 45 nm, and a linker length of 7 nm. The particle density is 1.2 times that of water. Results correspond to overtone 7.}
    \label{fig:Effect_angle}
\end{figure}

It is also interesting to consider the effect of the linker orientation, which in some cases  (large enough structures) can be somewhat experimentally controlled by tuning the advection due to injected flow. Also, ellipsoidal colloids (with recent QCM experiment \cite{2024sadowska}) could present a preferred orientation at large coverage. Figure \ref{fig:Effect_angle} show contour plots of the frequency and dissipation shifts for a linker tethering a sphere of radius $a=2.5\text{nm}$, with length $\ell =7 \text{nm}$ ($k=0.32\,k_BT/\text{nm}^2$ and $\xi=\xi_S$) oriented at varying azimuth $\varphi$ and polar angle $\theta$ [results are representative of the FtsZ protein of Fig. \ref{fig:levels}(a) using the mesoscopic model of Fig. \ref{fig:levels}(b)]. QCM signals reach a maximum at $\varphi=0^o$ and $180^o$ and around $\theta \sim 50^o$, corresponding to the linker substantially tilted in the flow plane ($xz$) (see Fig. \ref{fig:levels}). Interestingly, this effect could be used to increase the signal of DNA-tethered liposomes \cite{Prapp2020}, by increasing (above one) the Peclet number $\dot\gamma a^2/D$ associated with the shear rate $\dot\gamma$ of the injecting flow.

\begin{figure}
    \centering
    \includegraphics[width=0.8\linewidth]{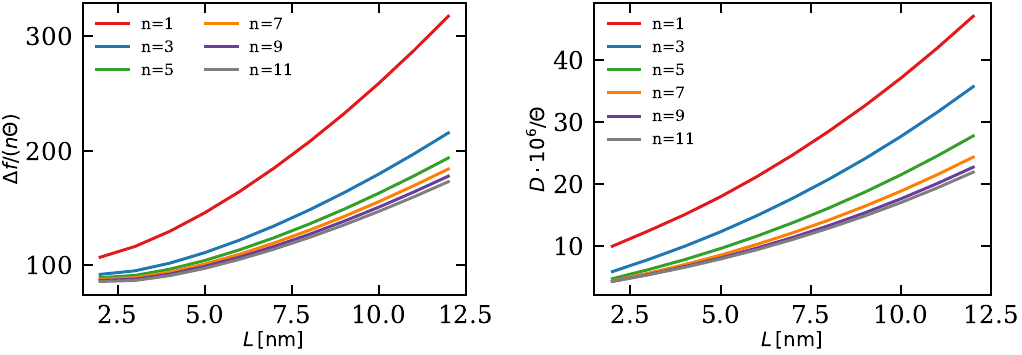}
    \caption{Frequency (left) and dissipation (right) shifts in the QCM for a sphere tethered to the wall as a function of linker length, for various overtones. The sphere is discretized as a filled icosahedron with a central blob. The spring constant is $k = k_BT$/nm$^2$, the friction parameter is $\xi = \xi_s$, and the particle density is 1.2 times that of water. The linker is oriented at an azimuthal angle of $45^\circ$ and a polar angle of $30^\circ$. The sphere has a radius of 2.5 nm, and the box width is 50 nm. }
    \label{fig:Effect_length}
\end{figure}

The length of the linker has a significant effect on the QCM signals, as already experimentally observed in previous works \cite{Prapp2020,Milioni2017}. As illustrated in Fig. \ref{fig:Effect_length}, both $\Delta f$ and $\Delta D$ increase with the linker length (as well as the acoustic ratio $n\Delta D/\Delta f$). These results correspond to a small globule of $a=2.5\text{nm}$ tethered to linkers of lengths ranging from $[2-12]\text{nm}$. In this regime (small $a$), the linker contribution is dominant (Fig. \ref{fig:Effect_head}). However, above a certain sphere radius (probably around $a\sim 10\text{nm}$), one should expect the sphere stresslet-induced impedance to overpower the linker-induced impedance. In such cases, and similarly to suspended particles (Fig. \ref{fgr:Suspended} and elsewhere \cite{Leshansky2024,Buscalioni_langmuir23,Prapp2020}), the impedance should decrease with the sphere's height. A decreasing impedance with height has already been experimentally and numerically confirmed in the case of DNA-tethered liposomes, with radii $a\in [15-100]\text{nm}$ \cite{Prapp2020}. The study of such a cross-over in the height response is left for future work.

\section{Application to force spectroscopy by AFM \label{sec:afm}}
\subsection{Spectra of Thermal Vibrations Measured with an AFM}
We now demonstrate how the proposed algorithm can quantitatively predict the spectrum of thermal vibrations of a microscopic particle, based on the calculation of the hydrodynamic resistance in the direction normal to the plane. We compare our numerical results with experimental data from \cite{AFM_spectra}, obtained in two different fluids—dodecane and oil—for particles with a radius of $27\, \mu$ m, without using any free parameters.
For completeness, we start by recalling the theory behind force spectroscopy in a thermal environment.

\subsubsection{Thermal vibration spectra}

Consider an experimentally accessible quantity which can be written as a linear combination of the bead position (CG) variables ${\br_i}$. For instance, let $\Phi_{\alpha}= \sum_i a_{i\alpha} r_{i\alpha}$ be some experimental CG variable, e.g. the overall height of a virus, some polymer extension,  the amplitude of a vesicle normal mode, etc. It can be written as $\Phi_{\alpha}=\Phi_{\alpha}^e+\phi_{\alpha}$, where $\phi_{\alpha} (t)$ is the fluctuation about the equilibrium value $\Phi^e_{\alpha}$. Let us consider a nanoscopic system in a thermal environment and obtain the power spectra of the vector $\boldsymbol{\Phi} = \Phi_{\alpha} \bx_{\alpha}$. This requires obtaining the Fourier transform of the time correlation
$\langle \phi_{\alpha}(t) \phi_{\beta}(0)\rangle = \sum_{ij} a_{i\alpha} a_{j\beta}\, \langle q_{i\alpha}(t) q_{j\beta}(0)\rangle$.
In matrix notation $\mathcal{A}_{ij}\equiv a_ia_j$ and $[\mathcal{C}_{qq}]_{ij}\equiv \langle \bq_i(t) \bq_j(0)\rangle$ so that 
$\mathcal{C}_{\phi\phi}(t)=\langle \phi(t) \phi(0)\rangle = \mathcal{A}:\mathcal{C}_{qq}$

The covariance matrix $\mathcal{C}$ can be obtained from the standard procedure. Starting from the stochastic dynamics in real time domain  $m_i \ddot q_i = \kmat_{ik} q_k - \int_0^t [\Gamma_{ik}(t-s) +\mathcal{R}_{ik}(t-s)] \dot q_k(s) ds + \tilde{F}_i$; multiply by $q_j(0)$ and take the time correlation $m_i \langle \ddot q_i(t) q_j(0)\rangle$. Then take Laplace transform and assign $s\rightarrow \ii \omega$ to get 
$$
\ii \omega \left[\ii \omega m_i \delta_{ik} + \frac{\kmat_{ik}}{\ii \omega} + \Gamma_{ik} +\mathcal{R}_{ik} \right] [\hat{\mathcal{C}}_{qq}]_{ik}(\omega)=  \left[\ii \omega m_i \delta_{ik} + \Gamma_{ik} +\mathcal{R}_{ik}\right] [\mathcal{C}_{qq}]_{kj}(0)
$$
Then use equipartition theorem for $\mathcal{C}_{qq}(0)=\langle Q Q^{T}\rangle = k_BT\, \kmat^{-1}$. Comparison with Eq. \ref{eq:buu} yields, 
$$
 \hat{\mathcal{C}}_{qq}(\omega)=  k_BT \,\left[\hmat + \res \right]^{-1} \left[{\bf M} - \frac{\ii}{\omega}\left(\boldsymbol{\Gamma} + \res\right)\right] \kmat^{-1}
$$
And the power spectra results in $S_{\Phi}(\omega)=\hat{\mathcal{C}}_{\phi}(\omega)=\mathcal{A}:\hat{\mathcal{C}}_{qq}(\omega)$

\subsubsection{Vibrational spectra of a colloid tethered to an AFM cantilever }

To illustrate this procedure, we consider the oscillation of the center of mass (CoM) of some 
structure, i.e. $Q_{\alpha} = \sum_i a_i q_{i\alpha}$ with $a_i=m_i/M$ and $M=\sum_i m_i$.  
The same protocol can be generalized to various quantities, such as the CoM velocity $U_{\alpha}$, angular rotation $\Omega_{\alpha}$, and any other linear quantity in $q_i$ describing the shape of the structure (e.g. the components of the spherical harmonics of a vibrating vesicle). We consider a spherical object tethered to some harmonic trap, which could be a laser trap or the tip of an AFM cantilever as illustrated in Fig. \ref{fig:levels}(b). We focus on the latter case as it was experimentally studied in Ref. \cite{AFM_spectra}: the thermal vibrational spectra of a spherical colloid or $27\mu\text{m}$ tethered to an AFM cantilever (Fig. \ref{fig:levels}b). 
The stochastic equation of motion for the fluctuating position of the colloid, expressed in terms of the excess mass diagonal matrix $M_e$ is, 
\begin{equation}
  M_e \ddot {\bf Q} = -{\bf K} {\bf Q} - \int_0^t \boldsymbol{\Xi}(t-s) \dot {\bf Q}(s) ds + {\bf E}
  \label{eq:fs}
\end{equation}

where $\Xi(t)=\boldsymbol{\xi}(t)+\res(t)$ is the total friction kernel which is decomposed into an internal friction $\boldsymbol{\xi}$ and the hydrodynamic contribution $\res_Q$, corresponding to the \emph{induced-force} exchanged by the particle and fluid (this explains the use of the excess mass $M_e$ in Eq. \ref{eq:fs}, see Refs. \cite{Mazur1974,2025_spectral_solver}). Note that for a rigid colloid $\boldsymbol{\xi}={\bf 0}$. Here ${\bf K}$ represents the trap elastic constant (generally, a $3\times3$ matrix) which for a AFM cantilever vibrating in $z$ direction is just ${\bf K}= K\,\bz\bz$.  The external forcing is indicated by ${\bf E}(t)$ and for a thermal environment ${\bf E}=\tilde{{\bf F}}$ corresponds to the thermal noise, to be considered later. Taking the Fourier transform in the above equation leads to 
$$
\left[-M_e\omega^2 {\bf I}_3 + {\bf K} - \ii \omega \boldsymbol{\Xi} \right] {\bf Q}(\omega) = {\bf E}(\omega)
$$
We now multiply and divide by the fluid displacement mass matrix $M_f$ and define $\boldsymbol{\mu}_e=M_f^{-1} M_e$
$$
M_f \left[ \boldsymbol{\Omega}^2- \boldsymbol{\mu}_e \omega^2 {\bf I}_3  - \ii \omega \boldsymbol{\Gamma} \right] {\bf Q}(\omega) = {\bf E}(\omega)
$$
where the eigenvalues of $ \boldsymbol{\Omega}^2\equiv M_f^{-1}{\bf K}$ provide the square of the natural frequency of the trap, and we have introduced the damping matrix (rate units) $\boldsymbol{\Gamma}=M_f^{-1} \boldsymbol{\Xi}$. The susceptibility is defined as
$$
\boldsymbol{\chi}(\omega) =M_f^{-1} \left[ \boldsymbol{\Omega}^2-\boldsymbol{\mu}_e \omega^2 {\bf I}_3  - \ii \omega \boldsymbol{\Gamma} \right]^{-1}
$$
so that $ {\bf Q}(\omega) = \boldsymbol{\chi}(\omega) {\bf E}(\omega)$ and multiplying by its complex conjugate and taking a thermal average,
$$
\langle {\bf Q}(\omega) {\bf Q}^{*}(\omega)\rangle = \boldsymbol{\chi}(\omega) \,\langle {\bf E}(\omega) {\bf E}^*(\omega) \,\rangle \boldsymbol{\chi}^*(\omega) 
$$
This expression is valid for a force vibration with \emph{known} imposed spectra (${\bf E}(t) = \int {\bf E}(\omega) e^{\ii \omega t} d\omega$) and also for vibrations induced by thermal noise, ${\bf E}=\tilde{{\bf F}}$. In this latter case, the vibrational problem is closed thanks to the Second Fluctuation Dissipation Relation (2FDT) \cite{Zwanzig2001}, which relates the noise time correlation and dissipative memory kernel as, 
$$
\langle \tilde{{\bf F}}(\omega) \tilde{{\bf F}}^*(\omega) \rangle = 2 k_BT \, \text{Re}\left[\boldsymbol{\Xi}(\omega)\right]
$$
This relation finally yields the vibration spectra 
$$
\boldsymbol{S}_Q(\omega) = \langle {\bf Q}(\omega) {\bf Q}^{*}(\omega)\rangle = 2 M k_B\,T \boldsymbol{\chi}(\omega) \,\text{Re}[\boldsymbol{\Gamma}]\boldsymbol{\chi}^*(\omega) 
$$
For vertical vibrations $Q_z=\chi_{z z} {\bf E}_{z}$ and this equation becomes a one-dimensional relation,
$$
S_Z(\omega) = \langle  Q_z(\omega) Q_z^{*}(\omega)\rangle = 2M k_B\,T |\chi_{zz}^2|(\omega) \,\text{Re}[\Gamma_{zz}] = 2k_BT \frac{\text{Im}\left[ \chi_{zz}\right]}{\omega}
$$

Finally, we demonstrate how the proposed algorithm can quantitatively predict the spectrum of thermal vibrations of a microscopic particle, based on the calculation of the hydrodynamic resistance in the direction normal to the plane. We compare our numerical results with experimental data from \cite{AFM_spectra}, obtained in two different fluids—dodecane and oil—for particles with a radius of $27,\mu$m, without using any free parameters.

\begin{figure}[ht]
\centering
\includegraphics[width=0.95\textwidth]{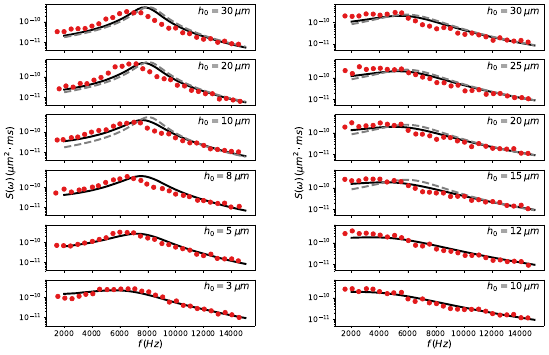}
\caption{
Spectrum of thermal vibrations of spherical microparticles of density $\rho_p = 2230\, \mathrm{kg}/\mathrm{m}^3$ with a radius of $27\,\mu$m, measured in dodecane (right panels) and oil (left panels), as a function of the particle’s height above the bottom surface. The setup is illustrated in Fig. \ref{fig:levels}(b). Red dots correspond to experimental measurements, black lines show the numerical predictions from our model, and the dashed gray line represents the theoretical prediction using the Mazur-Bedaux friction coefficient, Eq. \ref{eq:mb}. The physical parameters employed in the calculation are those reported in Ref.~\cite{AFM_spectra}: $k_{\text{AFM}} = 0.68\,\mathrm{N/m}$, $\eta_{\text{oil}} = 9.3\,\mathrm{mPa{\cdot}s}$, $\eta_{\text{dodecane}} = 1.34\,\mathrm{mPa{\cdot}s}$, $\rho_{\text{oil}} = 930\,\mathrm{kg/m^3}$, and $\rho_{\text{dodecane}} = 750\,\mathrm{kg/m^3}$. The hydrodynamic resistance was computed using the filled-sphere model with $T=7$ subdivisions of the outer icosphere ($h_g\approx0.04\,R$).
}
\label{fig:AFM_spectra}
\end{figure}

As shown in Fig.~\ref{fig:AFM_spectra}, the agreement between the numerical predictions and the experimental measurements is remarkably good across a wide range of heights—from very close to the wall up to distances greater than the particle radius. In this far-field regime, the predictions of our algorithm closely match those obtained using the Mazur-Bedaux expression, Eq.\ref{eq:mb}, which is valid for an isolated sphere far from any boundary. In both fluids, our model accurately captures the shape and amplitude of the fluctuation spectrum.

However, in some cases—particularly when the particle is far from the bottom surface—we observe a slight shift in the position of the spectral peak compared to the experimental data. Since our numerical results converge to the Mazur-Bedaux prediction in this limit, we believe that the observed discrepancies at large distances are unlikely to arise from limitations in the model itself. Instead, they may originate from subtle experimental effects. For instance, the presence of the AFM cantilever could alter the fluctuation dynamics, leading to deviations from the idealized configuration assumed in the theory.

\section{Concluding remarks \label{sec:con}}

This work presents a solver to solve small vibrations of viscoelastic structures in Newtonian fluids.  It uses a fully spectral solver (in space and time) for the oscillatory Stokes equations \cite{2025_spectral_solver} to connect the fluid and the structure, via the tensions generated (by the fluid and/or by external forces) on the viscoelastic networks representing the immersed objects. Two routes have been proposed, one based on calculating the mobility matrix of the structure, to solve its vibrational dynamics in post-processing, and a second one where the fluid-structure coupling is done iteratively, using the Anderson scheme \cite{1965_Anderson,2015_AndersonAcc} adapted to complex (phasor) variables. We have tested the scheme against two experimental setups: quartz-crystal microbalance (QCM) and force spectroscopy via atomic force microscopy (AFM). In the QCM scenario, we have first validated the scheme against accurate solutions \cite{Leshansky2023,Leshansky2024} for the QCM impedance of rigid spheres, either rigidly adsorbed or suspended over the oscillating substrate. For the case of viscoelastic linkers (flexible chains) tethering a spherical globule to the substrate, we have developed an analytical solution which treats the globular part as a hydrodynamic monopole (without internal stress). The code exactly reproduces the analytical solution for this simple model, and allows for insertion of the effect of hydrodynamic multipolar response of the globular domain (stresslet). Preliminary comparisons with experimental results on FtsZ show good agreement with this mesoscopic model. A point for further theoretical advancement is the treatment and the role of thermal fluctuations in these fast-oscillatory flows. In the case of force spectroscopy, we used the Second Fluctuation Dissipation relation to determine, via the mobility route of our solver, the vibrational spectra of colloids over a substrate. Comparison with experimental results is excellent, without \emph{any} adjustable parameter (contrary to what is usually done in the literature \cite{AFM_spectra}). These colloids are huge $27\mu\text{m}$, compared with proteins $2.5\text{nm}$, which indicates the code flexibility dealing with a broad spatial range of scales. Also, the role of fluctuations in the tether proteins case will deserve further study, probably by integrating the fluctuation-dissipation relations into the QCM theory. We find two non-dimensional parameters $\kappa = k/(\omega \xi_S)$ (with $\xi_S=6\pi\eta a$ the Stokes friction of the globule) and $\gamma = \xi/\xi_S$ the internal friction of the linker. For many tether proteins, the ratio $\kappa/\gamma = k/(\omega \xi) \sim O(10)$, meaning that the linker relation rate $k/\xi$ is not far from the imposed oscillation frequency $\omega \sim 10^7 \text{Hz}$. Moreover, the dynamic resonance takes place around $\kappa \sim 1$, close to the protein range, which implies a large sensitivity against mechanical parameters (protein height, linker length, stiffness, etc.) which can even alter the order of the overtones $n$ in dissipation ($\Delta D_n$ increasing or decreasing with $n$). In view of this complexity, a message for future QCM experiments is the need to report results for the \emph{full collection} of overtones, which, although probably not being consistent with the simple Sauerbrey theory, will provide essential information for advanced theory-experiment integrated approaches. Following its integration with AFM spectroscopy or QCM experiments, we expect this solver to be extremely useful to extract information on the internal viscoelasticity of biomolecular structures, either soft, as the coronavirus \cite{GarciaArribas2024}, rigid and yet deformable as the tobacco virus \cite{DiezMartinez2025} or thermally resistant as encapsulines  \cite{Berger2025}.

\section{Data Availability Statement }
The data that support the findings of this study
are openly available in github repository at 
\verb| github.com/orgs/ComplexFluidsUAM/PAPERS/Oscillatory_Viscoelastic|

\begin{acknowledgments}
We acknowledge funding from the Spanish Agencia Estatal de Investigacion (AEI) with grant numbers PID2024-158994OB-C41 and PDC 2021–121441-C21 (Prueba de Concepto program). Also AEI ‘María de Maeztu’ Program for Units of Excellence (grant R\&D CEX 2023–001316-M) and funding from HORIZON-EIC-2023 PATHFINDER program (grant ID 101130615 ‘FASTCOMET’).
\end{acknowledgments}

\appendix

\section{Optimal Lagrangian discretisation: blobs size and spatial distribution \label{app:Lagrange}}

A key aspect when dealing with immersed boundary (IB) method for fluid-structure interaction is the location of the beads or markers forming the structure \cite{vazquez2014multiblob,kallemov2016immersed,balboa-blaise,anna-broms}. We have revisited this problem here and confirmed that the spatial distribution of the beads is robust against the fluid dynamic regime (steady or oscillatory). While the geometric size $R_g$ of a structure (here a sphere for simplicity) is uniquely determined by the spatial location of the beads, relating the IB response to the continuum-media formulae leads to slightly different sizes of the IB structure for each "perturbation". For instance, in Ref. \cite{vazquez2014multiblob}, some of us investigated different "multiblob sizes" arising from a simple model of a rigid no-slip sphere which used 12 (or 13) beads in the vertices (and center) of an icosahedron. For instance, the hydrodynamic size $R_h$ is obtained from the center of mass (CoM) velocity $U_\alpha$ under a forced translation by a force $F_\alpha$, using the Stokes relation $R_h=U_\alpha/( 6\pi \eta  F_\alpha)$. Similarly, \cite{vazquez2014multiblob}, one can measure "effective sizes" from rotations (upon the application of a torque) or the application of a stresslet, or via the fluctuation-dissipation relations for random translations and/or rotations in a thermal solvent (described by the Landau-Lifshitz-Navier-Stokes equations). As shown by Donev and others \cite{kallemov2016immersed,balboa-blaise,anna-broms}, these differences in IB sizes vanish as the Lagrangian resolution (the number of beads per unit surface) is increased. 

\begin{figure}[ht]
\centering
\includegraphics[width=0.95\textwidth]{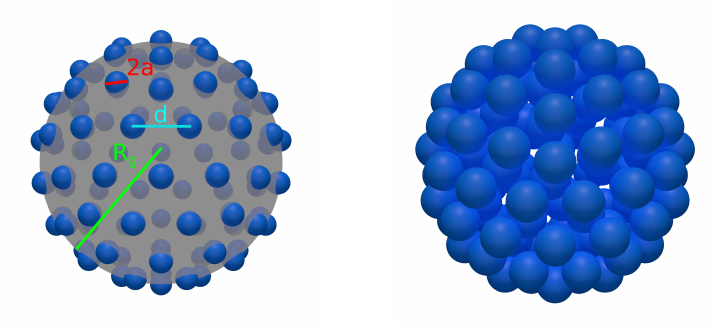}
\caption{Illustration of the two types of Lagrangian discretization employed in this work. (Left panel) A spherical shell with N = 92 beads located at the vertices of an icosphere. (Right panel) A filled sphere with 92 beads on the surface and an additional 55 beads inside, distributed as concentric icospheres with a progressively lower number of divisions.}
\label{fig:DiscretizationModels}
\end{figure}

In this work, we consider two different methods for discretizing a sphere (see Fig.~\ref{fig:DiscretizationModels}): a surface discretization (left panel), similar to those used in previous studies~\cite{anna-broms, balboa-blaise}, and a volumetric discretization (right panel) that allows for the representation of non-neutrally buoyant particles.

\begin{figure}[ht]
\centering
\includegraphics[width=0.95\textwidth]{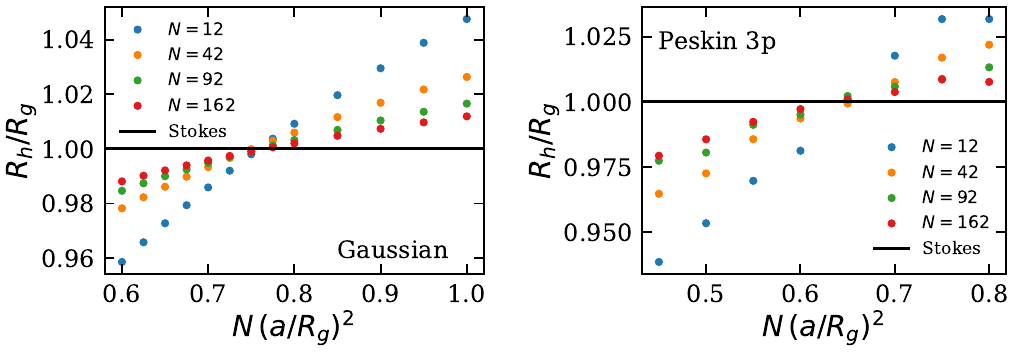}
\caption{Hydrodynamic size of the multiblob as a function of the hydrodynamic size of the beads, using different levels of discretization ranging from 12 to 162 beads. Two types of kernels were employed: a Gaussian kernel (left) and a Peskin kernel (right). Simulations were performed in a box with open boundary conditions and with dimensions $ L_x = L_y = 10 R_g $ and height $ L_z = 3 R_g $, with a penetration length $ \delta = 100 R_g $}
\label{fig:sizeStokes}
\end{figure}

In the surface discretization, the sphere is represented using $N$ beads placed according to an icosphere construction. An icosphere is a triangulated approximation of a sphere, built by recursively subdividing the faces of a regular icosahedron and projecting the resulting vertices onto the spherical surface. This method yields a highly regular discretization in which all beads have six neighbors, except for twelve exceptional vertices with five neighbors, as illustrated in Fig.~\ref{fig:sizeStokes} (left panel). A limitation of this approach is that it only allows specific numbers of beads, given by $N = 10T^2 + 2$, where $T$ is a natural number indicating the subdivision level~\cite{icosphere}.

Consistent with previous observations~\cite{anna-broms, balboa-blaise}, we find that the hydrodynamic response of the multiblob sphere depends not only on the spatial distribution of the beads but also on the hydrodynamic radius of each immersed boundary (IB) bead, denoted $a$~\cite{florentesis,Thesis-Pelaez}. This radius is connected to the mesh size $h$: for the 3-point Peskin kernel, one has $a = 0.91\,h$, while for a Gaussian kernel with standard deviation $\sigma$, the relation is $a = \sigma\, \sqrt{\pi}$~\cite{2014_keaveny}. Although $a$ appears independent of $h$ in the Gaussian case, its optimal value is in fact related to both the mesh size and the truncation tolerance $\epsilon$, as shown in Ref.~\cite{Thesis-Pelaez}:

\begin{eqnarray}\label{eqn:tolerance}
  \sigma(\epsilon) &=& \pi^{1/2} \min[0.55 - 0.11 \log_{10}(3\epsilon), 1.65] \, h \\
  r_c(\epsilon) &=& \left(-\sqrt{2} \log\left[-(2\pi \sigma^2)^{d/2} \epsilon\right]\right)^{1/2} \, \sigma
\end{eqnarray}

Here, $r_c$ denotes the cutoff radius such that $\phi(r_c) = \epsilon$, and we assume $\phi(r > r_c) = 0$.

To determine the optimal value of $a$ for a given number of beads, we measure the translational friction of the multiblob in the regime $\delta \gg R_g$, where the Stokes law is valid. Since the simulations are performed under periodic boundary conditions, we include the Hasimoto correction to the Stokes friction\cite{Hasimoto}. Specifically, we adjust the dimensionless group $N(a/R_g)^2$ so that the measured hydrodynamic radius $R_h$, obtained from

\begin{equation}\label{eq:hasimoto}
    \xi(L) = 6\pi\eta R_h\, \left(1 - 2.8373\frac{R_h}{L}\right)^{-1},
\end{equation}

matches the geometric radius, i.e., $R_h = R_g$.

Once this optimal value is determined, we validate it by comparing the numerical friction results with the theoretical Mazur–Bedeaux prediction in the regime where $\delta$ becomes comparable to $R_g$. Moreover, in Sec.~\ref{sec:qcm}, we use this same hydrodynamic radius to compute the QCM signal, finding perfect agreement with the analytical results derived by Leshansky et al.~\cite{Leshansky2023} for suspended particles.

Figure~\ref{fig:sizeStokes} shows how the hydrodynamic radius of the multiblob changes as a function of the dimensionless group $N(a/R_g)^2$, which is related to the surface coverage by $4\Theta = N(a/R_g)^2$, for different discretization levels ranging from $T=1$—corresponding to an icosahedrons—up to $T=4$, which corresponds to an icosphere with 162 beads. The same calculation was performed using both a Gaussian kernel (left panel) and a 3-point Peskin kernel (right panel).

Notably, all the trends for the ratio $R_h/R_g$ obtained at different discretization levels $T$ intersect approximately at a single "magic" value of the group $N(a/R_g)^2$, which also corresponds to the condition $R_h = R_g$. In the case of the Gaussian kernel, this optimal surface coverage occurs at $4\Theta^*_a \approx 0.76$, yielding $R_h/R_g = 1.000$ with a maximum deviation of only 0.1\%. For the 3-point Peskin kernel, the intersection occurs at $4\Theta^*_a \approx 0.65$, again with $R_h/R_g \approx 1.000$ and a slightly larger maximum deviation of 0.2\%. In both cases, this confirms that the hydrodynamic radius of the multiblob matches the geometric radius at these coverage values, making them optimal for accurate hydrodynamic representation. As done by \cite{anna-broms, balboa-blaise}, we want to determine the typical distance between the beads compared to the beads' radius. In the case of an icosphere, the bead separation can be approximated by (note that for more than 12 beads, the distance between beads is no longer uniform):

\begin{equation}
d = 2 \sin\left(\frac{\varrho}{T}\right)\, R_g
\end{equation}

where $\varrho = \arcsin((\varphi^2 + 1)^{-0.5}) \approx 0.553$ and $\varphi = (1 + \sqrt{5}) / 2$ denotes the golden ratio. This result follows from the expression for the radius of the sphere circumscribed around an icosahedron of edge length $d$, given by \cite{icosahedron} $R_g = d\sqrt{\varphi^2 + 1}\,/2$, combined with the fact that an icosphere is constructed by subdividing each edge of an icosahedron and projecting the resulting mesh onto a sphere. This result allows us to estimate the ratio between the typical distance between beads and their optimal hydrodynamic radius as:

\begin{equation}
    \frac{d}{a} \approx 2\sqrt{\frac{N}{4\Theta^*_a}}\sin(\varrho/T) = \sqrt{\frac{10\,T^2+2}{\Theta^*_a}}\sin(\varrho/T)
\end{equation}

In the case of an icosahedron ($T = 1$), we find that $d/a \approx 4.17$ for a Gaussian kernel while $d/a \approx 4.51$ for a 3-pt Peskin kernel, and this value decreases and asymptotically converges to approximately 4.01 and 4.33 as $T \rightarrow \infty$ for a Gaussian and a three points Peskin kernel repectively. This implies that optimal blob size avoids overlaps between neighboring blobs (for example, $d > 3\,h \approx 3.30\,a$ in the case of a 3-point Peskin kernel), which has the advantage of preserving the effective mass of the multiblob particle. However, this approach to discretizing the sphere presents significant limitations when dealing with particles that are not neutrally buoyant. In such cases, the entire mass of the sphere must be concentrated in the surface beads, which requires assigning them an artificially high density. This unrealistic distribution can lead to unphysical results. Additionally, due to the relatively large spacing between beads compared to their radius, for large spheres (i.e., when the inter-bead distance is comparable to the fluid penetration length), the surrounding fluid can infiltrate the interior of the particle, resulting in spurious flows that may also introduce errors in the calculations.

To overcome the limitations of surface-only discretizations, we propose an alternative strategy in which the interior of the sphere is also filled with beads (see Fig.~\ref{fig:DiscretizationModels}, right panel). This volumetric discretization is achieved by inserting several concentric spherical shells, each discretized using icospheres. The shells are generated by progressively decreasing the subdivision level of the icosahedron toward the center of the sphere, where a single central bead is finally placed. This construction yields a more uniform mass distribution and reduces the emergence of spurious flows, since the entire volume is constrained to move as a rigid body.

In this model, the total number of beads required to discretize the volume depends on the number of subdivisions $ T $ of the outermost shell, and is given by:

\begin{equation}
    N(T) = 1 + \sum_{T_i=1}^T (10\, T_i^2 + 2) = 1 + 2T + \frac{5}{3}T(T+1)(2T+1)
\end{equation}

Unlike the previous model, in this case, the hydrodynamic radius of the beads is not a free parameter. Since this model is intended to represent non-neutrally buoyant materials, we impose that both the total mass of the multiblob particle and the density of each individual bead match the corresponding properties of the target material. This constraint leads to a closed-form expression for the bead radius $a$:

\begin{equation}
    \frac{4}{3}\pi R_g^3\rho = N \vol\rho \quad \Rightarrow \quad \vol = \frac{4}{3N}\pi R_g^3
\end{equation}

For a Gaussian kernel, where $ \vol = 8a^3 $, we obtain:

\begin{equation}
    a_G = \sqrt[3]{\frac{\pi}{6N}} R_g
\end{equation}

While for a Peskin 3-point kernel, where $ \vol = 8h^3 = 8 (a_P/h)^3 $, the resulting expression is:

\begin{equation}
    a_P = \sqrt[3]{\frac{\pi}{6N}} \cdot 0.91 R_g
\end{equation}

Finally, it is important to emphasize that the radii at which each spherical shell is placed are selected such that the surface bead density, defined as $N_{\text{sur}} / (4\pi R^2)$, remains constant across all shells. Here, $N_{\text{sur}}$ denotes the number of beads on a given shell. As we did in the case of a single spherical shell, we can also estimate the typical distance between beads in this volumetric model. In this case of a Gaussian kernel, it is approximately given by:

\begin{equation}
    \frac{d}{a_G} \approx \sqrt[3]{\frac{6N}{\pi}} \sin\left(\frac{\varrho}{T}\right) = 2\sin\left(\frac{\varrho}{T}\right) \sqrt[3]{\frac{6 + 12T + 10T(T+1)(2T+1)}{\pi}}
\end{equation}

While in the case of a 3-pt Peskin kernel, we reach the same expression but divided by 0.91.

In the case of a pure icosahedron (T=1), the typical bead spacing is $d = 3.06a_G$, while in the limit as $T \rightarrow \infty$, this distance approaches $d = 2.04a_G$. 

\begin{figure}[ht]
\centering
\includegraphics[width=0.95\textwidth]{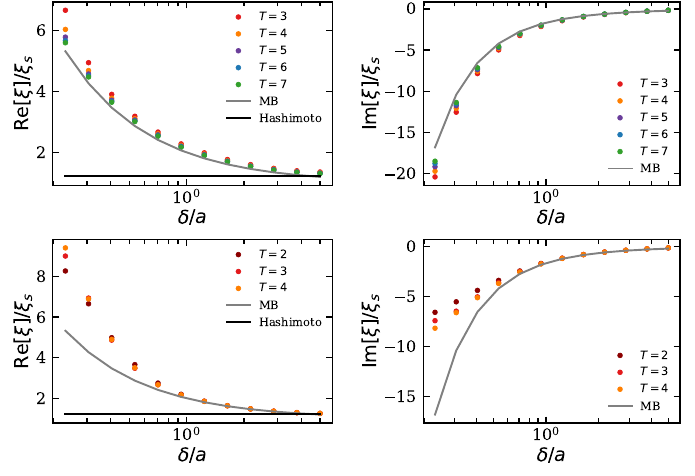}
\caption{Real (left panels) and imaginary (right panels) components of the friction as a function of the penetration length, computed using volumetric discretization (top panels) and surface discretization (bottom panels) at various levels of resolution. Dots represent the numerical results for different discretization levels, grey lines indicate the theoretical Mazur–Bedeaux prediction (Eq.~\ref{eq:MB_friction}), and the black line shows the Stokes friction corrected for periodic boundary conditions using the Hasimoto expression (Eq.~\ref{eq:hasimoto}). In all cases, simulations were performed in a box of size $L_x = L_y = 15a$, with open boundary conditions at $z = -H$. The box height was set to $L_z = 2H = 4a$, and a Gaussian kernel was used throughout.}
\label{fig:MB}
\end{figure}

As a test of the accuracy of the two discretization strategies described above, we now analyze their performance under the action of an oscillatory force in an unbounded fluid, in a regime where the penetration length is comparable to the sphere radius. In this case, the hydrodynamic resistance depends on the forcing frequency, so the standard Stokes relation no longer applies. Instead, the friction is described by the Mazur–Bedeaux expression\cite{Mazur1974}, 

\begin{equation}\label{eq:MB_friction}
    \xi_{MB}=\xi_S\,\left[1+\idel a +\frac{1}{3}\,(\idel a)^2\right]
\end{equation}

where $\idel = (1-i)/\delta$. Figure~\ref{fig:MB} shows the results obtained using both discretization strategies -volumetric (top panels) and surface (bottom panels)- evaluated at different levels of resolution. The effective friction is plotted as a function of the penetration length and compared with the theoretical prediction given by the Mazur–Bedeaux friction.

In the case of the volumetric discretization (top panels), we observe excellent agreement between the numerical results and the theoretical prediction given by the Mazur–Bedeaux friction. However, as the $\delta$ decreases, a finer discretization becomes necessary to maintain this level of accuracy. For example, with a discretization level $T=3$, the grid spacing is $h = 0.09a$, while for $T=7$ it is $h = 0.04a$. When the penetration length becomes comparable, the resolution is insufficient to fully capture the flow within the boundary layer, resulting in a slight deviation from the theoretical curve.

On the other hand, for the surface discretization (bottom panels), the results are in good agreement with the Mazur–Bedeaux prediction for $\delta/a \geq 1$. However, significant discrepancies arise when the penetration length becomes small. We attribute this to the large inter-blob spacing relative to both the blob radius and the penetration length. In such cases, fluid can infiltrate the interior of the particle, effectively altering the physical nature of the object. Instead of a solid, compact rigid sphere—as assumed in the Mazur–Bedeaux model—the system behaves more like a porous structure, leading to deviations in the measured friction.

\section{Iterative Scheme: Alternating Anderson-Jacobi Acceleration \label{app:Anderson}}

In this appendix, we describe in detail the iterative algorithm implemented to efficiently determine the induced forces exerted on the fluid through an iterative scheme. Our implementation follows the approach outlined in Ref.~\cite{2015_AndersonAcc}.

We consider a generic fixed-point equation of the form $\bfind = g(\bfind)$. The simplest algorithm to solve for $\bfind$ consists of starting from an initial guess $\bfind_0$ and repeatedly applying the function $g$:

\begin{equation}
    \bfind_{n+1} = g(\bfind_n) 
\end{equation}

until the convergence condition is reached:
\begin{equation} \label{eq:error_fixed}
    \frac{||\bfind_n - g(\bfind_n)||^2}{||\bfind_n||^2} < \epsilon
\end{equation}

where $\epsilon$ is the prescribed error tolerance. This simple algorithm may fail to converge depending on the properties of the function $g$. One way to improve its convergence is to introduce a relaxation parameter $\beta \in (0,1]$, such that the next iterate is computed as:

\begin{equation} \label{eq:relaxed_fixed}
    \bfind_{n+1} = \beta\, g(\bfind_n) + (1 - \beta)\bfind_n
\end{equation}

In this way, by choosing an appropriate value of $\beta$, convergence can be achieved in cases where the original iteration fails. However, in many situations, even with small values of $\beta$, convergence may still be very slow or not occur at all.

A more robust strategy to improve convergence is the Anderson acceleration method. Unlike the standard fixed-point scheme, which only uses the most recent iterate, Anderson acceleration leverages an arbitrary number of previous iterates to construct an optimal update direction.

To apply this technique, the fixed-point problem is reformulated in residual form as:
\[
f(\bfind) = g(\bfind) - \bfind
\]

At each iteration step, we store the differences between successive values of $\bfind$ and their corresponding residuals in the following matrices:

\begin{align*}
\tilde \bfind_n &= 
\begin{bmatrix}
\bfind_{n-m+1} - \bfind_{n-m}, & \cdots &, \bfind_n - \bfind_{n-1}
\end{bmatrix} \\
\tilde F_n &=
\begin{bmatrix}
f(\bfind_{n-m+1}) - f(\bfind_{n-m}), & \cdots &, f(\bfind_n) - f(\bfind_{n-1})
\end{bmatrix}
\end{align*}
where $m$ is the number of previous steps retained in the history.

Next, we compute a vector $\Gamma_n \in \mathbb{C}^m$ that minimizes the residual in the least-squares sense:
\[
\Gamma_n = \arg\min_{\Gamma \in \mathbb{C}^m} \left\| \tilde F_n \Gamma - f(\bfind_n) \right\|_2
\]

This minimization is carried out using a QR decomposition of $\tilde F_n$, resulting in an orthogonal matrix $Q$ and an upper triangular matrix $R$ such that:

\[
\tilde F_n = QR \quad \Rightarrow \quad R \Gamma_n = Q^H f(\bfind_n)
\]

where $Q^H$ the conjugate transpose of $Q$.

Once $\Gamma_n$ is obtained, the next value of $\bfind$ is computed as:

\begin{equation} \label{eq:anderson}
\bfind_{n+1} = \bfind_n + \beta f(\bfind_n) - [\tilde \bfind_n + \beta \tilde F_n] \Gamma_n
\end{equation}

Following the strategy proposed in Ref.~\cite{2015_AndersonAcc}, we have implemented an algorithm that alternates between relaxed fixed-point iterations (Equation~\ref{eq:relaxed_fixed}) and Anderson acceleration steps (Equation~\ref{eq:anderson}) to further enhance convergence.

Typically, the super-vector $\bfind$ can contain a large number of elements, and computing the error (Eq.~\ref{eq:error_fixed}) involves two reduction operations on the GPU, which can be relatively expensive. For this reason, the error is not evaluated after every iteration, but rather every $m$ iterations.

Furthermore, to improve the robustness of the algorithm, if the error increases compared to the last evaluation a specified number of times, the relaxation parameter $\beta$ is slightly reduced. This allows the solution to approach the fixed point more smoothly and helps promote convergence in challenging scenarios.

The algorithm parameters used in the computations presented in this work were: $m \in [10, 100]$, $\beta \in [10^{-7}, 10^{-4}]$, $\epsilon=10^{-4}$. The error is evaluated every $m$ iterations, and if it increases in 5 consecutive evaluations, the value of $\beta$ is scaled by a factor of 0.75. Relaxed fixed-point iterations and Anderson acceleration steps are alternated with a 1:1 ratio.

\section{Convergence \label{app:Conv}}

In this section, we analyze the numerical error of the iterative scheme as a function of the structural properties (specifically, rigidity) and the error tolerance defined in Equation~\ref{eq:error_fixed}.

We begin by evaluating how the relative error ($RE$) in the computed wall impedance during a QCM experiment varies with the error tolerance $\epsilon$, considering different values of bond stiffness.
To estimate the relative error, we take as a reference value ($Z_r$) the result obtained with a tight tolerance of $\epsilon = 10^{-6}$ and compute $RE$ as:

\begin{equation}
RE(\epsilon) = \frac{||Z(\epsilon) - Z_r||}{||Z_r||}
\end{equation}

\begin{figure}
\centering
\includegraphics[width=0.95\linewidth]{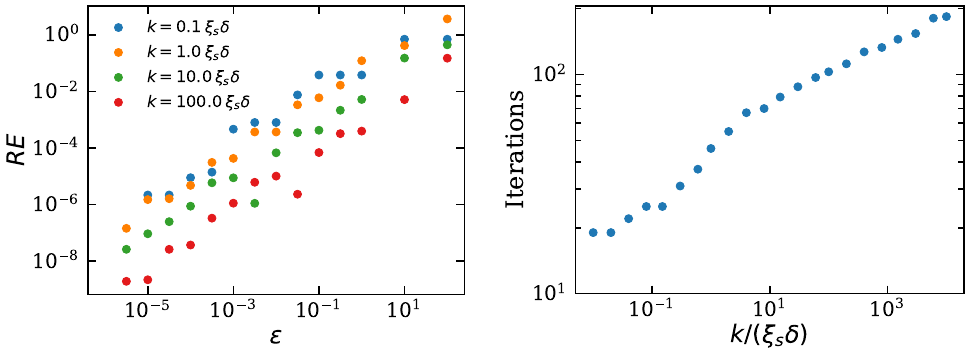}
\caption{(Left panel) Relative error in wall impedance in a QCM experiment as a function of error tolerance, for various values of $\kappa$ ranging from highly elastic to very stiff.
(Right panel) Number of iterations required to achieve convergence with an error tolerance of $10^{-4}$, plotted against bond strength.
In both panels, a spherical shell of radius $0.5\delta$ positioned at a height of $0.65\delta$ above the wall was used.}
\label{fig:errors}
\end{figure}

The right panel of Fig. \ref{fig:errors} shows that the relative error increases with tolerance approximately following a power law. Notably, with a tolerance of $\epsilon = 10^{-2}$, the relative error remains below 1\% across all cases.

In addition, we examined how the number of iterations required to reach convergence varies with the spring stiffness.
As shown in the right panel, the number of iterations increases with stiffness; however, in all analyzed cases, convergence was achieved in fewer than 200 iterations. This number is significantly lower than the number of time steps that would be required to perform a full time-resolved simulation of a QCM experiment.

%


\begin{thebibliography}{85}%
\makeatletter
\providecommand \@ifxundefined [1]{%
 \@ifx{#1\undefined}
}%
\providecommand \@ifnum [1]{%
 \ifnum #1\expandafter \@firstoftwo
 \else \expandafter \@secondoftwo
 \fi
}%
\providecommand \@ifx [1]{%
 \ifx #1\expandafter \@firstoftwo
 \else \expandafter \@secondoftwo
 \fi
}%
\providecommand \natexlab [1]{#1}%
\providecommand \enquote  [1]{``#1''}%
\providecommand \bibnamefont  [1]{#1}%
\providecommand \bibfnamefont [1]{#1}%
\providecommand \citenamefont [1]{#1}%
\providecommand \href@noop [0]{\@secondoftwo}%
\providecommand \href [0]{\begingroup \@sanitize@url \@href}%
\providecommand \@href[1]{\@@startlink{#1}\@@href}%
\providecommand \@@href[1]{\endgroup#1\@@endlink}%
\providecommand \@sanitize@url [0]{\catcode `\\12\catcode `\$12\catcode
  `\&12\catcode `\#12\catcode `\^12\catcode `\_12\catcode `\%12\relax}%
\providecommand \@@startlink[1]{}%
\providecommand \@@endlink[0]{}%
\providecommand \url  [0]{\begingroup\@sanitize@url \@url }%
\providecommand \@url [1]{\endgroup\@href {#1}{\urlprefix }}%
\providecommand \urlprefix  [0]{URL }%
\providecommand \Eprint [0]{\href }%
\providecommand \doibase [0]{https://doi.org/}%
\providecommand \selectlanguage [0]{\@gobble}%
\providecommand \bibinfo  [0]{\@secondoftwo}%
\providecommand \bibfield  [0]{\@secondoftwo}%
\providecommand \translation [1]{[#1]}%
\providecommand \BibitemOpen [0]{}%
\providecommand \bibitemStop [0]{}%
\providecommand \bibitemNoStop [0]{.\EOS\space}%
\providecommand \EOS [0]{\spacefactor3000\relax}%
\providecommand \BibitemShut  [1]{\csname bibitem#1\endcsname}%
\let\auto@bib@innerbib\@empty
\bibitem [{\citenamefont {Gopalakrishna}\ \emph {et~al.}(2021)\citenamefont
  {Gopalakrishna}, \citenamefont {Langhoff}, \citenamefont {Brenner},\ and\
  \citenamefont {Johannsmann}}]{Johannsmann2021}%
  \BibitemOpen
  \bibfield  {author} {\bibinfo {author} {\bibfnamefont {S.}~\bibnamefont
  {Gopalakrishna}}, \bibinfo {author} {\bibfnamefont {A.}~\bibnamefont
  {Langhoff}}, \bibinfo {author} {\bibfnamefont {G.}~\bibnamefont {Brenner}},\
  and\ \bibinfo {author} {\bibfnamefont {D.}~\bibnamefont {Johannsmann}},\
  }\bibfield  {title} {\bibinfo {title} {Soft viscoelastic particles in contact
  with a quartz crystal microbalance (qcm): A frequency-domain lattice
  boltzmann simulation},\ }\href {https://doi.org/10.1021/acs.analchem.1c01612}
  {\bibfield  {journal} {\bibinfo  {journal} {Analytical Chemistry}\ }\textbf
  {\bibinfo {volume} {93}},\ \bibinfo {pages} {10229} (\bibinfo {year}
  {2021})},\ \Eprint
  {https://arxiv.org/abs/https://doi.org/10.1021/acs.analchem.1c01612}
  {https://doi.org/10.1021/acs.analchem.1c01612} \BibitemShut {NoStop}%
\bibitem [{\citenamefont {de~Beer}\ \emph {et~al.}(2008)\citenamefont
  {de~Beer}, \citenamefont {van~den Ende},\ and\ \citenamefont
  {Mugele}}]{AFMhydro_APL2008}%
  \BibitemOpen
  \bibfield  {author} {\bibinfo {author} {\bibfnamefont {S.}~\bibnamefont
  {de~Beer}}, \bibinfo {author} {\bibfnamefont {D.}~\bibnamefont {van~den
  Ende}},\ and\ \bibinfo {author} {\bibfnamefont {F.}~\bibnamefont {Mugele}},\
  }\bibfield  {title} {\bibinfo {title} {{Atomic force microscopy cantilever
  dynamics in liquid in the presence of tip sample interaction}},\ }\href@noop
  {} {\bibfield  {journal} {\bibinfo  {journal} {Applied Physics Letters}\
  }\textbf {\bibinfo {volume} {93}},\ \bibinfo {pages} {253106} (\bibinfo
  {year} {2008})}\BibitemShut {NoStop}%
\bibitem [{\citenamefont {Calzado-Mart{\'{i}}n}\ \emph
  {et~al.}(2016)\citenamefont {Calzado-Mart{\'{i}}n}, \citenamefont {Encinar},
  \citenamefont {Tamayo}, \citenamefont {Calleja},\ and\ \citenamefont
  {Paulo}}]{CalzadoMartin2016}%
  \BibitemOpen
  \bibfield  {author} {\bibinfo {author} {\bibfnamefont {A.}~\bibnamefont
  {Calzado-Mart{\'{i}}n}}, \bibinfo {author} {\bibfnamefont {M.}~\bibnamefont
  {Encinar}}, \bibinfo {author} {\bibfnamefont {J.}~\bibnamefont {Tamayo}},
  \bibinfo {author} {\bibfnamefont {M.}~\bibnamefont {Calleja}},\ and\ \bibinfo
  {author} {\bibfnamefont {{\'{A}}.~S.}\ \bibnamefont {Paulo}},\ }\bibfield
  {title} {\bibinfo {title} {Effect of actin organization on the stiffness of
  living breast cancer cells revealed by peak-force modulation atomic force
  microscopy},\ }\href@noop {} {\bibfield  {journal} {\bibinfo  {journal} {ACS
  Nano}\ }\textbf {\bibinfo {volume} {10}},\ \bibinfo {pages} {3365} (\bibinfo
  {year} {2016})}\BibitemShut {NoStop}%
\bibitem [{\citenamefont {Heenan}\ \emph {et~al.}(2021)\citenamefont {Heenan},
  \citenamefont {Jacobson}, \citenamefont {Woodside}, \citenamefont {Gaub},\
  and\ \citenamefont {Perkins}}]{Heenan2021}%
  \BibitemOpen
  \bibfield  {author} {\bibinfo {author} {\bibfnamefont {P.}~\bibnamefont
  {Heenan}}, \bibinfo {author} {\bibfnamefont {D.}~\bibnamefont {Jacobson}},
  \bibinfo {author} {\bibfnamefont {M.}~\bibnamefont {Woodside}}, \bibinfo
  {author} {\bibfnamefont {H.}~\bibnamefont {Gaub}},\ and\ \bibinfo {author}
  {\bibfnamefont {T.~T.}\ \bibnamefont {Perkins}},\ }\bibfield  {title}
  {\bibinfo {title} {Modulation of a protein-folding landscape revealed by
  afm-based force spectroscopy notwithstanding instrumental limitations},\
  }\href@noop {} {\bibfield  {journal} {\bibinfo  {journal} {PNAS}\ }\textbf
  {\bibinfo {volume} {118}},\ \bibinfo {pages} {e2015728118} (\bibinfo {year}
  {2021})}\BibitemShut {NoStop}%
\bibitem [{\citenamefont {Zhong}\ \emph {et~al.}(2021)\citenamefont {Zhong},
  \citenamefont {Rösch}, \citenamefont {Viereck}, \citenamefont {Schilling},\
  and\ \citenamefont {Ludwig}}]{2021_covid_harmonics}%
  \BibitemOpen
  \bibfield  {author} {\bibinfo {author} {\bibfnamefont {J.}~\bibnamefont
  {Zhong}}, \bibinfo {author} {\bibfnamefont {E.~L.}\ \bibnamefont {Rösch}},
  \bibinfo {author} {\bibfnamefont {T.}~\bibnamefont {Viereck}}, \bibinfo
  {author} {\bibfnamefont {M.}~\bibnamefont {Schilling}},\ and\ \bibinfo
  {author} {\bibfnamefont {F.}~\bibnamefont {Ludwig}},\ }\bibfield  {title}
  {\bibinfo {title} {Toward rapid and sensitive detection of sars-cov-2 with
  functionalized magnetic nanoparticles},\ }\href@noop {} {\bibfield  {journal}
  {\bibinfo  {journal} {ACS Sensors}\ }\textbf {\bibinfo {volume} {6}},\
  \bibinfo {pages} {976} (\bibinfo {year} {2021})}\BibitemShut {NoStop}%
\bibitem [{\citenamefont {Du}\ \emph {et~al.}(2022)\citenamefont {Du},
  \citenamefont {Cui}, \citenamefont {Sun}, \citenamefont {Zhang},
  \citenamefont {Bai},\ and\ \citenamefont
  {Yoshida}}]{2022_susceptibility_empirical_biosensing}%
  \BibitemOpen
  \bibfield  {author} {\bibinfo {author} {\bibfnamefont {Z.}~\bibnamefont
  {Du}}, \bibinfo {author} {\bibfnamefont {Y.}~\bibnamefont {Cui}}, \bibinfo
  {author} {\bibfnamefont {Y.}~\bibnamefont {Sun}}, \bibinfo {author}
  {\bibfnamefont {H.}~\bibnamefont {Zhang}}, \bibinfo {author} {\bibfnamefont
  {S.}~\bibnamefont {Bai}},\ and\ \bibinfo {author} {\bibfnamefont
  {T.}~\bibnamefont {Yoshida}},\ }\bibfield  {title} {\bibinfo {title}
  {Empirical expression of ac susceptibility of magnetic nanoparticles and
  potential application in biosensing},\ }\href
  {https://doi.org/10.1109/TNB.2021.3126905} {\bibfield  {journal} {\bibinfo
  {journal} {IEEE Transactions on NanoBioscience}\ }\textbf {\bibinfo {volume}
  {21}},\ \bibinfo {pages} {496} (\bibinfo {year} {2022})}\BibitemShut
  {NoStop}%
\bibitem [{\citenamefont {Aranda}\ \emph {et~al.}(2008)\citenamefont {Aranda},
  \citenamefont {Riske}, \citenamefont {Lipowsky},\ and\ \citenamefont
  {Dimova}}]{Aranda2008}%
  \BibitemOpen
  \bibfield  {author} {\bibinfo {author} {\bibfnamefont {S.}~\bibnamefont
  {Aranda}}, \bibinfo {author} {\bibfnamefont {K.~A.}\ \bibnamefont {Riske}},
  \bibinfo {author} {\bibfnamefont {R.}~\bibnamefont {Lipowsky}},\ and\
  \bibinfo {author} {\bibfnamefont {R.}~\bibnamefont {Dimova}},\ }\bibfield
  {title} {\bibinfo {title} {Morphological transitions of vesicles induced by
  alternating electric fields},\ }\href@noop {} {\bibfield  {journal} {\bibinfo
   {journal} {Biophysical Journal}\ }\textbf {\bibinfo {volume} {95}},\
  \bibinfo {pages} {L19} (\bibinfo {year} {2008})}\BibitemShut {NoStop}%
\bibitem [{\citenamefont {Vlahovska}(2015)}]{Vlahovska2015}%
  \BibitemOpen
  \bibfield  {author} {\bibinfo {author} {\bibfnamefont {P.~M.}\ \bibnamefont
  {Vlahovska}},\ }\bibfield  {title} {\bibinfo {title} {Voltage-morphology
  coupling in biomimetic membranes: dynamics of giant vesicles in applied
  electric fields},\ }\href@noop {} {\bibfield  {journal} {\bibinfo  {journal}
  {Soft Matter}\ }\textbf {\bibinfo {volume} {11}},\ \bibinfo {pages} {7232}
  (\bibinfo {year} {2015})}\BibitemShut {NoStop}%
\bibitem [{\citenamefont {Lazanas}\ and\ \citenamefont
  {Prodromidis}(2023)}]{2023Lanzanas}%
  \BibitemOpen
  \bibfield  {author} {\bibinfo {author} {\bibfnamefont {A.~C.}\ \bibnamefont
  {Lazanas}}\ and\ \bibinfo {author} {\bibfnamefont {M.~I.}\ \bibnamefont
  {Prodromidis}},\ }\bibfield  {title} {\bibinfo {title} {Electrochemical
  impedance spectroscopy-a tutorial},\ }\href
  {https://doi.org/10.1021/acsmeasuresciau.2c00070} {\bibfield  {journal}
  {\bibinfo  {journal} {ACS Measurement Science Au}\ }\textbf {\bibinfo
  {volume} {3}},\ \bibinfo {pages} {162} (\bibinfo {year} {2023})}\BibitemShut
  {NoStop}%
\bibitem [{\citenamefont {Balboa~Usabiaga}\ and\ \citenamefont
  {Delgado-Buscalioni}(2013)}]{balboaUltrasound13}%
  \BibitemOpen
  \bibfield  {author} {\bibinfo {author} {\bibfnamefont {F.}~\bibnamefont
  {Balboa~Usabiaga}}\ and\ \bibinfo {author} {\bibfnamefont {R.}~\bibnamefont
  {Delgado-Buscalioni}},\ }\bibfield  {title} {\bibinfo {title} {Minimal model
  for acoustic forces on brownian particles},\ }\href@noop {} {\bibfield
  {journal} {\bibinfo  {journal} {Phys. Rev. E}\ }\textbf {\bibinfo {volume}
  {88}},\ \bibinfo {pages} {063304} (\bibinfo {year} {2013})}\BibitemShut
  {NoStop}%
\bibitem [{\citenamefont {Bruus}(2012)}]{Bruus2012}%
  \BibitemOpen
  \bibfield  {author} {\bibinfo {author} {\bibfnamefont {H.}~\bibnamefont
  {Bruus}},\ }\bibfield  {title} {\bibinfo {title} {Acoustofluidics 2:
  Perturbation theory and ultrasound resonance modes},\ }\href@noop {}
  {\bibfield  {journal} {\bibinfo  {journal} {Lab. Chip}\ }\textbf {\bibinfo
  {volume} {12}},\ \bibinfo {pages} {20} (\bibinfo {year} {2012})}\BibitemShut
  {NoStop}%
\bibitem [{\citenamefont {Fernández-Mateo}\ \emph {et~al.}(2021)\citenamefont
  {Fernández-Mateo}, \citenamefont {García-Sánchez}, \citenamefont {Calero},
  \citenamefont {Morgan},\ and\ \citenamefont {Ramos}}]{Ramos_2021}%
  \BibitemOpen
  \bibfield  {author} {\bibinfo {author} {\bibfnamefont {R.}~\bibnamefont
  {Fernández-Mateo}}, \bibinfo {author} {\bibfnamefont {P.}~\bibnamefont
  {García-Sánchez}}, \bibinfo {author} {\bibfnamefont {V.}~\bibnamefont
  {Calero}}, \bibinfo {author} {\bibfnamefont {H.}~\bibnamefont {Morgan}},\
  and\ \bibinfo {author} {\bibfnamefont {A.}~\bibnamefont {Ramos}},\ }\bibfield
   {title} {\bibinfo {title} {Stationary electro-osmotic flow driven by ac
  fields around charged dielectric spheres},\ }\href
  {https://doi.org/10.1017/jfm.2021.650} {\bibfield  {journal} {\bibinfo
  {journal} {J. Fluid Mech.}\ }\textbf {\bibinfo {volume} {924}},\ \bibinfo
  {pages} {R2} (\bibinfo {year} {2021})}\BibitemShut {NoStop}%
\bibitem [{\citenamefont {Bruus}(2007)}]{Bruus2007}%
  \BibitemOpen
  \bibfield  {author} {\bibinfo {author} {\bibfnamefont {H.}~\bibnamefont
  {Bruus}},\ }\href@noop {} {\emph {\bibinfo {title} {Theoretical
  Microfluidics}}},\ \bibinfo {series} {Oxford Master Series in Physics},
  Vol.~\bibinfo {volume} {18}\ (\bibinfo  {publisher} {Oxford University
  Press},\ \bibinfo {year} {2007})\BibitemShut {NoStop}%
\bibitem [{\citenamefont {Huang}\ \emph {et~al.}(2022)\citenamefont {Huang},
  \citenamefont {Louis}, \citenamefont {Bresol{\'\i}-Obach}, \citenamefont
  {Kudo}, \citenamefont {Camacho}, \citenamefont {Scheblykin}, \citenamefont
  {Sugiyama}, \citenamefont {Hofkens},\ and\ \citenamefont
  {Masuhara}}]{huang2022primeval}%
  \BibitemOpen
  \bibfield  {author} {\bibinfo {author} {\bibfnamefont {C.-H.}\ \bibnamefont
  {Huang}}, \bibinfo {author} {\bibfnamefont {B.}~\bibnamefont {Louis}},
  \bibinfo {author} {\bibfnamefont {R.}~\bibnamefont {Bresol{\'\i}-Obach}},
  \bibinfo {author} {\bibfnamefont {T.}~\bibnamefont {Kudo}}, \bibinfo {author}
  {\bibfnamefont {R.}~\bibnamefont {Camacho}}, \bibinfo {author} {\bibfnamefont
  {I.~G.}\ \bibnamefont {Scheblykin}}, \bibinfo {author} {\bibfnamefont
  {T.}~\bibnamefont {Sugiyama}}, \bibinfo {author} {\bibfnamefont
  {J.}~\bibnamefont {Hofkens}},\ and\ \bibinfo {author} {\bibfnamefont
  {H.}~\bibnamefont {Masuhara}},\ }\bibfield  {title} {\bibinfo {title} {The
  primeval optical evolving matter by optical binding inside and outside the
  photon beam},\ }\href@noop {} {\bibfield  {journal} {\bibinfo  {journal}
  {Nature Communications}\ }\textbf {\bibinfo {volume} {13}} (\bibinfo {year}
  {2022})}\BibitemShut {NoStop}%
\bibitem [{\citenamefont {Palacios-Alonso}\ \emph {et~al.}(2025)\citenamefont
  {Palacios-Alonso}, \citenamefont {Shams}, \citenamefont {Ozel-Okcu},
  \citenamefont {Sanz-de Diego}, \citenamefont {Teran},\ and\ \citenamefont
  {Delgado-Buscalioni}}]{2025_palacios}%
  \BibitemOpen
  \bibfield  {author} {\bibinfo {author} {\bibfnamefont {P.}~\bibnamefont
  {Palacios-Alonso}}, \bibinfo {author} {\bibfnamefont {M.~M.}\ \bibnamefont
  {Shams}}, \bibinfo {author} {\bibfnamefont {S.}~\bibnamefont {Ozel-Okcu}},
  \bibinfo {author} {\bibfnamefont {E.}~\bibnamefont {Sanz-de Diego}}, \bibinfo
  {author} {\bibfnamefont {F.~J.}\ \bibnamefont {Teran}},\ and\ \bibinfo
  {author} {\bibfnamefont {R.}~\bibnamefont {Delgado-Buscalioni}},\ }\bibfield
  {title} {\bibinfo {title} {Fast and accurate characterization of
  bioconjugated particles and solvent properties by a general nonlinear
  analytical relationship for the ac magnetic hysteresis area},\ }\href@noop {}
  {\bibfield  {journal} {\bibinfo  {journal} {Nanoscale}\ ,\ } (\bibinfo {year}
  {2025})}\BibitemShut {NoStop}%
\bibitem [{\citenamefont {Sanz-de Diego}\ \emph {et~al.}(2024)\citenamefont
  {Sanz-de Diego}, \citenamefont {Aires}, \citenamefont {Palacios-Alonso},
  \citenamefont {Cabrera}, \citenamefont {Silvestri}, \citenamefont
  {Vequi-Suplicy}, \citenamefont {Artés-Ibáñez}, \citenamefont
  {Requejo-Isidro}, \citenamefont {Delgado-Buscalioni}, \citenamefont
  {Pellegrino}, \citenamefont {Cortajarena},\ and\ \citenamefont
  {Terán}}]{2024_Elena}%
  \BibitemOpen
  \bibfield  {author} {\bibinfo {author} {\bibfnamefont {E.}~\bibnamefont
  {Sanz-de Diego}}, \bibinfo {author} {\bibfnamefont {A.}~\bibnamefont
  {Aires}}, \bibinfo {author} {\bibfnamefont {P.}~\bibnamefont
  {Palacios-Alonso}}, \bibinfo {author} {\bibfnamefont {D.}~\bibnamefont
  {Cabrera}}, \bibinfo {author} {\bibfnamefont {N.}~\bibnamefont {Silvestri}},
  \bibinfo {author} {\bibfnamefont {C.~C.}\ \bibnamefont {Vequi-Suplicy}},
  \bibinfo {author} {\bibfnamefont {E.~J.}\ \bibnamefont {Artés-Ibáñez}},
  \bibinfo {author} {\bibfnamefont {J.}~\bibnamefont {Requejo-Isidro}},
  \bibinfo {author} {\bibfnamefont {R.}~\bibnamefont {Delgado-Buscalioni}},
  \bibinfo {author} {\bibfnamefont {T.}~\bibnamefont {Pellegrino}}, \bibinfo
  {author} {\bibfnamefont {A.~L.}\ \bibnamefont {Cortajarena}},\ and\ \bibinfo
  {author} {\bibfnamefont {F.~J.}\ \bibnamefont {Terán}},\ }\bibfield  {title}
  {\bibinfo {title} {Multiparametric modulation of magnetic transduction for
  biomolecular sensing in liquids},\ }\href@noop {} {\bibfield  {journal}
  {\bibinfo  {journal} {Nanoscale}\ }\textbf {\bibinfo {volume} {16}},\
  \bibinfo {pages} {4082} (\bibinfo {year} {2024})}\BibitemShut {NoStop}%
\bibitem [{\citenamefont {Et-Thakafy}\ \emph {et~al.}(2017)\citenamefont
  {Et-Thakafy}, \citenamefont {Delorme}, \citenamefont {Gaillard},
  \citenamefont {M{\'{e}}riadec}, \citenamefont {Artzner}, \citenamefont
  {Lopez},\ and\ \citenamefont {Guyomarc'h}}]{2017-liposomes}%
  \BibitemOpen
  \bibfield  {author} {\bibinfo {author} {\bibfnamefont {O.}~\bibnamefont
  {Et-Thakafy}}, \bibinfo {author} {\bibfnamefont {N.}~\bibnamefont {Delorme}},
  \bibinfo {author} {\bibfnamefont {C.}~\bibnamefont {Gaillard}}, \bibinfo
  {author} {\bibfnamefont {C.}~\bibnamefont {M{\'{e}}riadec}}, \bibinfo
  {author} {\bibfnamefont {F.}~\bibnamefont {Artzner}}, \bibinfo {author}
  {\bibfnamefont {C.}~\bibnamefont {Lopez}},\ and\ \bibinfo {author}
  {\bibfnamefont {F.}~\bibnamefont {Guyomarc'h}},\ }\bibfield  {title}
  {\bibinfo {title} {{Mechanical Properties of Membranes Composed of Gel-Phase
  or Fluid-Phase Phospholipids Probed on Liposomes by Atomic Force
  Spectroscopy}},\ }\href@noop {} {\bibfield  {journal} {\bibinfo  {journal}
  {Langmuir}\ }\textbf {\bibinfo {volume} {33}},\ \bibinfo {pages} {5117}
  (\bibinfo {year} {2017})}\BibitemShut {NoStop}%
\bibitem [{\citenamefont {Carrey}\ \emph {et~al.}(2011)\citenamefont {Carrey},
  \citenamefont {Mehdaoui},\ and\ \citenamefont
  {Respaud}}]{2011_Carrey_empiricalEQ_hyperthermia}%
  \BibitemOpen
  \bibfield  {author} {\bibinfo {author} {\bibfnamefont {J.}~\bibnamefont
  {Carrey}}, \bibinfo {author} {\bibfnamefont {B.}~\bibnamefont {Mehdaoui}},\
  and\ \bibinfo {author} {\bibfnamefont {M.}~\bibnamefont {Respaud}},\
  }\bibfield  {title} {\bibinfo {title} {{Simple models for dynamic hysteresis
  loop calculations of magnetic single-domain nanoparticles: Application to
  magnetic hyperthermia optimization}},\ }\href@noop {} {\bibfield  {journal}
  {\bibinfo  {journal} {Journal of Applied Physics}\ }\textbf {\bibinfo
  {volume} {109}},\ \bibinfo {pages} {083921} (\bibinfo {year}
  {2011})}\BibitemShut {NoStop}%
\bibitem [{\citenamefont {Puerto-Belda}\ \emph {et~al.}(2024)\citenamefont
  {Puerto-Belda}, \citenamefont {Ruz}, \citenamefont {Mill{\'a}}, \citenamefont
  {Cano}, \citenamefont {Yubero}, \citenamefont {Garc{\'i}a}, \citenamefont
  {Kosaka}, \citenamefont {Calleja},\ and\ \citenamefont
  {Tamayo}}]{Puerto-Belda2024}%
  \BibitemOpen
  \bibfield  {author} {\bibinfo {author} {\bibfnamefont {V.}~\bibnamefont
  {Puerto-Belda}}, \bibinfo {author} {\bibfnamefont {J.~J.}\ \bibnamefont
  {Ruz}}, \bibinfo {author} {\bibfnamefont {C.}~\bibnamefont {Mill{\'a}}},
  \bibinfo {author} {\bibfnamefont {{\'A}.}~\bibnamefont {Cano}}, \bibinfo
  {author} {\bibfnamefont {M.~L.}\ \bibnamefont {Yubero}}, \bibinfo {author}
  {\bibfnamefont {S.}~\bibnamefont {Garc{\'i}a}}, \bibinfo {author}
  {\bibfnamefont {P.~M.}\ \bibnamefont {Kosaka}}, \bibinfo {author}
  {\bibfnamefont {M.}~\bibnamefont {Calleja}},\ and\ \bibinfo {author}
  {\bibfnamefont {J.}~\bibnamefont {Tamayo}},\ }\bibfield  {title} {\bibinfo
  {title} {Measuring vibrational modes in living human cells},\ }\href@noop {}
  {\bibfield  {journal} {\bibinfo  {journal} {PRX Life}\ }\textbf {\bibinfo
  {volume} {2}},\ \bibinfo {pages} {013003} (\bibinfo {year}
  {2024})}\BibitemShut {NoStop}%
\bibitem [{\citenamefont {Martín-González}\ \emph {et~al.}(2021)\citenamefont
  {Martín-González}, \citenamefont {Ibáñez-Freire}, \citenamefont {Álvaro
  Ortega-Esteban}, \citenamefont {Laguna-Castro}, \citenamefont {Martín},
  \citenamefont {Valbuena}, \citenamefont {Delgado-Buscalioni},\ and\
  \citenamefont {de~Pablo}}]{MartinGonzalez2021}%
  \BibitemOpen
  \bibfield  {author} {\bibinfo {author} {\bibfnamefont {N.}~\bibnamefont
  {Martín-González}}, \bibinfo {author} {\bibfnamefont {P.}~\bibnamefont
  {Ibáñez-Freire}}, \bibinfo {author} {\bibnamefont {Álvaro
  Ortega-Esteban}}, \bibinfo {author} {\bibfnamefont {M.}~\bibnamefont
  {Laguna-Castro}}, \bibinfo {author} {\bibfnamefont {C.~S.}\ \bibnamefont
  {Martín}}, \bibinfo {author} {\bibfnamefont {A.}~\bibnamefont {Valbuena}},
  \bibinfo {author} {\bibfnamefont {R.}~\bibnamefont {Delgado-Buscalioni}},\
  and\ \bibinfo {author} {\bibfnamefont {P.~J.}\ \bibnamefont {de~Pablo}},\
  }\bibfield  {title} {\bibinfo {title} {Long-range cooperative disassembly and
  aging during adenovirus uncoating},\ }\href@noop {} {\bibfield  {journal}
  {\bibinfo  {journal} {Phys.l Rev. X}\ }\textbf {\bibinfo {volume} {11}},\
  \bibinfo {pages} {021025} (\bibinfo {year} {2021})}\BibitemShut {NoStop}%
\bibitem [{\citenamefont {Patel}\ \emph {et~al.}(2021)\citenamefont {Patel},
  \citenamefont {Jha},\ and\ \citenamefont {K.}}]{Patel2021}%
  \BibitemOpen
  \bibfield  {author} {\bibinfo {author} {\bibfnamefont {A.}~\bibnamefont
  {Patel}}, \bibinfo {author} {\bibfnamefont {H.~S.}\ \bibnamefont {Jha}},\
  and\ \bibinfo {author} {\bibfnamefont {S.}~\bibnamefont {K.}},\ }\bibfield
  {title} {\bibinfo {title} {Deconvolution of dissipative pathways for the
  interpretation of tapping-mode atomic force microscopy from phase-contrast},\
  }\href@noop {} {\bibfield  {journal} {\bibinfo  {journal} {Communications
  Physics}\ }\textbf {\bibinfo {volume} {4}} (\bibinfo {year}
  {2021})}\BibitemShut {NoStop}%
\bibitem [{\citenamefont {Yongabi}\ \emph {et~al.}(2020)\citenamefont
  {Yongabi}, \citenamefont {Khorshid}, \citenamefont {Gennaro}, \citenamefont
  {Jooken}, \citenamefont {Duw{\'e}}, \citenamefont {Deschaume}, \citenamefont
  {Losada-P{\'e}rez}, \citenamefont {Dedecker}, \citenamefont {Bartic},
  \citenamefont {W{\"u}bbenhorst},\ and\ \citenamefont
  {Wagner}}]{2020QCM_cells}%
  \BibitemOpen
  \bibfield  {author} {\bibinfo {author} {\bibfnamefont {D.}~\bibnamefont
  {Yongabi}}, \bibinfo {author} {\bibfnamefont {M.}~\bibnamefont {Khorshid}},
  \bibinfo {author} {\bibfnamefont {A.}~\bibnamefont {Gennaro}}, \bibinfo
  {author} {\bibfnamefont {S.}~\bibnamefont {Jooken}}, \bibinfo {author}
  {\bibfnamefont {S.}~\bibnamefont {Duw{\'e}}}, \bibinfo {author}
  {\bibfnamefont {O.}~\bibnamefont {Deschaume}}, \bibinfo {author}
  {\bibfnamefont {P.}~\bibnamefont {Losada-P{\'e}rez}}, \bibinfo {author}
  {\bibfnamefont {P.}~\bibnamefont {Dedecker}}, \bibinfo {author}
  {\bibfnamefont {C.}~\bibnamefont {Bartic}}, \bibinfo {author} {\bibfnamefont
  {M.}~\bibnamefont {W{\"u}bbenhorst}},\ and\ \bibinfo {author} {\bibfnamefont
  {P.}~\bibnamefont {Wagner}},\ }\bibfield  {title} {\bibinfo {title} {Qcm-d
  study of time-resolved cell adhesion and detachment: Effect of surface free
  energy on eukaryotes and prokaryotes},\ }\href
  {https://doi.org/10.1021/acsami.0c00353} {\bibfield  {journal} {\bibinfo
  {journal} {ACS Applied Materials \& Interfaces}\ }\textbf {\bibinfo {volume}
  {12}},\ \bibinfo {pages} {18258} (\bibinfo {year} {2020})},\ \bibinfo {note}
  {pMID: 32223273},\ \Eprint
  {https://arxiv.org/abs/https://doi.org/10.1021/acsami.0c00353}
  {https://doi.org/10.1021/acsami.0c00353} \BibitemShut {NoStop}%
\bibitem [{\citenamefont {Bingen}\ \emph {et~al.}(2008)\citenamefont {Bingen},
  \citenamefont {Wang}, \citenamefont {Steinmetz}, \citenamefont {Rodahl},\
  and\ \citenamefont {Richter}}]{Bingen2008}%
  \BibitemOpen
  \bibfield  {author} {\bibinfo {author} {\bibfnamefont {P.}~\bibnamefont
  {Bingen}}, \bibinfo {author} {\bibfnamefont {G.}~\bibnamefont {Wang}},
  \bibinfo {author} {\bibfnamefont {N.~F.}\ \bibnamefont {Steinmetz}}, \bibinfo
  {author} {\bibfnamefont {M.}~\bibnamefont {Rodahl}},\ and\ \bibinfo {author}
  {\bibfnamefont {R.~P.}\ \bibnamefont {Richter}},\ }\bibfield  {title}
  {\bibinfo {title} {{Solvation effects in the quartz crystal microbalance with
  dissipation monitoring response to biomolecular adsorption. A
  phenomenological approach}},\ }\href@noop {} {\bibfield  {journal} {\bibinfo
  {journal} {Anal. Chem.}\ }\textbf {\bibinfo {volume} {80}},\ \bibinfo {pages}
  {8880} (\bibinfo {year} {2008})}\BibitemShut {NoStop}%
\bibitem [{\citenamefont {Adeel~Afzal}\ \emph {et~al.}(2017)\citenamefont
  {Adeel~Afzal}, \citenamefont {Mujahi}, \citenamefont {Schirhagl},
  \citenamefont {Bajwa}, \citenamefont {Latif},\ and\ \citenamefont
  {Feroz}}]{Feroz2017}%
  \BibitemOpen
  \bibfield  {author} {\bibinfo {author} {\bibfnamefont {A.}~\bibnamefont
  {Adeel~Afzal}}, \bibinfo {author} {\bibfnamefont {A.}~\bibnamefont {Mujahi}},
  \bibinfo {author} {\bibfnamefont {R.}~\bibnamefont {Schirhagl}}, \bibinfo
  {author} {\bibfnamefont {A.~Z.}\ \bibnamefont {Bajwa}}, \bibinfo {author}
  {\bibfnamefont {U.}~\bibnamefont {Latif}},\ and\ \bibinfo {author}
  {\bibfnamefont {S.}~\bibnamefont {Feroz}},\ }\bibfield  {title} {\bibinfo
  {title} {Gravimetric viral diagnostics: Qcm based biosensors for early
  detection of viruses},\ }\href@noop {} {\bibfield  {journal} {\bibinfo
  {journal} {Chemosensors}\ }\textbf {\bibinfo {volume} {5}},\ \bibinfo {pages}
  {7} (\bibinfo {year} {2017})}\BibitemShut {NoStop}%
\bibitem [{\citenamefont {Bratek-Skicki}\ \emph {et~al.}(2021)\citenamefont
  {Bratek-Skicki}, \citenamefont {Sadowska}, \citenamefont
  {Maciejewska-Prończuk},\ and\ \citenamefont {Adamczyk}}]{Adamczyk2021}%
  \BibitemOpen
  \bibfield  {author} {\bibinfo {author} {\bibfnamefont {A.}~\bibnamefont
  {Bratek-Skicki}}, \bibinfo {author} {\bibfnamefont {M.}~\bibnamefont
  {Sadowska}}, \bibinfo {author} {\bibfnamefont {J.}~\bibnamefont
  {Maciejewska-Prończuk}},\ and\ \bibinfo {author} {\bibfnamefont
  {Z.}~\bibnamefont {Adamczyk}},\ }\bibfield  {title} {\bibinfo {title}
  {Nanoparticle and bioparticle deposition kinetics: Quartz microbalance
  measurements},\ }\bibfield  {journal} {\bibinfo  {journal} {Nanomaterials}\
  }\textbf {\bibinfo {volume} {11}},\ \href
  {https://doi.org/10.3390/nano11010145} {10.3390/nano11010145} (\bibinfo
  {year} {2021})\BibitemShut {NoStop}%
\bibitem [{\citenamefont {Adamczyk}\ and\ \citenamefont
  {Sadowska}(2020)}]{Sadowska2020}%
  \BibitemOpen
  \bibfield  {author} {\bibinfo {author} {\bibfnamefont {Z.}~\bibnamefont
  {Adamczyk}}\ and\ \bibinfo {author} {\bibfnamefont {M.}~\bibnamefont
  {Sadowska}},\ }\bibfield  {title} {\bibinfo {title} {{Hydrodynamic Solvent
  Coupling Effects in Quartz Crystal Microbalance Measurements of Nanoparticle
  Deposition Kinetics}},\ }\href@noop {} {\bibfield  {journal} {\bibinfo
  {journal} {Analytical Chemistry}\ }\textbf {\bibinfo {volume} {92}},\
  \bibinfo {pages} {3896} (\bibinfo {year} {2020})}\BibitemShut {NoStop}%
\bibitem [{\citenamefont {Johannsmann}(2015)}]{Johannsmann2015}%
  \BibitemOpen
  \bibfield  {author} {\bibinfo {author} {\bibfnamefont {D.}~\bibnamefont
  {Johannsmann}},\ }\href@noop {} {\emph {\bibinfo {title} {{The Quartz Crystal
  Microbalance in Soft Matter Research, Fundamentals and modeling}}}}\
  (\bibinfo  {publisher} {Springer},\ \bibinfo {year} {2015})\BibitemShut
  {NoStop}%
\bibitem [{\citenamefont {Schofield}\ and\ \citenamefont
  {Delgado-Buscalioni}(2021)}]{BuscalioniSM21}%
  \BibitemOpen
  \bibfield  {author} {\bibinfo {author} {\bibfnamefont {M.~M.}\ \bibnamefont
  {Schofield}}\ and\ \bibinfo {author} {\bibfnamefont {R.}~\bibnamefont
  {Delgado-Buscalioni}},\ }\bibfield  {title} {\bibinfo {title} {Quantitative
  description of the response of finite size adsorbates on a quartz crystal
  microbalance in liquids using analytical hydrodynamics},\ }\href@noop {}
  {\bibfield  {journal} {\bibinfo  {journal} {Soft Matter}\ }\textbf {\bibinfo
  {volume} {17}},\ \bibinfo {pages} {8160} (\bibinfo {year}
  {2021})}\BibitemShut {NoStop}%
\bibitem [{\citenamefont {Wolny}\ \emph {et~al.}(2010)\citenamefont {Wolny},
  \citenamefont {Spatz},\ and\ \citenamefont {Richter}}]{Wolny2010}%
  \BibitemOpen
  \bibfield  {author} {\bibinfo {author} {\bibfnamefont {P.~M.}\ \bibnamefont
  {Wolny}}, \bibinfo {author} {\bibfnamefont {J.~P.}\ \bibnamefont {Spatz}},\
  and\ \bibinfo {author} {\bibfnamefont {R.~P.}\ \bibnamefont {Richter}},\
  }\bibfield  {title} {\bibinfo {title} {On the adsorption behavior of
  biotin-binding proteins on gold and silica},\ }\href@noop {} {\bibfield
  {journal} {\bibinfo  {journal} {Langmuir}\ }\textbf {\bibinfo {volume}
  {26}},\ \bibinfo {pages} {1029} (\bibinfo {year} {2010})}\BibitemShut
  {NoStop}%
\bibitem [{\citenamefont {Louis}\ \emph {et~al.}(2023)\citenamefont {Louis},
  \citenamefont {Huang}, \citenamefont {Camacho}, \citenamefont {Scheblykin},
  \citenamefont {Sugiyama}, \citenamefont {Kudo}, \citenamefont {Melendez},
  \citenamefont {Delgado-Buscalioni}, \citenamefont {Masuhara}, \citenamefont
  {Hofkens},\ and\ \citenamefont {Bresol{\'\i}-Obach}}]{louis2023unravelling}%
  \BibitemOpen
  \bibfield  {author} {\bibinfo {author} {\bibfnamefont {B.}~\bibnamefont
  {Louis}}, \bibinfo {author} {\bibfnamefont {C.-H.}\ \bibnamefont {Huang}},
  \bibinfo {author} {\bibfnamefont {R.}~\bibnamefont {Camacho}}, \bibinfo
  {author} {\bibfnamefont {I.~G.}\ \bibnamefont {Scheblykin}}, \bibinfo
  {author} {\bibfnamefont {T.}~\bibnamefont {Sugiyama}}, \bibinfo {author}
  {\bibfnamefont {T.}~\bibnamefont {Kudo}}, \bibinfo {author} {\bibfnamefont
  {M.}~\bibnamefont {Melendez}}, \bibinfo {author} {\bibfnamefont
  {R.}~\bibnamefont {Delgado-Buscalioni}}, \bibinfo {author} {\bibfnamefont
  {H.}~\bibnamefont {Masuhara}}, \bibinfo {author} {\bibfnamefont
  {J.}~\bibnamefont {Hofkens}},\ and\ \bibinfo {author} {\bibfnamefont
  {R.}~\bibnamefont {Bresol{\'\i}-Obach}},\ }\bibfield  {title} {\bibinfo
  {title} {Unravelling 3d dynamics and hydrodynamics during incorporation of
  dielectric particles to an optical trapping site},\ }\href@noop {} {\bibfield
   {journal} {\bibinfo  {journal} {ACS Nano}\ }\textbf {\bibinfo {volume}
  {17}},\ \bibinfo {pages} {3797} (\bibinfo {year} {2023})}\BibitemShut
  {NoStop}%
\bibitem [{\citenamefont {Monago}\ \emph {et~al.}(2025)\citenamefont {Monago},
  \citenamefont {Torre}, \citenamefont {Delgado-Buscalioni},\ and\
  \citenamefont {Español}}]{2025Monago}%
  \BibitemOpen
  \bibfield  {author} {\bibinfo {author} {\bibfnamefont {C.}~\bibnamefont
  {Monago}}, \bibinfo {author} {\bibfnamefont {J.~A. d.~l.}\ \bibnamefont
  {Torre}}, \bibinfo {author} {\bibfnamefont {R.}~\bibnamefont
  {Delgado-Buscalioni}},\ and\ \bibinfo {author} {\bibfnamefont
  {P.}~\bibnamefont {Español}},\ }\bibfield  {title} {\bibinfo {title}
  {Unraveling internal friction in a coarse-grained protein model},\
  }\href@noop {} {\bibfield  {journal} {\bibinfo  {journal} {J. Chem. Phys.}\
  }\textbf {\bibinfo {volume} {162}},\ \bibinfo {pages} {114115} (\bibinfo
  {year} {2025})}\BibitemShut {NoStop}%
\bibitem [{\citenamefont {Gillissen}\ \emph {et~al.}(2018)\citenamefont
  {Gillissen}, \citenamefont {Jackman}, \citenamefont {Tabaei},\ and\
  \citenamefont {Cho}}]{Gillissen2018}%
  \BibitemOpen
  \bibfield  {author} {\bibinfo {author} {\bibfnamefont {J.~J.~J.}\
  \bibnamefont {Gillissen}}, \bibinfo {author} {\bibfnamefont {J.~A.}\
  \bibnamefont {Jackman}}, \bibinfo {author} {\bibfnamefont {S.~R.}\
  \bibnamefont {Tabaei}},\ and\ \bibinfo {author} {\bibfnamefont {N.-J.}\
  \bibnamefont {Cho}},\ }\bibfield  {title} {\bibinfo {title} {{A Numerical
  Study on the Effect of Particle Surface Coverage on the Quartz Crystal
  Microbalance Response}},\ }\href
  {https://doi.org/10.1021/acs.analchem.7b04607} {\bibfield  {journal}
  {\bibinfo  {journal} {Analytical Chemistry}\ }\textbf {\bibinfo {volume}
  {90}},\ \bibinfo {pages} {2238} (\bibinfo {year} {2018})}\BibitemShut
  {NoStop}%
\bibitem [{\citenamefont {Delgado-Buscalioni}(2024)}]{Buscalioni_langmuir23}%
  \BibitemOpen
  \bibfield  {author} {\bibinfo {author} {\bibfnamefont {R.}~\bibnamefont
  {Delgado-Buscalioni}},\ }\bibfield  {title} {\bibinfo {title} {Coverage
  effects in quartz crystal microbalance measurements with suspended and
  adsorbed nanoparticles},\ }\href
  {https://doi.org/10.1021/acs.langmuir.3c02792} {\bibfield  {journal}
  {\bibinfo  {journal} {Langmuir}\ }\textbf {\bibinfo {volume} {40}},\ \bibinfo
  {pages} {580} (\bibinfo {year} {2024})}\BibitemShut {NoStop}%
\bibitem [{\citenamefont {Fouxon}\ \emph {et~al.}(2023)\citenamefont {Fouxon},
  \citenamefont {Rubinstein},\ and\ \citenamefont {Leshansky}}]{Leshansky2023}%
  \BibitemOpen
  \bibfield  {author} {\bibinfo {author} {\bibfnamefont {I.}~\bibnamefont
  {Fouxon}}, \bibinfo {author} {\bibfnamefont {B.~Y.}\ \bibnamefont
  {Rubinstein}},\ and\ \bibinfo {author} {\bibfnamefont {A.~M.}\ \bibnamefont
  {Leshansky}},\ }\bibfield  {title} {\bibinfo {title} {Excess shear force
  exerted on an oscillating plate due to a nearby particle},\ }\href
  {https://doi.org/10.1103/PhysRevFluids.8.054104} {\bibfield  {journal}
  {\bibinfo  {journal} {Phys. Rev. Fluids}\ }\textbf {\bibinfo {volume} {8}},\
  \bibinfo {pages} {054104} (\bibinfo {year} {2023})}\BibitemShut {NoStop}%
\bibitem [{\citenamefont {Leshansky}\ \emph {et~al.}(2024)\citenamefont
  {Leshansky}, \citenamefont {Rubinstein}, \citenamefont {Fouxon},
  \citenamefont {Johannsmann}, \citenamefont {Sadowska},\ and\ \citenamefont
  {Adamczyk}}]{Leshansky2024}%
  \BibitemOpen
  \bibfield  {author} {\bibinfo {author} {\bibfnamefont {A.}~\bibnamefont
  {Leshansky}}, \bibinfo {author} {\bibfnamefont {B.}~\bibnamefont
  {Rubinstein}}, \bibinfo {author} {\bibfnamefont {I.}~\bibnamefont {Fouxon}},
  \bibinfo {author} {\bibfnamefont {D.}~\bibnamefont {Johannsmann}}, \bibinfo
  {author} {\bibfnamefont {M.}~\bibnamefont {Sadowska}},\ and\ \bibinfo
  {author} {\bibfnamefont {Z.}~\bibnamefont {Adamczyk}},\ }\bibfield  {title}
  {\bibinfo {title} {Quartz crystal microbalance frequency response to discrete
  adsorbates in liquids},\ }\href
  {https://doi.org/10.1021/acs.analchem.4c00968} {\bibfield  {journal}
  {\bibinfo  {journal} {Analytical chemistry}\ }\textbf {\bibinfo {volume}
  {96}} (\bibinfo {year} {2024})}\BibitemShut {NoStop}%
\bibitem [{\citenamefont {Sun}\ \emph {et~al.}(2023)\citenamefont {Sun},
  \citenamefont {Du}, \citenamefont {Zhang}, \citenamefont {Wang},
  \citenamefont {Sasayama},\ and\ \citenamefont
  {Yoshida}}]{2023_Yoshida_simultaneous}%
  \BibitemOpen
  \bibfield  {author} {\bibinfo {author} {\bibfnamefont {Y.}~\bibnamefont
  {Sun}}, \bibinfo {author} {\bibfnamefont {Z.}~\bibnamefont {Du}}, \bibinfo
  {author} {\bibfnamefont {H.}~\bibnamefont {Zhang}}, \bibinfo {author}
  {\bibfnamefont {H.}~\bibnamefont {Wang}}, \bibinfo {author} {\bibfnamefont
  {T.}~\bibnamefont {Sasayama}},\ and\ \bibinfo {author} {\bibfnamefont
  {T.}~\bibnamefont {Yoshida}},\ }\bibfield  {title} {\bibinfo {title}
  {Simultaneous estimation of magnetic moment and brownian relaxation time
  distributions of magnetic nanoparticles based on magnetic particle
  spectroscopy for biosensing application},\ }\href@noop {} {\bibfield
  {journal} {\bibinfo  {journal} {Nanoscale}\ }\textbf {\bibinfo {volume}
  {15}},\ \bibinfo {pages} {16089} (\bibinfo {year} {2023})}\BibitemShut
  {NoStop}%
\bibitem [{\citenamefont {Zwanzig}(1961)}]{Zwanzig1961}%
  \BibitemOpen
  \bibfield  {author} {\bibinfo {author} {\bibfnamefont {R.}~\bibnamefont
  {Zwanzig}},\ }\bibfield  {title} {\bibinfo {title} {Memory effects in
  irreversible thermodynamics},\ }\href@noop {} {\bibfield  {journal} {\bibinfo
   {journal} {Physical Review}\ }\textbf {\bibinfo {volume} {124}},\ \bibinfo
  {pages} {983} (\bibinfo {year} {1961})}\BibitemShut {NoStop}%
\bibitem [{\citenamefont {Hij{\'{o}}n}\ \emph {et~al.}(2010)\citenamefont
  {Hij{\'{o}}n}, \citenamefont {Espa{\~{n}}ol}, \citenamefont
  {Vanden-Eijnden},\ and\ \citenamefont {Delgado-Buscalioni}}]{Hijon_2010}%
  \BibitemOpen
  \bibfield  {author} {\bibinfo {author} {\bibfnamefont {C.}~\bibnamefont
  {Hij{\'{o}}n}}, \bibinfo {author} {\bibfnamefont {P.}~\bibnamefont
  {Espa{\~{n}}ol}}, \bibinfo {author} {\bibfnamefont {E.}~\bibnamefont
  {Vanden-Eijnden}},\ and\ \bibinfo {author} {\bibfnamefont {R.}~\bibnamefont
  {Delgado-Buscalioni}},\ }\bibfield  {title} {\bibinfo {title} {{Mori-Zwanzig
  formalism as a practical computational tool}},\ }\href
  {https://doi.org/10.1039/b902479b} {\bibfield  {journal} {\bibinfo  {journal}
  {Faraday discussions}\ }\textbf {\bibinfo {volume} {144}},\ \bibinfo {pages}
  {301} (\bibinfo {year} {2010})}\BibitemShut {NoStop}%
\bibitem [{\citenamefont {García-Arribas}\ \emph {et~al.}(2024)\citenamefont
  {García-Arribas}, \citenamefont {Ibáñez-Freire}, \citenamefont {Carlero},
  \citenamefont {Palacios-Alonso}, \citenamefont {Cantero-Reviejo},
  \citenamefont {Ares}, \citenamefont {López-Polín}, \citenamefont {Yan},
  \citenamefont {Wang}, \citenamefont {Sarkar}, \citenamefont {Chhowalla},
  \citenamefont {Oksanen}, \citenamefont {Martín-Benito}, \citenamefont
  {de~Pablo},\ and\ \citenamefont {Delgado-Buscalioni}}]{GarciaArribas2024}%
  \BibitemOpen
  \bibfield  {author} {\bibinfo {author} {\bibfnamefont {A.~B.}\ \bibnamefont
  {García-Arribas}}, \bibinfo {author} {\bibfnamefont {P.}~\bibnamefont
  {Ibáñez-Freire}}, \bibinfo {author} {\bibfnamefont {D.}~\bibnamefont
  {Carlero}}, \bibinfo {author} {\bibfnamefont {P.}~\bibnamefont
  {Palacios-Alonso}}, \bibinfo {author} {\bibfnamefont {M.}~\bibnamefont
  {Cantero-Reviejo}}, \bibinfo {author} {\bibfnamefont {P.}~\bibnamefont
  {Ares}}, \bibinfo {author} {\bibfnamefont {G.}~\bibnamefont {López-Polín}},
  \bibinfo {author} {\bibfnamefont {H.}~\bibnamefont {Yan}}, \bibinfo {author}
  {\bibfnamefont {Y.}~\bibnamefont {Wang}}, \bibinfo {author} {\bibfnamefont
  {S.}~\bibnamefont {Sarkar}}, \bibinfo {author} {\bibfnamefont
  {M.}~\bibnamefont {Chhowalla}}, \bibinfo {author} {\bibfnamefont {H.~M.}\
  \bibnamefont {Oksanen}}, \bibinfo {author} {\bibfnamefont {J.}~\bibnamefont
  {Martín-Benito}}, \bibinfo {author} {\bibfnamefont {P.~J.}\ \bibnamefont
  {de~Pablo}},\ and\ \bibinfo {author} {\bibfnamefont {R.}~\bibnamefont
  {Delgado-Buscalioni}},\ }\bibfield  {title} {\bibinfo {title} {Broad
  adaptability of coronavirus adhesion revealed from the complementary surface
  affinity of membrane and spikes},\ }\href
  {https://doi.org/10.1002/advs.202404186} {\bibfield  {journal} {\bibinfo
  {journal} {Advanced Science}\ }\textbf {\bibinfo {volume} {11}},\ \bibinfo
  {pages} {2404186} (\bibinfo {year} {2024})}\BibitemShut {NoStop}%
\bibitem [{\citenamefont {Ayton}\ and\ \citenamefont {Voth}(2004)}]{2004Voth}%
  \BibitemOpen
  \bibfield  {author} {\bibinfo {author} {\bibfnamefont {G.~S.}\ \bibnamefont
  {Ayton}}\ and\ \bibinfo {author} {\bibfnamefont {G.~A.}\ \bibnamefont
  {Voth}},\ }\bibfield  {title} {\bibinfo {title} {{Mesoscopic Lateral
  Diffusion in Lipid Bilayers}},\ }\href
  {http://www.sciencedirect.com/science/article/pii/S0006349504737972}
  {\bibfield  {journal} {\bibinfo  {journal} {Biophysical Journal}\ }\textbf
  {\bibinfo {volume} {87}},\ \bibinfo {pages} {3299} (\bibinfo {year}
  {2004})}\BibitemShut {NoStop}%
\bibitem [{\citenamefont {Peláez}\ \emph {et~al.}(2025)\citenamefont
  {Peláez}, \citenamefont {Palacios-Alonso},\ and\ \citenamefont
  {Delgado-Buscalioni}}]{2025_spectral_solver}%
  \BibitemOpen
  \bibfield  {author} {\bibinfo {author} {\bibfnamefont {R.~P.}\ \bibnamefont
  {Peláez}}, \bibinfo {author} {\bibfnamefont {P.}~\bibnamefont
  {Palacios-Alonso}},\ and\ \bibinfo {author} {\bibfnamefont {R.}~\bibnamefont
  {Delgado-Buscalioni}},\ }\bibfield  {title} {\bibinfo {title} {Spectral
  solver for the oscillatory stokes frequency-based equation in doubly periodic
  confined domains},\ }\href {https://doi.org/10.1017/jfm.2025.279} {\bibfield
  {journal} {\bibinfo  {journal} {J. Fluid Mech.}\ }\textbf {\bibinfo {volume}
  {1010}},\ \bibinfo {pages} {A57} (\bibinfo {year} {2025})}\BibitemShut
  {NoStop}%
\bibitem [{\citenamefont {Hashemi}\ \emph {et~al.}(2023)\citenamefont
  {Hashemi}, \citenamefont {Peláez}, \citenamefont {Natesh}, \citenamefont
  {Sprinkle}, \citenamefont {Maxian}, \citenamefont {Gan},\ and\ \citenamefont
  {Donev}}]{2023aleksdp}%
  \BibitemOpen
  \bibfield  {author} {\bibinfo {author} {\bibfnamefont {A.}~\bibnamefont
  {Hashemi}}, \bibinfo {author} {\bibfnamefont {R.~P.}\ \bibnamefont
  {Peláez}}, \bibinfo {author} {\bibfnamefont {S.}~\bibnamefont {Natesh}},
  \bibinfo {author} {\bibfnamefont {B.}~\bibnamefont {Sprinkle}}, \bibinfo
  {author} {\bibfnamefont {O.}~\bibnamefont {Maxian}}, \bibinfo {author}
  {\bibfnamefont {Z.}~\bibnamefont {Gan}},\ and\ \bibinfo {author}
  {\bibfnamefont {A.}~\bibnamefont {Donev}},\ }\bibfield  {title} {\bibinfo
  {title} {{Computing hydrodynamic interactions in confined doubly periodic
  geometries in linear time}},\ }\href@noop {} {\bibfield  {journal} {\bibinfo
  {journal} {J. Chem. Phys.}\ }\textbf {\bibinfo {volume} {158}},\ \bibinfo
  {pages} {154101} (\bibinfo {year} {2023})}\BibitemShut {NoStop}%
\bibitem [{\citenamefont {Pérez~Peláez}(2022)}]{Thesis-Pelaez}%
  \BibitemOpen
  \bibfield  {author} {\bibinfo {author} {\bibfnamefont {R.}~\bibnamefont
  {Pérez~Peláez}},\ }{\selectlanguage {english}\emph {\bibinfo {title}
  {Complex fluids in the {Gpu} era. {Algorithms} and simulations}}},\ \href
  {https://repositorio.uam.es/handle/10486/703353} {\bibinfo {type} {{Doctoral
  Thesis}}},\ \bibinfo  {school} {Universidad Autónoma de Madrid} (\bibinfo
  {year} {2022})\BibitemShut {NoStop}%
\bibitem [{\citenamefont {V{\'{a}}zquez-Quesada}\ \emph
  {et~al.}(2020)\citenamefont {V{\'{a}}zquez-Quesada}, \citenamefont
  {Mel{\'{e}}ndez-Schofield}, \citenamefont {Tsortos}, \citenamefont
  {Mateos-Gil}, \citenamefont {Milioni}, \citenamefont {Gizeli},\ and\
  \citenamefont {Delgado\~Buscalioni}}]{Prapp2020}%
  \BibitemOpen
  \bibfield  {author} {\bibinfo {author} {\bibfnamefont {A.}~\bibnamefont
  {V{\'{a}}zquez-Quesada}}, \bibinfo {author} {\bibfnamefont {M.}~\bibnamefont
  {Mel{\'{e}}ndez-Schofield}}, \bibinfo {author} {\bibfnamefont
  {A.}~\bibnamefont {Tsortos}}, \bibinfo {author} {\bibfnamefont
  {P.}~\bibnamefont {Mateos-Gil}}, \bibinfo {author} {\bibfnamefont
  {D.}~\bibnamefont {Milioni}}, \bibinfo {author} {\bibfnamefont
  {E.}~\bibnamefont {Gizeli}},\ and\ \bibinfo {author} {\bibfnamefont
  {R.}~\bibnamefont {Delgado\~Buscalioni}},\ }\bibfield  {title} {\bibinfo
  {title} {{Hydrodynamics of Quartz-Crystal-Microbalance DNA Sensors Based on
  Liposome Amplifiers}},\ }\href@noop {} {\bibfield  {journal} {\bibinfo
  {journal} {Phys. Rev. Applied}\ }\textbf {\bibinfo {volume} {13}},\ \bibinfo
  {pages} {64059} (\bibinfo {year} {2020})}\BibitemShut {NoStop}%
\bibitem [{\citenamefont {Zhang}\ \emph {et~al.}(2023)\citenamefont {Zhang},
  \citenamefont {Bertin}, \citenamefont {Essink}, \citenamefont {Zhang},
  \citenamefont {Fares}, \citenamefont {Shen}, \citenamefont {Bickel},
  \citenamefont {Salez},\ and\ \citenamefont {Maali}}]{AFM_spectra}%
  \BibitemOpen
  \bibfield  {author} {\bibinfo {author} {\bibfnamefont {Z.}~\bibnamefont
  {Zhang}}, \bibinfo {author} {\bibfnamefont {V.}~\bibnamefont {Bertin}},
  \bibinfo {author} {\bibfnamefont {M.~H.}\ \bibnamefont {Essink}}, \bibinfo
  {author} {\bibfnamefont {H.}~\bibnamefont {Zhang}}, \bibinfo {author}
  {\bibfnamefont {N.}~\bibnamefont {Fares}}, \bibinfo {author} {\bibfnamefont
  {Z.}~\bibnamefont {Shen}}, \bibinfo {author} {\bibfnamefont {T.}~\bibnamefont
  {Bickel}}, \bibinfo {author} {\bibfnamefont {T.}~\bibnamefont {Salez}},\ and\
  \bibinfo {author} {\bibfnamefont {A.}~\bibnamefont {Maali}},\ }\bibfield
  {title} {\bibinfo {title} {Unsteady drag force on an immersed sphere
  oscillating near a wall},\ }\href {https://doi.org/10.1017/jfm.2023.987}
  {\bibfield  {journal} {\bibinfo  {journal} {J. Fluid Mech.}\ }\textbf
  {\bibinfo {volume} {977}},\ \bibinfo {pages} {A21} (\bibinfo {year}
  {2023})}\BibitemShut {NoStop}%
\bibitem [{\citenamefont {Español}\ and\ \citenamefont
  {Donev}(2015)}]{2015PepAleks}%
  \BibitemOpen
  \bibfield  {author} {\bibinfo {author} {\bibfnamefont {P.}~\bibnamefont
  {Español}}\ and\ \bibinfo {author} {\bibfnamefont {A.}~\bibnamefont
  {Donev}},\ }\bibfield  {title} {\bibinfo {title} {Coupling a nano-particle
  with isothermal fluctuating hydrodynamics: Coarse-graining from microscopic
  to mesoscopic dynamics},\ }\href@noop {} {\bibfield  {journal} {\bibinfo
  {journal} {J. Chem. Phys.}\ }\textbf {\bibinfo {volume} {143}},\ \bibinfo
  {pages} {234104} (\bibinfo {year} {2015})}\BibitemShut {NoStop}%
\bibitem [{\citenamefont {Camargo}\ \emph {et~al.}(2019)\citenamefont
  {Camargo}, \citenamefont {de~la Torre}, \citenamefont {Delgado-Buscalioni},
  \citenamefont {Chejne},\ and\ \citenamefont {Español}}]{2019Camargo}%
  \BibitemOpen
  \bibfield  {author} {\bibinfo {author} {\bibfnamefont {D.}~\bibnamefont
  {Camargo}}, \bibinfo {author} {\bibfnamefont {J.~A.}\ \bibnamefont {de~la
  Torre}}, \bibinfo {author} {\bibfnamefont {R.}~\bibnamefont
  {Delgado-Buscalioni}}, \bibinfo {author} {\bibfnamefont {F.}~\bibnamefont
  {Chejne}},\ and\ \bibinfo {author} {\bibfnamefont {P.}~\bibnamefont
  {Español}},\ }\bibfield  {title} {\bibinfo {title} {Boundary conditions
  derived from a microscopic theory of hydrodynamics near solids},\ }\href
  {https://doi.org/10.1063/1.5088354} {\bibfield  {journal} {\bibinfo
  {journal} {J. Chem. Phys.}\ }\textbf {\bibinfo {volume} {150}},\ \bibinfo
  {pages} {144104} (\bibinfo {year} {2019})}\BibitemShut {NoStop}%
\bibitem [{\citenamefont {Li}\ \emph {et~al.}(2014)\citenamefont {Li},
  \citenamefont {Bian}, \citenamefont {Caswell},\ and\ \citenamefont
  {Karniadakis}}]{Li2014}%
  \BibitemOpen
  \bibfield  {author} {\bibinfo {author} {\bibfnamefont {Z.}~\bibnamefont
  {Li}}, \bibinfo {author} {\bibfnamefont {X.}~\bibnamefont {Bian}}, \bibinfo
  {author} {\bibfnamefont {B.}~\bibnamefont {Caswell}},\ and\ \bibinfo {author}
  {\bibfnamefont {G.~E.}\ \bibnamefont {Karniadakis}},\ }\bibfield  {title}
  {\bibinfo {title} {Construction of dissipative particle dynamics models for
  complex fluids via the mori–zwanzig formulation},\ }\href@noop {}
  {\bibfield  {journal} {\bibinfo  {journal} {Soft Matter}\ }\textbf {\bibinfo
  {volume} {10}},\ \bibinfo {pages} {8659} (\bibinfo {year}
  {2014})}\BibitemShut {NoStop}%
\bibitem [{\citenamefont {Grabert}(1982)}]{Grabert1982}%
  \BibitemOpen
  \bibfield  {author} {\bibinfo {author} {\bibfnamefont {H.}~\bibnamefont
  {Grabert}},\ }\href {https://doi.org/10.1007/BFb0044591} {\emph {\bibinfo
  {title} {Projection Operator Techniques in Nonequilibrium Statistical
  Mechanics}}},\ \bibinfo {series} {Springer Tracts in Modern Physics},
  Vol.~\bibinfo {volume} {95}\ (\bibinfo  {publisher} {Springer-Verlag},\
  \bibinfo {year} {1982})\BibitemShut {NoStop}%
\bibitem [{\citenamefont {Zwanzig}(2001)}]{Zwanzig2001}%
  \BibitemOpen
  \bibfield  {author} {\bibinfo {author} {\bibfnamefont {R.}~\bibnamefont
  {Zwanzig}},\ }\href@noop {} {\emph {\bibinfo {title} {Nonequilibrium
  Statistical Mechanics}}}\ (\bibinfo  {publisher} {Oxford University Press},\
  \bibinfo {year} {2001})\BibitemShut {NoStop}%
\bibitem [{\citenamefont {Mazur}\ \emph {et~al.}(1974)\citenamefont {Mazur},
  \citenamefont {Bedeaux},\ and\ \citenamefont {Mazur}}]{Mazur1974}%
  \BibitemOpen
  \bibfield  {author} {\bibinfo {author} {\bibfnamefont {P.}~\bibnamefont
  {Mazur}}, \bibinfo {author} {\bibfnamefont {D.}~\bibnamefont {Bedeaux}},\
  and\ \bibinfo {author} {\bibfnamefont {P.}~\bibnamefont {Mazur}},\ }\bibfield
   {title} {\bibinfo {title} {{A generalization of Fax{\'{e}}n's theorem to
  nonsteady motion of a sphere through an incompressible fluid in arbitrary
  flow}},\ }\href@noop {} {\bibfield  {journal} {\bibinfo  {journal} {Physica}\
  }\textbf {\bibinfo {volume} {76}},\ \bibinfo {pages} {235} (\bibinfo {year}
  {1974})}\BibitemShut {NoStop}%
\bibitem [{\citenamefont {Peskin}(2002)}]{2002_Peskin_IBM}%
  \BibitemOpen
  \bibfield  {author} {\bibinfo {author} {\bibfnamefont {C.~S.}\ \bibnamefont
  {Peskin}},\ }\bibfield  {title} {\bibinfo {title} {The immersed boundary
  method},\ }\href {https://doi.org/10.1017/S0962492902000077} {\bibfield
  {journal} {\bibinfo  {journal} {Acta Numerica}\ }\textbf {\bibinfo {volume}
  {11}},\ \bibinfo {pages} {479–517} (\bibinfo {year} {2002})}\BibitemShut
  {NoStop}%
\bibitem [{\citenamefont {Pelaez}(2024)}]{uammd}%
  \BibitemOpen
  \bibfield  {author} {\bibinfo {author} {\bibfnamefont {R.~P.}\ \bibnamefont
  {Pelaez}},\ }\href@noop {} {\bibinfo {title} {The uammd library
  repository}},\ \bibinfo {howpublished}
  {\url{https://github.com/RaulPPelaez/UAMMD}} (\bibinfo {year}
  {2024})\BibitemShut {NoStop}%
\bibitem [{\citenamefont {Vazquez-Quesada}\ \emph {et~al.}(2014)\citenamefont
  {Vazquez-Quesada}, \citenamefont {{Balboa Usabiaga}},\ and\ \citenamefont
  {Delgado-Buscalioni}}]{vazquez2014multiblob}%
  \BibitemOpen
  \bibfield  {author} {\bibinfo {author} {\bibfnamefont {A.}~\bibnamefont
  {Vazquez-Quesada}}, \bibinfo {author} {\bibfnamefont {F.}~\bibnamefont
  {{Balboa Usabiaga}}},\ and\ \bibinfo {author} {\bibfnamefont
  {R.}~\bibnamefont {Delgado-Buscalioni}},\ }\bibfield  {title} {\bibinfo
  {title} {{A multiblob approach to colloidal hydrodynamics with inherent
  lubrication}},\ }\href@noop {} {\bibfield  {journal} {\bibinfo  {journal}
  {The Journal of chemical physics}\ }\textbf {\bibinfo {volume} {141}},\
  \bibinfo {pages} {204102} (\bibinfo {year} {2014})}\BibitemShut {NoStop}%
\bibitem [{\citenamefont {Broms}(2024)}]{broms2024accuracy}%
  \BibitemOpen
  \bibfield  {author} {\bibinfo {author} {\bibfnamefont {A.}~\bibnamefont
  {Broms}},\ }\emph {\bibinfo {title} {Accuracy, efficiency and robustness for
  rigid particle simulations in Stokes flow}},\ \href@noop {} {\bibinfo {type}
  {Phd dissertation}},\ \bibinfo  {school} {KTH Royal Institute of Technology}
  (\bibinfo {year} {2024})\BibitemShut {NoStop}%
\bibitem [{\citenamefont {Broms}\ \emph {et~al.}(2023)\citenamefont {Broms},
  \citenamefont {Sandberg},\ and\ \citenamefont {Tornberg}}]{anna-broms}%
  \BibitemOpen
  \bibfield  {author} {\bibinfo {author} {\bibfnamefont {A.}~\bibnamefont
  {Broms}}, \bibinfo {author} {\bibfnamefont {M.}~\bibnamefont {Sandberg}},\
  and\ \bibinfo {author} {\bibfnamefont {A.-K.}\ \bibnamefont {Tornberg}},\
  }\bibfield  {title} {\bibinfo {title} {A locally corrected multiblob method
  with hydrodynamically matched grids for the stokes mobility problem},\
  }\href@noop {} {\bibfield  {journal} {\bibinfo  {journal} {J. Comp. Phys.}\
  }\textbf {\bibinfo {volume} {487}},\ \bibinfo {pages} {112172} (\bibinfo
  {year} {2023})}\BibitemShut {NoStop}%
\bibitem [{\citenamefont {Usabiaga}\ \emph {et~al.}(2016)\citenamefont
  {Usabiaga}, \citenamefont {Kallemov}, \citenamefont {Delmotte}, \citenamefont
  {Bhalla}, \citenamefont {Griffith},\ and\ \citenamefont
  {Donev}}]{balboa-blaise}%
  \BibitemOpen
  \bibfield  {author} {\bibinfo {author} {\bibfnamefont {F.~B.}\ \bibnamefont
  {Usabiaga}}, \bibinfo {author} {\bibfnamefont {B.}~\bibnamefont {Kallemov}},
  \bibinfo {author} {\bibfnamefont {B.}~\bibnamefont {Delmotte}}, \bibinfo
  {author} {\bibfnamefont {A.~P.~S.}\ \bibnamefont {Bhalla}}, \bibinfo {author}
  {\bibfnamefont {B.~E.}\ \bibnamefont {Griffith}},\ and\ \bibinfo {author}
  {\bibfnamefont {A.}~\bibnamefont {Donev}},\ }\bibfield  {title} {\bibinfo
  {title} {Hydrodynamics of suspensions of passive and active rigid particles:
  a rigid multiblob approach},\ }\href
  {https://doi.org/10.2140/camcos.2016.11.217} {\bibfield  {journal} {\bibinfo
  {journal} {Communications in Applied Mathematics and Computational Science}\
  }\textbf {\bibinfo {volume} {11}},\ \bibinfo {pages} {217} (\bibinfo {year}
  {2016})}\BibitemShut {NoStop}%
\bibitem [{\citenamefont {Helfrich}(1973)}]{Helfrich1973}%
  \BibitemOpen
  \bibfield  {author} {\bibinfo {author} {\bibfnamefont {W.}~\bibnamefont
  {Helfrich}},\ }\bibfield  {title} {\bibinfo {title} {Elastic properties of
  lipid bilayers: Theory and possible experiments},\ }\href
  {https://doi.org/10.1515/znc-1973-11-1209} {\bibfield  {journal} {\bibinfo
  {journal} {Zeitschrift für Naturforschung C}\ }\textbf {\bibinfo {volume}
  {28}},\ \bibinfo {pages} {693} (\bibinfo {year} {1973})}\BibitemShut
  {NoStop}%
\bibitem [{\citenamefont {Ou-Yang}\ and\ \citenamefont
  {Helfrich}(1987)}]{OuYang1987}%
  \BibitemOpen
  \bibfield  {author} {\bibinfo {author} {\bibfnamefont {Z.-C.}\ \bibnamefont
  {Ou-Yang}}\ and\ \bibinfo {author} {\bibfnamefont {W.}~\bibnamefont
  {Helfrich}},\ }\bibfield  {title} {\bibinfo {title} {Instability and
  deformation of a spherical vesicle by pressure},\ }\href
  {https://doi.org/10.1103/PhysRevLett.59.2486} {\bibfield  {journal} {\bibinfo
   {journal} {Phys. Rev. Lett.}\ }\textbf {\bibinfo {volume} {59}},\ \bibinfo
  {pages} {2486} (\bibinfo {year} {1987})}\BibitemShut {NoStop}%
\bibitem [{\citenamefont {Maxian}\ \emph {et~al.}(2021)\citenamefont {Maxian},
  \citenamefont {Peláez}, \citenamefont {Greengard},\ and\ \citenamefont
  {Donev}}]{Maxian2021}%
  \BibitemOpen
  \bibfield  {author} {\bibinfo {author} {\bibfnamefont {O.}~\bibnamefont
  {Maxian}}, \bibinfo {author} {\bibfnamefont {R.~P.}\ \bibnamefont {Peláez}},
  \bibinfo {author} {\bibfnamefont {L.}~\bibnamefont {Greengard}},\ and\
  \bibinfo {author} {\bibfnamefont {A.}~\bibnamefont {Donev}},\ }\bibfield
  {title} {\bibinfo {title} {{A fast spectral method for electrostatics in
  doubly periodic slit channels}},\ }\href@noop {} {\bibfield  {journal}
  {\bibinfo  {journal} {J. Chem. Phys.}\ }\textbf {\bibinfo {volume} {154}},\
  \bibinfo {pages} {204107} (\bibinfo {year} {2021})}\BibitemShut {NoStop}%
\bibitem [{spa()}]{sparse_matrix}%
  \BibitemOpen
  \bibinfo {title} {3. sparse matrices},\ in\ \href@noop {} {\emph {\bibinfo
  {booktitle} {Iterative Methods for Sparse Linear Systems}}},\ pp.\ \bibinfo
  {pages} {73--101}\BibitemShut {NoStop}%
\bibitem [{\citenamefont {Pratapa}\ \emph {et~al.}(2015)\citenamefont
  {Pratapa}, \citenamefont {Suryanarayana},\ and\ \citenamefont
  {Pask}}]{2015_AndersonAcc}%
  \BibitemOpen
  \bibfield  {author} {\bibinfo {author} {\bibfnamefont {P.~P.}\ \bibnamefont
  {Pratapa}}, \bibinfo {author} {\bibfnamefont {P.}~\bibnamefont
  {Suryanarayana}},\ and\ \bibinfo {author} {\bibfnamefont {J.~E.}\
  \bibnamefont {Pask}},\ }\bibfield  {title} {\bibinfo {title} {Anderson
  acceleration of the jacobi iterative method: An efficient alternative to
  krylov methods for large, sparse linear systems},\ }\bibfield  {journal}
  {\bibinfo  {journal} {J. Comp. Phys.}\ }\textbf {\bibinfo {volume} {306}},\
  \href {https://doi.org/10.1016/j.jcp.2015.11.018} {10.1016/j.jcp.2015.11.018}
  (\bibinfo {year} {2015})\BibitemShut {NoStop}%
\bibitem [{\citenamefont {Felderhof}(2005)}]{Felderhof2005}%
  \BibitemOpen
  \bibfield  {author} {\bibinfo {author} {\bibfnamefont {B.~U.}\ \bibnamefont
  {Felderhof}},\ }\bibfield  {title} {\bibinfo {title} {{Effect of the wall on
  the velocity autocorrelation function and long-time tail of Brownian
  motion}},\ }\href@noop {} {\bibfield  {journal} {\bibinfo  {journal} {J.
  Phys. Chem. B}\ }\textbf {\bibinfo {volume} {109}},\ \bibinfo {pages} {21406}
  (\bibinfo {year} {2005})}\BibitemShut {NoStop}%
\bibitem [{\citenamefont {Stein}\ \emph {et~al.}(2016)\citenamefont {Stein},
  \citenamefont {Guy},\ and\ \citenamefont {Thomases}}]{2016Stein}%
  \BibitemOpen
  \bibfield  {author} {\bibinfo {author} {\bibfnamefont {D.~B.}\ \bibnamefont
  {Stein}}, \bibinfo {author} {\bibfnamefont {R.~D.}\ \bibnamefont {Guy}},\
  and\ \bibinfo {author} {\bibfnamefont {B.}~\bibnamefont {Thomases}},\
  }\bibfield  {title} {\bibinfo {title} {Immersed boundary smooth extension: A
  high-order method for solving pde on arbitrary smooth domains using fourier
  spectral methods},\ }\href@noop {} {\bibfield  {journal} {\bibinfo  {journal}
  {J. Comp. Phys.}\ }\textbf {\bibinfo {volume} {304}},\ \bibinfo {pages} {252}
  (\bibinfo {year} {2016})}\BibitemShut {NoStop}%
\bibitem [{\citenamefont {Das}\ and\ \citenamefont
  {Pappu}(2013)}]{das2013conformations}%
  \BibitemOpen
  \bibfield  {author} {\bibinfo {author} {\bibfnamefont {R.~K.}\ \bibnamefont
  {Das}}\ and\ \bibinfo {author} {\bibfnamefont {R.~V.}\ \bibnamefont
  {Pappu}},\ }\bibfield  {title} {\bibinfo {title} {Conformations of
  intrinsically disordered proteins are influenced by linear sequence
  distributions of oppositely charged residues},\ }\href
  {https://doi.org/10.1073/pnas.1304749110} {\bibfield  {journal} {\bibinfo
  {journal} {Proceedings of the National Academy of Sciences}\ }\textbf
  {\bibinfo {volume} {110}},\ \bibinfo {pages} {13392–13397} (\bibinfo {year}
  {2013})}\BibitemShut {NoStop}%
\bibitem [{\citenamefont {Buske}\ and\ \citenamefont
  {Levin}(2023)}]{buske2023evolved}%
  \BibitemOpen
  \bibfield  {author} {\bibinfo {author} {\bibfnamefont {P.~J.}\ \bibnamefont
  {Buske}}\ and\ \bibinfo {author} {\bibfnamefont {P.~A.}\ \bibnamefont
  {Levin}},\ }\bibfield  {title} {\bibinfo {title} {Evolved sequence features
  within the intrinsically disordered tail influence ftsz assembly and
  bacterial cell division},\ }\href {https://doi.org/10.1073/pnas.2309383120}
  {\bibfield  {journal} {\bibinfo  {journal} {Proceedings of the National
  Academy of Sciences}\ }\textbf {\bibinfo {volume} {120}},\ \bibinfo {pages}
  {e2309383120} (\bibinfo {year} {2023})}\BibitemShut {NoStop}%
\bibitem [{\citenamefont {Milioni}\ \emph {et~al.}(2017)\citenamefont
  {Milioni}, \citenamefont {Tsortos}, \citenamefont {Velez},\ and\
  \citenamefont {Gizeli}}]{Milioni2017}%
  \BibitemOpen
  \bibfield  {author} {\bibinfo {author} {\bibfnamefont {D.}~\bibnamefont
  {Milioni}}, \bibinfo {author} {\bibfnamefont {A.}~\bibnamefont {Tsortos}},
  \bibinfo {author} {\bibfnamefont {M.}~\bibnamefont {Velez}},\ and\ \bibinfo
  {author} {\bibfnamefont {E.}~\bibnamefont {Gizeli}},\ }\bibfield  {title}
  {\bibinfo {title} {{Extracting the Shape and Size of Biomolecules Attached to
  a Surface as Suspended Discrete Nanoparticles}},\ }\href@noop {} {\bibfield
  {journal} {\bibinfo  {journal} {Analytical Chemistry}\ }\textbf {\bibinfo
  {volume} {89}},\ \bibinfo {pages} {4198} (\bibinfo {year}
  {2017})}\BibitemShut {NoStop}%
\bibitem [{\citenamefont {Mateos-Gil}\ \emph {et~al.}(2016)\citenamefont
  {Mateos-Gil}, \citenamefont {Vélez}, \citenamefont {Tsortos},\ and\
  \citenamefont {Gizeli}}]{mateos2016monitoring}%
  \BibitemOpen
  \bibfield  {author} {\bibinfo {author} {\bibfnamefont {P.}~\bibnamefont
  {Mateos-Gil}}, \bibinfo {author} {\bibfnamefont {M.}~\bibnamefont {Vélez}},
  \bibinfo {author} {\bibfnamefont {E.}~\bibnamefont {Tsortos}},\ and\ \bibinfo
  {author} {\bibfnamefont {E.}~\bibnamefont {Gizeli}},\ }\bibfield  {title}
  {\bibinfo {title} {Monitoring structural changes in intrinsically disordered
  proteins using qcm-d: application to the bacterial cell division protein
  zipa},\ }\href {https://doi.org/10.1039/C6CC02789A} {\bibfield  {journal}
  {\bibinfo  {journal} {Chemical Communications}\ }\textbf {\bibinfo {volume}
  {52}},\ \bibinfo {pages} {6541} (\bibinfo {year} {2016})}\BibitemShut
  {NoStop}%
\bibitem [{Note1()}]{Note1}%
  \BibitemOpen
  \bibinfo {note} {The inertia contribution $(1/3)\protect \tmspace
  +\thinmuskip {.1667em}(\protect \tilde {\alpha }a)^2$ corresponds to the
  fluid-particle induced force (particle excess mass inertia), one needs to add
  the acceleration of the displaced fluid to get inertia contribution appearing
  in the total-drag friction coefficient, $(1/9)\protect \tmspace +\thinmuskip
  {.1667em}(\protect \tilde {\alpha }a)^2$ \cite {Mazur1974}.}\BibitemShut
  {Stop}%
\bibitem [{\citenamefont {Simha}\ \emph {et~al.}(2018)\citenamefont {Simha},
  \citenamefont {Mo},\ and\ \citenamefont {Morrison}}]{morrison2018}%
  \BibitemOpen
  \bibfield  {author} {\bibinfo {author} {\bibfnamefont {A.}~\bibnamefont
  {Simha}}, \bibinfo {author} {\bibfnamefont {J.}~\bibnamefont {Mo}},\ and\
  \bibinfo {author} {\bibfnamefont {P.~J.}\ \bibnamefont {Morrison}},\
  }\bibfield  {title} {\bibinfo {title} {{Unsteady stokes flow near boundaries:
  the point-particle approximation and the method of reflections}},\
  }\href@noop {} {\bibfield  {journal} {\bibinfo  {journal} {J. Fluid Mach.}\
  }\textbf {\bibinfo {volume} {841}},\ \bibinfo {pages} {883} (\bibinfo {year}
  {2018})}\BibitemShut {NoStop}%
\bibitem [{\citenamefont {Felderhof}(2009)}]{Felderhof2009}%
  \BibitemOpen
  \bibfield  {author} {\bibinfo {author} {\bibfnamefont {B.~U.}\ \bibnamefont
  {Felderhof}},\ }\bibfield  {title} {\bibinfo {title} {{Flow of a viscous
  incompressible fluid after a sudden point impulse near a wall}},\ }\href@noop
  {} {\bibfield  {journal} {\bibinfo  {journal} {J. Fluid Mech.}\ }\textbf
  {\bibinfo {volume} {629}},\ \bibinfo {pages} {425} (\bibinfo {year}
  {2009})}\BibitemShut {NoStop}%
\bibitem [{\citenamefont {Buske}\ and\ \citenamefont
  {Levin}(2022)}]{buske2022connecting}%
  \BibitemOpen
  \bibfield  {author} {\bibinfo {author} {\bibfnamefont {P.~J.}\ \bibnamefont
  {Buske}}\ and\ \bibinfo {author} {\bibfnamefont {P.~A.}\ \bibnamefont
  {Levin}},\ }\bibfield  {title} {\bibinfo {title} {Connecting sequence
  features within the disordered c-terminal linker of *bacillus subtilis* ftsz
  to functions and bacterial cell division},\ }\href
  {https://doi.org/10.1073/pnas.2211178119} {\bibfield  {journal} {\bibinfo
  {journal} {Proceedings of the National Academy of Sciences}\ }\textbf
  {\bibinfo {volume} {119}},\ \bibinfo {pages} {e2211178119} (\bibinfo {year}
  {2022})}\BibitemShut {NoStop}%
\bibitem [{\citenamefont {Ohashi}\ \emph {et~al.}(2007)\citenamefont {Ohashi},
  \citenamefont {Galiacy}, \citenamefont {Briscoe},\ and\ \citenamefont
  {Erickson}}]{ohashi2007experimental}%
  \BibitemOpen
  \bibfield  {author} {\bibinfo {author} {\bibfnamefont {T.}~\bibnamefont
  {Ohashi}}, \bibinfo {author} {\bibfnamefont {S.~D.}\ \bibnamefont {Galiacy}},
  \bibinfo {author} {\bibfnamefont {G.}~\bibnamefont {Briscoe}},\ and\ \bibinfo
  {author} {\bibfnamefont {H.~P.}\ \bibnamefont {Erickson}},\ }\bibfield
  {title} {\bibinfo {title} {An experimental study of gfp-based fret, with
  application to intrinsically unstructured proteins},\ }\href
  {https://doi.org/10.1110/ps.072845607} {\bibfield  {journal} {\bibinfo
  {journal} {Protein Science}\ }\textbf {\bibinfo {volume} {16}},\ \bibinfo
  {pages} {1429} (\bibinfo {year} {2007})}\BibitemShut {NoStop}%
\bibitem [{\citenamefont {Cohan}\ \emph {et~al.}(2019)\citenamefont {Cohan},
  \citenamefont {Ruff},\ and\ \citenamefont {Pappu}}]{cohan2019information}%
  \BibitemOpen
  \bibfield  {author} {\bibinfo {author} {\bibfnamefont {M.~C.}\ \bibnamefont
  {Cohan}}, \bibinfo {author} {\bibfnamefont {K.~M.}\ \bibnamefont {Ruff}},\
  and\ \bibinfo {author} {\bibfnamefont {R.~V.}\ \bibnamefont {Pappu}},\
  }\bibfield  {title} {\bibinfo {title} {Information theoretic measures for
  quantifying sequence–ensemble relationships of intrinsically disordered
  proteins},\ }\href {https://doi.org/10.1093/protein/gzz014} {\bibfield
  {journal} {\bibinfo  {journal} {Protein Engineering, Design \& Selection}\
  }\textbf {\bibinfo {volume} {32}},\ \bibinfo {pages} {191} (\bibinfo {year}
  {2019})}\BibitemShut {NoStop}%
\bibitem [{\citenamefont {Tsortos}\ \emph {et~al.}(2016)\citenamefont
  {Tsortos}, \citenamefont {Papadakis},\ and\ \citenamefont
  {Gizeli}}]{tsortos2016hydrodynamic}%
  \BibitemOpen
  \bibfield  {author} {\bibinfo {author} {\bibfnamefont {A.}~\bibnamefont
  {Tsortos}}, \bibinfo {author} {\bibfnamefont {G.}~\bibnamefont {Papadakis}},\
  and\ \bibinfo {author} {\bibfnamefont {E.}~\bibnamefont {Gizeli}},\
  }\bibfield  {title} {\bibinfo {title} {{On the Hydrodynamic Nature of DNA
  Acoustic Sensing}},\ }\href@noop {} {\bibfield  {journal} {\bibinfo
  {journal} {Analytical Chemistry}\ }\textbf {\bibinfo {volume} {88}},\
  \bibinfo {pages} {6472} (\bibinfo {year} {2016})}\BibitemShut {NoStop}%
\bibitem [{\citenamefont {Sadowska}\ \emph {et~al.}(2024)\citenamefont
  {Sadowska}, \citenamefont {Nattich-Rak}, \citenamefont {Morga}, \citenamefont
  {Adamczyk}, \citenamefont {Basinska}, \citenamefont {Mickiewicz},\ and\
  \citenamefont {Gadzinowski}}]{2024sadowska}%
  \BibitemOpen
  \bibfield  {author} {\bibinfo {author} {\bibfnamefont {M.}~\bibnamefont
  {Sadowska}}, \bibinfo {author} {\bibfnamefont {M.}~\bibnamefont
  {Nattich-Rak}}, \bibinfo {author} {\bibfnamefont {M.}~\bibnamefont {Morga}},
  \bibinfo {author} {\bibfnamefont {Z.}~\bibnamefont {Adamczyk}}, \bibinfo
  {author} {\bibfnamefont {T.}~\bibnamefont {Basinska}}, \bibinfo {author}
  {\bibfnamefont {D.}~\bibnamefont {Mickiewicz}},\ and\ \bibinfo {author}
  {\bibfnamefont {M.}~\bibnamefont {Gadzinowski}},\ }\bibfield  {title}
  {\bibinfo {title} {Anisotropic particle deposition kinetics from quartz
  crystal microbalance measurements: Beyond the sphere paradigm},\ }\href@noop
  {} {\bibfield  {journal} {\bibinfo  {journal} {Langmuir}\ }\textbf {\bibinfo
  {volume} {40}},\ \bibinfo {pages} {7907} (\bibinfo {year}
  {2024})}\BibitemShut {NoStop}%
\bibitem [{\citenamefont {Anderson}(1965)}]{1965_Anderson}%
  \BibitemOpen
  \bibfield  {author} {\bibinfo {author} {\bibfnamefont {D.~G.}\ \bibnamefont
  {Anderson}},\ }\bibfield  {title} {\bibinfo {title} {Iterative procedures for
  nonlinear integral equations},\ }\href@noop {} {\bibfield  {journal}
  {\bibinfo  {journal} {J. ACM}\ }\textbf {\bibinfo {volume} {12}},\ \bibinfo
  {pages} {547–560} (\bibinfo {year} {1965})}\BibitemShut {NoStop}%
\bibitem [{\citenamefont {Diez~Martínez}\ \emph {et~al.}(2025)\citenamefont
  {Diez~Martínez}, \citenamefont {Ibáñez~Freire}, \citenamefont
  {Delgado~Buscalioni}, \citenamefont {Reguera}, \citenamefont {Bittner},\ and\
  \citenamefont {de~Pablo}}]{DiezMartinez2025}%
  \BibitemOpen
  \bibfield  {author} {\bibinfo {author} {\bibfnamefont {A.}~\bibnamefont
  {Diez~Martínez}}, \bibinfo {author} {\bibfnamefont {P.}~\bibnamefont
  {Ibáñez~Freire}}, \bibinfo {author} {\bibfnamefont {R.}~\bibnamefont
  {Delgado~Buscalioni}}, \bibinfo {author} {\bibfnamefont {D.}~\bibnamefont
  {Reguera}}, \bibinfo {author} {\bibfnamefont {A.~M.}\ \bibnamefont
  {Bittner}},\ and\ \bibinfo {author} {\bibfnamefont {P.~J.}\ \bibnamefont
  {de~Pablo}},\ }\bibfield  {title} {\bibinfo {title} {The tubular cavity of
  tobacco mosaic virus shields mechanical stress and regulates disassembly},\
  }\href@noop {} {\bibfield  {journal} {\bibinfo  {journal} {Acta
  Biomaterialia}\ }\textbf {\bibinfo {volume} {198}},\ \bibinfo {pages} {356}
  (\bibinfo {year} {2025})}\BibitemShut {NoStop}%
\bibitem [{\citenamefont {Berger}\ \emph {et~al.}(2025)\citenamefont {Berger},
  \citenamefont {Lewis}, \citenamefont {Gao}, \citenamefont {Knoops},
  \citenamefont {López-Iglesias}, \citenamefont {Peters},\ and\ \citenamefont
  {Ravelli}}]{Berger2025}%
  \BibitemOpen
  \bibfield  {author} {\bibinfo {author} {\bibfnamefont {C.}~\bibnamefont
  {Berger}}, \bibinfo {author} {\bibfnamefont {C.}~\bibnamefont {Lewis}},
  \bibinfo {author} {\bibfnamefont {Y.}~\bibnamefont {Gao}}, \bibinfo {author}
  {\bibfnamefont {K.}~\bibnamefont {Knoops}}, \bibinfo {author} {\bibfnamefont
  {C.}~\bibnamefont {López-Iglesias}}, \bibinfo {author} {\bibfnamefont
  {P.}~\bibnamefont {Peters}},\ and\ \bibinfo {author} {\bibfnamefont
  {R.}~\bibnamefont {Ravelli}},\ }\bibfield  {title} {\bibinfo {title} {In situ
  and in vitro cryo-em reveal structures of mycobacterial encapsulin assembly
  intermediates},\ }\href@noop {} {\bibfield  {journal} {\bibinfo  {journal}
  {Communications Biology}\ }\textbf {\bibinfo {volume} {8}},\ \bibinfo {pages}
  {245} (\bibinfo {year} {2025})}\BibitemShut {NoStop}%
\bibitem [{\citenamefont {Kallemov}\ \emph {et~al.}(2016)\citenamefont
  {Kallemov}, \citenamefont {Bhalla}, \citenamefont {Griffith},\ and\
  \citenamefont {Donev}}]{kallemov2016immersed}%
  \BibitemOpen
  \bibfield  {author} {\bibinfo {author} {\bibfnamefont {B.}~\bibnamefont
  {Kallemov}}, \bibinfo {author} {\bibfnamefont {A.~P.~S.}\ \bibnamefont
  {Bhalla}}, \bibinfo {author} {\bibfnamefont {B.~E.}\ \bibnamefont
  {Griffith}},\ and\ \bibinfo {author} {\bibfnamefont {A.}~\bibnamefont
  {Donev}},\ }\bibfield  {title} {\bibinfo {title} {An immersed boundary method
  for rigid bodies},\ }\href@noop {} {\bibfield  {journal} {\bibinfo  {journal}
  {Communications in Applied Mathematics and Computational Science}\ }\textbf
  {\bibinfo {volume} {11}},\ \bibinfo {pages} {79} (\bibinfo {year}
  {2016})}\BibitemShut {NoStop}%
\bibitem [{\citenamefont {Dahl}\ \emph {et~al.}(2014)\citenamefont {Dahl},
  \citenamefont {Dahl},\ and\ \citenamefont {Larsen}}]{icosphere}%
  \BibitemOpen
  \bibfield  {author} {\bibinfo {author} {\bibfnamefont {V.~A.}\ \bibnamefont
  {Dahl}}, \bibinfo {author} {\bibfnamefont {A.~B.}\ \bibnamefont {Dahl}},\
  and\ \bibinfo {author} {\bibfnamefont {R.}~\bibnamefont {Larsen}},\
  }\bibfield  {title} {\bibinfo {title} {Surface detection using round cut},\
  }in\ \href {https://doi.org/10.1109/3DV.2014.60} {\emph {\bibinfo {booktitle}
  {2014 2nd International Conference on 3D Vision}}},\ Vol.~\bibinfo {volume}
  {2}\ (\bibinfo {year} {2014})\ pp.\ \bibinfo {pages} {82--89}\BibitemShut
  {NoStop}%
\bibitem [{\citenamefont {Usabiaga}(2014)}]{florentesis}%
  \BibitemOpen
  \bibfield  {author} {\bibinfo {author} {\bibfnamefont {F.~B.}\ \bibnamefont
  {Usabiaga}},\ }\emph {\bibinfo {title} {{Minimal models for finite particles
  in fluctuating hydrodynamics}}},\ \href@noop {} {Ph.D. thesis},\ \bibinfo
  {school} {Universidad Autonoma de Madrid} (\bibinfo {year}
  {2014})\BibitemShut {NoStop}%
\bibitem [{\citenamefont {Keaveny}(2014)}]{2014_keaveny}%
  \BibitemOpen
  \bibfield  {author} {\bibinfo {author} {\bibfnamefont {E.~E.}\ \bibnamefont
  {Keaveny}},\ }\bibfield  {title} {\bibinfo {title} {Fluctuating
  force-coupling method for simulations of colloidal suspensions},\ }\href@noop
  {} {\bibfield  {journal} {\bibinfo  {journal} {J. Comp. Phys.}\ }\textbf
  {\bibinfo {volume} {269}},\ \bibinfo {pages} {61} (\bibinfo {year}
  {2014})}\BibitemShut {NoStop}%
\bibitem [{\citenamefont {Hasimoto}(1959)}]{Hasimoto}%
  \BibitemOpen
  \bibfield  {author} {\bibinfo {author} {\bibfnamefont {H.}~\bibnamefont
  {Hasimoto}},\ }\bibfield  {title} {\bibinfo {title} {On the periodic
  fundamental solutions of the stokes equations and their application to
  viscous flow past a cubic array of spheres},\ }\href
  {https://doi.org/10.1017/S0022112059000222} {\bibfield  {journal} {\bibinfo
  {journal} {J. Fluid Mech.}\ }\textbf {\bibinfo {volume} {5}},\ \bibinfo
  {pages} {317–328} (\bibinfo {year} {1959})}\BibitemShut {NoStop}%
\bibitem [{\citenamefont {MacLean}(2007)}]{icosahedron}%
  \BibitemOpen
  \bibfield  {author} {\bibinfo {author} {\bibfnamefont {K.~J.~M.}\
  \bibnamefont {MacLean}},\ }\href@noop {} {\emph {\bibinfo {title} {A
  Geometric Analysis of the Platonic Solids and Other Semi-Regular
  Polyhedra}}}\ (\bibinfo  {publisher} {Loving Healing Press},\ \bibinfo
  {address} {Ann Arbor, MI},\ \bibinfo {year} {2007})\BibitemShut {NoStop}%
\end{thebibliography}

\end{document}